\documentclass[aps,prd,reprint,superscriptaddress,showpacs,nofootinbib]{revtex4-1}

\usepackage{hyperref}
\usepackage{amsmath}
\usepackage{amssymb}
\usepackage{graphicx}
\usepackage{color}
\usepackage{subfig}
\usepackage{bm}
\usepackage{gensymb}
\usepackage{bigints}
\usepackage[usenames,dvipsnames,svgnames,table]{xcolor}
\usepackage{pifont}
\usepackage{aas_macros}
\usepackage{orcidlink}
\usepackage{physics}

\usepackage[labelfont=bf]{caption}

\hypersetup{
    colorlinks=true,
    linkcolor=red,
    filecolor=magenta,
    urlcolor=Sepia,
    citecolor=blue,
   pdftitle={Sharelatex Example},
}

\numberwithin{equation}{section}

\begin{document}


\title{Stability criterion for white dwarfs in Palatini $f(R)$ gravity}

\author{Lupamudra Sarmah\orcidlink{0000-0003-1651-9563}}
\email[E-mail: ]{lupamudrasarmah.phy20@itbhu.ac.in}
\affiliation{Department of Physics, Indian Institute of Technology (BHU), Varanasi 221005, India}
\author{Surajit Kalita\orcidlink{0000-0002-3818-6037}}
\email[E-mail: ]{surajitk@iisc.ac.in}
\affiliation{Department of Physics, Indian Institute of Science, Bangalore 560012, India}
\author{Aneta Wojnar\orcidlink{0000-0002-1545-1483}}
\thanks{Corresponding author}
\email[E-mail: ]{aneta.magdalena.wojnar@ut.ee}
\affiliation{Laboratory of Theoretical Physics, Institute of Physics, University of Tartu, W. Ostwaldi 1, Tartu 50411, Estonia}



\begin{abstract}
Recent observations of several peculiar over- and under-luminous type Ia supernovae infer indirect evidence for the violation of the Chandrasekhar mass-limit by suggesting the existence of super- and sub-Chandrasekhar limiting mass white dwarfs. In an attempt to explain these phenomena in the context of general relativistic extensions, we study these objects in Palatini $f(R)$ gravity. We obtain the super- and sub-Chandrasekhar limiting masses as well as the dynamical instability criteria for white dwarfs in the given gravitational theory. We further demonstrate that the conventional positivity condition $\pdv*{M}{\rho_\text{c}}>0$ with $M$ being the WD's mass with central density $\rho_\text{c}$, is also a valid criterion for stability in Palatini gravity.
\end{abstract}


\maketitle


\section{Introduction}\label{Introduction}

It is well known that General Relativity (GR) is by far the most adequate theory of gravitation. It successfully fulfills the shortcomings of Newtonian gravity to explain the various interesting phenomena like perihelion precession of Mercury, gravitational lensing, the physics of compact objects, and even predicts the gravitational waves~\citep{2004sgig.book.....C}. Despite its triumphs, some recent cosmological observations and corresponding theoretical arguments indicate the need for replacing GR with a more consistent theory. Quite recently, the study of observational cosmology provides the evidence that the universe has undergone two phases of acceleration; the first one is the exponential expansion of the early universe, called inflation~\citep{1981PhRvD..23..347G,2003A&G....44a..32C}, and the second phase occurs at a later time of its evolution~\citep{1998AJ....116.1009R,1999ApJ...517..565P,1999PhRvD..60h1301H}. The $\Lambda$CDM model, which is derived from GR with the cold dark matter and the cosmological constant introduced, was then adequately able to explain these observations at a large scale. Nevertheless, the model lacks an explanation for the cosmological constant problem and could not explain the small scale structures efficiently~\citep{2017Galax...5...17D}. Another questionable issue is the lack of an effective quantum theory of gravity. The two incredible theories of modern times, GR and quantum field theory, successfully dominates their domains of gravitational non-inertial systems and small-scale regimes, respectively. One of the reasons for not being able to unify these two theories is that the quantum gravitational effects become dominant only at the Planck scale due to the weak interaction of gravity. Therefore, these issues serve as a few of the motivations to go beyond GR, towards the extended theories of gravity~\citep{2011PhR...509..167C,2021arXiv210512582S}. 

Over the years, there have been proposals for a generalized and modified theory of gravity such as the scalar-tensor theory~\citep{2004cstg.book.....F,2016CQGra..33iLT01N}. A simple form of scalar-tensor theory is the $f(R)$ gravity~\cite{1970MNRAS.150....1B} with $R$ being the scalar curvature, which is based on generalizing the Lagrangian of the Einstein-Hilbert action. Instead of using an action linear in $R$ as in GR, $f(R)$ theory considers an action in which the Lagrangian density is an arbitrary function of $R$. Based on the form of $f(R)$ and the values of the model parameters, this theory has been extensively used to study various phenomena, including inflation~\citep{2020EPJP..135..576B,2018PhRvD..97f4001O,2019PhRvD..99f4049O}, dark energy problem~\citep{2012AnP...524..545C,2007IJGMM..04..115N}, gravitational waves~\citep{2019PhRvD..99l4050K,2021ApJ...909...65K}, compact objects like neutron stars~\citep{2013JCAP...12..040A,2014PhRvD..89j3509A,2014PhRvD..89f4019G,2016PhRvD..93b3501C,2016PhRvD..94f3008A}, and also the recently inferred super- and sub-Chandrasekhar limiting mass white dwarfs (WDs)~\citep{2018JCAP...09..007K,2015JCAP...05..045D,2015JCAP...01..001A,2016IJMPS..4160130A}. The so-called metric approach to $f(R)$ gravity, as studied in the works mentioned above, leads to the fourth-order field equations for the metric,\footnote{however, when transformed to the scalar-tensor representation, one deals with the second order differential equations for the metric, and additional equations for the Ricci scalar $R$, carrying dynamical properties of this object.} which can impose some practical difficulties to work with. Moreover, as shown by Chiba~\citep{2003PhLB..575....1C}, this theory might not be compatible with the solar system test if the scalar field is very light. Furthermore, the metric $f(R)$ gravity is flawed with the scalar curvature instability, which can change the gravitational field of a body greatly~\citep{2003PhLB..573....1D}.

On the other hand, there is another approach to the $f(R)$ gravity, called Palatini formalism. In this framework, the assumption on the metric dependence of the connection is waived, therefore one deals with the pair of independent objects (metric $g_{\mu\nu}$, connection $\Gamma^{\alpha}_{\beta\gamma}$)~\citep{2007PhRvD..75f3509F,2006CQGra..23.1253S}. Such an approach yields a second-order field equation that is not only compatible with the solar system test~\cite{toniato2020palatini} but also gives the correct Newtonian limit~\citep{1994CQGra..11.1505F}. Unlike the metric formalism, no such instabilities, as mentioned in~\citep{2003PhLB..573....1D}, arise in the Palatini approach. This is because the additional scalar degree of freedom arising due to the generalization of the Lagrangian density is not dynamical in nature~\citep{2007PhLB..645..389S}. In metric $f(R)$ gravity, the square of the mass of scalar mode is given by $m^2 = 1/3\left\{f'(R_0)/f''(R_0)-R_0\right\}$ with $R_0$ being the background Ricci scalar~\cite{2016PhRvD..93l4071K}. Hence, $m^2$ might be negative depending on the $f(R)$ form and values of model parameter, which gives rise to the ghost mode. In Palatini gravity, such massive dynamical scalar mode does not exist.

However, there have been some disagreements in the past regarding the Newtonian limit of Palatini $f(R)$ gravity. According to Meng and Wang~\citep{2004GReGr..36.1947M}, the correct Newtonian limit is always achieved in those models where the action contains inverse powers of $R$ and the weak field expansion yields a de-Sitter vacuum solution. However, it was shown that a $f(R)$ theory with a pole of order $n$ in $R=0$ and $f''(R_0)\neq0$ does not give a good Newtonian limit~\citep{2004PhRvD..70d3505D}. They further proposed that those $f(R)$ theories with a singular $f(R)$, satisfying the condition $f''(R_0)=0$, are worthwhile to study. These disagreements were settled in~\citep{2006GReGr..38.1407S}, where it was shown that the Palatini gravity models with negative powers of $R$, as well as their generalizations those include the positive powers, give rise to correct Newtonian limit, provided the coefficients of these powers are sufficiently small. Since the Palatini approach is more general and comparatively easier to work with, several phenomena are being studied using this model recently. Cosmological theories based on the modified gravity using Palatini variational principle has been used to study inflation and cosmic acceleration~\citep{2006PhRvD..73f3515S,2004GReGr..36.1765N,2007PhRvD..75f3509F,2006A&A...454..707A,2016EPJC...76..567S,2016JCAP...01..040B,2020JCAP...07..003B,2020PhRvD.102d4029J,2021PhRvD.104b3521G}. Similarly, Palatani $f(R)$ gravity has been used to study neutron stars and the alterations in the maximum mass-limit of WDs~\citep{2021EPJC...81..888H,2021IJGMM..1840006W,2017JCAP...10..004B}. 

Depending on the battle between self-gravity and thermodynamics, the end state of a star can either lead to the formation of a compact object like WD, neutron star, black hole, or it may lead to an explosion dispersing all matter into space and leaving behind nothing. WDs are compact stellar remnants, supported by the electron degeneracy pressure~\citep{1986bhwd.book.....S}. The end state of a progenitor star with mass $(10\pm2) M_\odot$ is a WD~\citep{2018MNRAS.480.1547L}. In WDs, the outward electron degeneracy pressure balances the inward gravitational force, which arises due to the Pauli exclusion principle, and thereby it maintains a stable equilibrium condition. However, beyond a certain mass, the electron degeneracy pressure is no longer sufficient to stop the star from collapsing under gravity. Taking into account relativistic effects into the degenerate electron equation of state (EoS), Chandrasekhar made the remarkable discovery that the mass of a non-rotating and non-magnetized WD cannot exceed approximately $1.44M_\odot$~\cite{1935MNRAS..95..207C}. This is known as the Chandrasekhar mass-limit. If a WD in a binary system slowly accretes matter from the companion such that its mass is over this mass-limit, the pressure balance no longer sustains. In such a situation, the WD explodes releasing a tremendous amount of energy in the form of a type Ia supernova (SN\,Ia)~\citep{1997Sci...276.1378N}. Due to this fixed critical mass, the peak luminosities of SNe\,Ia are consistent, and thus they are often used as a standard candle~\citep{2018PhRvD..97h3505W}. However, several over-luminous~\citep{2006Natur.443..308H,2010ApJ...713.1073S,2009ApJ...707L.118Y,2011MNRAS.410..585S} and under-luminous~\citep{1992AJ....104.1543F,1998AJ....116.2431T,2001PASP..113..308M,2008MNRAS.385...75T} SNe\,Ia have been observed lately, which are proposed to be originated from super- and sub-Chandrasekhar mass WDs, respectively. This suggests that the Chandrasekhar mass-limit of WDs may not be unique. Over the years, it has been extensively studied on the grounds of modified gravity~\citep{2015JCAP...05..045D,2018JCAP...09..007K,2015IJMPD..2444026D,2021IJGMM..1840006W}. Earlier Mukhopadhyay and collaborators used the metric formalism of $f(R)$ gravity~\citep{2018JCAP...09..007K,2015JCAP...05..045D}, and by choosing suitable values of the model parameter, they were able to obtain super- as well as sub-Chandrasekhar limiting mass WDs. On the other hand, some of us obtained the modified hydrostatic equilibrium equations for polytropic WDs using Palatini $f(R)$ gravity in the Newtonian limit~\citep{2021IJGMM..1840006W}, both in the Einstein and Jordan frames, and studied the mass-limit of WDs for various model parameters. In addition, other modified gravity theories like the $f(R,T)$ gravity theory has been used to study the equilibrium configuration as well as the physical properties of WDs~\citep{2017EPJC...77..871C,2021AIPC.2320e0029U}. This model was again able to provide an explanation to the violation of the Chandrasekhar mass-limit in WDs.

In this paper, we consider the Newtonian limit of Palatini $f(R)$ gravity to study the mass--radius relation of the WDs and their corresponding stability analysis in the case of degenerate EoS for electrons. We also check whether the standard stability criterion is valid for the considered model of gravity. 

This paper is organized as follows. In~\S\ref{Sec: 2} and~\S\ref{Sec: 3}, we recall the formalism of Palatini $f(R)$ gravity in the Einstein as well as the Jordan frames, and the hydrostatic equilibrium equations for the WDs in the Newtonian regime in both these frames, respectively. We also discuss the modified equations for radial oscillations in order to examine the stability of the WDs in this gravity model. In~\S\ref{Sec:4}, we present a discussion on the numerical results concerning the mass--radius relations and stability analysis of the modified gravity induced WDs. We conclude our work in~\S\ref{Sec:5} while in Appendix~\S\ref{appen}, we recall the relativistic stellar equations.

\section{Palatini $f(R)$ gravity}\label{Sec: 2}

Let us begin by briefly describing Palatini $f(R)$ gravity and the corresponding field equations. As already mentioned, this formalism considers the metric and the connection to be independent of each other. Therefore, to obtain the field equations, the action must be varied with respect to both variables. Even though the form of the action resembles $f(R)$ gravity in the metric formalism, the Riemann and Ricci tensors no longer depend on the metric; instead, they are constructed with the independent connection. More specifically, one can denote $R_{\mu\nu}\equiv R_{\mu\nu}(\Gamma)$, where $\Gamma$ represents the connection. We now recall the main properties of the so-called {\it game of frames} \cite{2021PhRvD.103b4022S} and the current interpretation, and therefore, for the reader's convenience, we provide the most relevant equations in both cases.

\subsection{Jordan frame formulation}

The generalized action in Palatini $f(R)$ gravity is given by
\begin{equation}\label{1}
S=\frac{1}{2\kappa^2}\int\sqrt{-g}f(R)\dd[4]{x} + S_m(g_{\mu\nu},\psi),
\end{equation}
where $\kappa^2=-8\pi G/c^4$, $g=\text{det}(g_{\mu\nu})$, and $S_m$ is the matter action which depends on the metric $g_{\mu\nu}$ and matter field $\psi$, and is independent of the connection. The Ricci scalar appearing in Equation~\eqref{1} is built of two structures, $g_{\mu\nu}$ and $\Gamma$, that is, $R=g^{\mu\nu}R_{\mu\nu}(\Gamma)$.

Varying Equation~\eqref{1} with respect to $g_{\mu\nu}$ gives the following modified field equations~\citep{2010LRR....13....3D}
\begin{equation}\label{2}
f'(R)R_{\mu\nu}-\frac{1}{2}f(R)g_{\mu\nu}=\kappa^2T_{\mu\nu},
\end{equation}
where $f'(R) = \dv*{f(R)}{R}$, while $T_{\mu\nu}$ is the energy-momentum tensor, given by
$$T_{\mu\nu}=-\frac{2}{\sqrt{-g}}\frac{\delta S_m}{\delta g^{\mu\nu}},$$
which is further assumed to have the perfect-fluid form. On the other hand, the result of variation with respect to $\Gamma$ can be written in the following form:
\begin{equation}\label{3}
\nabla_{\lambda}\left(\sqrt{-g}f'(R)g^{\mu\nu}\right)=0,
\end{equation}
where $\nabla_{\lambda}$ is the covariant derivative ruled by $\Gamma$.
Defining a new metric tensor
$ \bar g_{\mu\nu} = f'(R)g_{\mu\nu}$
allows to rewrite Equation~\eqref{3} as
\begin{equation}
\nabla_\lambda(\sqrt{-\bar g}\bar g^{\mu\nu})=0,\label{con2}
\end{equation}
providing that the connection $\Gamma$ is Levi-Civita with respect to $\bar g_{\mu\nu}$. It results that the independent connection is an auxiliary field which can be integrated out. Therefore, all physical degrees of freedom are given by the metric tensor $g$. It is quite evident that one may obtain the standard GR equations by choosing the linear $f(R)$. Then it will turn out from Equation~\eqref{3} that $\Gamma$ is the Levi-Civita connection of the metric $g$.

In order to present some useful interpretation of the field equations, taking the trace of Equation~\eqref{2} with respect to $g$, we obtain
\begin{equation}\label{4}
f'(R)R-2f(R)= \kappa^2 g^{\mu\nu}T_{\mu\nu} = \kappa^2 T,
\end{equation}
which provides the structural equation ($T$ is the trace of the energy-momentum tensor). It is worthwhile to note that, unlike metric formalism, no kinetic term, such as $\Box f'(R)$, arises in Equation~\eqref{4}. This ensures that the oscillatory mode appearing in the metric formalism does not exist in the Palatini approach. In case of vacuum or pure radiation ($T=0$), the theory reduces to the Einstein vacuum solution with the cosmological constant, independently of the $f(R)$ form \citep{1994CQGra..11.1505F}.

In this paper, we will work with the simplest extension of the GR - that is, with the Starobinsky model~\citep{1980PhLB...91...99S}, given by
\begin{equation}\label{quadratic}
    f(R)=R+\alpha R^2,
\end{equation}
where $\alpha R^2$ is the higher-order correction to the GR with $\alpha$ being the model parameter. When its sign is specified, it will provide both the regimes of mass-limit in WDs. 

\subsection{Scalar-tensor representation and Einstein frame}

Let us now briefly describe the Palatini gravity in the Einstein frame. Firstly we use the fact that the theory possesses a scalar-tensor representation, however the scalar field appearing there, as already discussed, does not carries any extra degree of freedom \cite{2019EPJC...79..335K,2018PhRvD..97b1503A}. Provided $f''(R)\neq 0$\footnote{the linear Lagrangian is excluded in that case.}, the action in Equation~\eqref{1} can be rewritten in a mathematically equivalent form, given by~\citep{2010RvMP...82..451S,2017EPJC...77..603S,2017EPJC...77..406S}
\begin{align}\label{6}
S(g_{\mu\nu},\Gamma^\lambda_{\rho\sigma},\chi)&=\frac{1}{2\kappa^2}\int \dd[4]{x} \sqrt{-g}\left[f'(\chi)(R-\chi) + f(\chi)\right] \nonumber \\&+ S_m(g_{\mu\nu},\psi),
\end{align}
where $\chi$ is a new scalar field. Redefining it via $\Phi=f'(\chi)$ with the constraint $\chi=R$, one may rewrite action~\eqref{6} into the form of Palatini-Brans-Dicke gravity as
\cite{2017EPJC...77..406S} 
\begin{equation}\label{7}
S(g_{\mu\nu},\Gamma^\lambda_{\rho\sigma},\Phi)=\frac{1}{2\kappa^2}\int \dd[4]{x} \sqrt{-g}\left[\Phi R-U(\Phi)\right]+ S_m(g_{\mu\nu},\psi),
\end{equation}
where $U(\Phi)=\chi(\Phi)\Phi-f(\chi(\Phi))$. Performing now the conformal transformation of the metric $g$, one writes the action in the Einstein frame as
\begin{equation}\label{11}
S(\Bar{g}_{\mu\nu},\Phi)=\frac{1}{2\kappa^2}\int\dd[4]{x}\sqrt{-\Bar g}\left[\Bar{R}-\Bar{U}(\Phi)\right]+S_m(\Phi^{-1}\Bar{g}_{\mu\nu},\psi),
\end{equation}
for which the field equations are obtained by variation with respect to $\bar g$ and $\Phi$
\begin{align}
\Bar{R}_{\mu\nu}-\frac{1}{2}\Bar{g}_{\mu\nu}\Bar{R}&=\Bar{T}_{\mu\nu}-\frac{1}{2}\Bar{g}_{\mu\nu}\Bar{U}(\Phi)\label{8},\\
0&=\Phi\Bar{R}-\left(\Phi^2\Bar{U}(\Phi)\right)',
\end{align}
where prime denotes here the derivative with respect to $\Phi$. Moreover, it can be shown that the following relations are true: for the conformal metric
 $\Bar{g}_{\mu\nu}=g_{\mu\nu}\Phi$, one is equipped with $\Bar{R}_{\mu\nu}=R_{\mu\nu}$, $\Bar{R}=\Bar{g}^{\mu\nu}\Bar{R}_{\mu\nu}=\Phi^{-1}R$, $\Bar{g}_{\mu\nu}\Bar{R}=g_{\mu\nu}R$, $\Bar{U}(\Phi)=U(\Phi)/\Phi^2$, and $\Bar{T}_{\mu\nu}=\Phi^{-1}T_{\mu\nu}$. 
 The structural equation in this case is given by
\begin{equation}\label{10}
\Phi\Bar{U}'(\Phi)+\Bar{T}=0,
\end{equation}
where $\Bar{T}=\Bar{g}^{\mu\nu}\Bar{T}_{\mu\nu}$ and again it demonstrates the non-dynamical aspect of $\Phi$ (recall that $\Phi(R(\Gamma))$). This representation is useful as it allows to study particular physical problems represented by simpler equations, whose solutions afterwards can easily be transformed to the physical Jordan frame \cite{2018PhRvD..97b1503A,2019PhRvD..99d4040A}. 

\section{Stellar structure equations and corresponding stability analysis}\label{Sec: 3}
In this section, we will recall the hydrostatic equilibrium equations in the Newtonian regime, given in the Einstein and Jordan frames. The full relativistic equations used here were obtained in~\cite{2018EPJC...78..421W}, and for the reader's convenience, we recall them in the Appendix~\S\ref{appen}. Thereafter, we derive the modified radial oscillation equations for the stability analysis of the Palatini $f(R)$ gravity induced WDs. 

\subsection{Stellar structure equations}\label{subsec 3.1}

The Newtonian limit approximation is applicable to systems exhibiting weak gravitation and slowly varying or static gravitational field. In order to study WDs in Newtonian regime, one considers $p(r)\ll\rho(r)$, $4\pi r^3 p(r)\ll M(r)$ and $2GM(r)/r\ll 1$, where $p(r)$ is the pressure, $\rho(r)$ is the density, and $M(r)$ is the mass of the object at a radius $r$. Therefore, in the Einstein frame, from Equations~\eqref{3.5} and~\eqref{3.6}, the pressure-balance and mass-estimate equations are given by~\citep{2019EPJC...79...51W}
\begin{equation}\label{3.12}
    \dv{p}{\Tilde{r}}=-\frac{GM\rho}{\Phi\Tilde{r}^2}
\end{equation}
and
\begin{equation}\label{3.13}
    \dv{M}{\Tilde{r}}=4\pi \Tilde{r}^2\rho.
\end{equation}
Let us comment that $p$ and $\rho$ are the physical quantities, while only $\tilde r$ indicates that we are dealing with the Einstein frame's radial coordinate (see detailed discussion in~\citep{2015JCAP...10..040M,2019EPJC...79...51W,2020EPJC...80..313S}). The $p$ and $\rho$ in the case of our WDs model are related to each other by the Chandrasekhar EoS for degenerate electrons, given by~\cite{1935MNRAS..95..207C}
\begin{equation}\label{Chandrasekhar EoS}
\begin{aligned}
p &= \frac{\pi m_\text{e}^4 c^5}{3 h^3}\left[x_\text{F}\left(2x_\text{F}^2-3\right)\sqrt{x_\text{F}^2+1}+3\sinh^{-1}x_\text{F}\right],\\
\rho &= \frac{8\pi \mu_\text{e} m_\text{H}(m_\text{e}c)^3}{3h^3}x_\text{F}^3,
\end{aligned}
\end{equation}  
where $x_\text{F} = p_\text{F}/m_\text{e}c$, $p_\text{F}$ is the Fermi momentum, $m_\text{e}$ is the mass of electron, $h$ is the Planck's constant, $\mu_\text{e}$ is the mean molecular weight per electron and $m_\text{H}$ is the mass of hydrogen atom. For our work, we choose $\mu_\text{e}=2$ indicating the carbon-oxygen WD.
Moreover, in the Newtonian limit, $T\approx-\rho c^2$. Therefore, from Equation~\eqref{quadratic}, $\Phi=f'(R(T))$ becomes
\begin{equation}\label{3.15}
    \Phi=1+2\alpha\kappa^2 c^2\rho.
\end{equation}
Thus, Equations~\eqref{3.12} and~\eqref{3.13} are the hydrostatic balance equations for WDs in the Newtonian limit in the Einstein frame.

One can also obtain the corresponding hydrostatic balance equations in the Jordan frame, through the conformal transformation, $\Tilde{r}^2=\Phi r^2$, which are given by~\cite{2021IJGMM..1840006W}
\begin{equation}\label{3.16}
    \dv{p}{r}=-\frac{GM\rho}{\Phi^{\frac{3}{2}} r^2}\left(1+\frac{1}{2}r\frac{\Phi'}{\Phi}\right)
\end{equation}
and
\begin{equation}\label{3.17}
    \dv{M}{r}=4\pi r^2\rho \Phi^{\frac{3}{2}}\left(1+\frac{1}{2}r\frac{\Phi'}{\Phi}\right).
\end{equation}
These are the Newtonian hydrostatic equilibrium equations in the Jordan frame. Here, prime (`$'$') is the derivative with respect to the coordinate $r$ while $\Phi'$ can be obtained from Equation~\eqref{3.15}.

\subsection{Modified equations for radial oscillations}\label{subsec 3.2}

In the Newtonian gravity, a non-rotating, non-magnetized finite temperature star, whose matter content is given by polytrope (with the polytropic parameter $\gamma$), is unstable against adiabatic radial perturbations if $\gamma<4/3$. Later, Chandrasekhar showed that due to the strong gravity in the framework of GR, instability may arise at a larger value of $\gamma$~\citep{1964ApJ...140..417C}. Similarly, stability analysis for compact stars has been done in the framework of modified gravity theory~\citep{2021JCAP...04..064M,2020JCAP...11..048P}, demonstrating that this critical value of $\gamma$ differs, and also depends on the model's parameter~\cite{2020arXiv200100388W} in the case of non-relativistic regime. In more general context, it was also shown that in the case of a relativistic star, one deals with the similar to GR stability condition, that is, it depends on an EoS, but also on the $f(R)$ model in Palatini gravity~\cite{2018EPJC...78..421W}.

Let us now discuss in details the stability problem. A star is stable under radial perturbations if the frequencies of normal modes are real. However, Pretel et al. considered the same stability equations for GR and solved them to examine the stability of neutron stars in modified gravity~\citep{2020JCAP...11..048P}. Moreover, in GR, $\pdv*{M}{\rho_\text{c}}>0$ provides the necessary stability criterion, where $M$ is the mass of and $\rho_\text{c}$ is the central density of the star. In this work, we will also examine whether this criterion is valid for Palatani $f(R)$ gravity with the given EoS~\eqref{Chandrasekhar EoS}. To do so, one needs to derive and solve the modified equations for radial oscillations. Since we are interested in the WDs in the Newtonian regime, we do not take the relativistic effects into account.

When a non-rotating, spherically symmetric star in hydrostatic and thermal equilibrium is subjected to a small radial perturbation, it will cause oscillations in the radial direction such that a fluid element located at $r_0$ will be displaced to $r_0+\delta r(t,r_0)$ maintaining its spherical symmetry. Here, $\delta r$ is the Langrangian perturbation of the WD's radius. In our discussion of stability analysis, the radial oscillations are assumed to be adiabatic in nature, such that any heat exchange mechanism is ignored~\citep{2004sipp.book.....H,1986bhwd.book.....S}. Even though such an adiabatic approximation significantly simplifies the analysis and gives accurate values of amplitude within the star, it does not provide any information about the thermodynamics of the star. 

The radial oscillation equations can be derived using the Eulerian as well as the Langrangian formalisms. The Eulerian perturbation $(\Delta f)$ and the Langrangian perturbation $(\delta f)$ of a variable $f$ are related by
\begin{equation}\label{3.18}
\delta f = \Delta f + \dv{f_0}{r}\delta r.
\end{equation}
In general, Langrangian formalism is convenient while dealing with the systems bearing one degree of freedom. Since we assume spherically symmetric WDs in this work, we derive the equations for radial oscillations utilizing the Langrangian formalism. Considering small oscillations about the equilibrium position, the perturbed radius $\Tilde{r}(t,\Tilde{r}_0)$, density $\rho(t,\Tilde{r}_0)$, and pressure $p(t,\Tilde{r}_0)$ are given by
\begin{align}\label{3.19}
    \Tilde{r}(t,\Tilde{r}_0)&=\Tilde{r}_0\left[1+\frac{\delta \Tilde{r}(t,\Tilde{r}_0)}{\Tilde{r}_0}\right],\\
\label{3.20}
\rho(t,\Tilde{r}_0)&=\rho_0\left[1+\frac{\delta\rho(t,\Tilde{r}_0)}{\rho_0}\right],\\
\label{3.21}
p(t,\Tilde{r}_0)&=p_0\left[1+\frac{\delta p(t,\Tilde{r}_0)}{p_0}\right],
\end{align}
where $\delta \Tilde{r}(t,\Tilde{r}_0)$, $\delta\rho(t,\Tilde{r}_0)$, and $\delta p(t,\Tilde{r}_0)$ are the Langrangian perturbation in the radius, density, and pressure, respectively. The subscript zero in the above relations denote the quantities in the static state. Moreover, we assume the perturbations to be small enough such that, $|\delta \Tilde{r}/\Tilde{r}_0|\ll 1$, $|\delta\rho/\rho_0|\ll 1$, and $|\delta p/p_0|\ll 1$, and hence we can apply the linear theory by preserving only the linear terms, neglecting the higher-order ones. Since the adiabatic approximation is assumed, the mechanical structure of the star can be described by the mass--radius relation. Now, the mass conservation and the conservation of momentum equation are given by
\begin{equation}\label{3.22}
\pdv{M}{\Tilde{r}}=4\pi\Tilde{r}^2\rho
\end{equation}
and
\begin{equation}\label{3.23}
    \rho\dv{\Vec{\Tilde{v}}}{t}=-\left(\Tilde{\grad}p+\rho\Tilde{\grad}\psi\right),
\end{equation}
where $\Vec{\Tilde{v}}$ is the fluid velocity and $\psi$ is the gravitation potential such that, $\Tilde{\grad}p=-\rho\Tilde{\grad}\psi$ in equilibrium. Note that the `Tilde' in the above equations denote the quantities in the Einstein frame and $\dv*{t}$ is given by 
\begin{equation*}
    \dv{t}\equiv\pdv{t}+\left(\Vec{v}\vdot\Tilde{\grad}\right).
\end{equation*}
Since the quantities are now functions of both $r_0$ and time $t$, we explicitly introduce partial derivatives. Perturbing Equations~\eqref{3.22} and~\eqref{3.23} and replacing $\Tilde{r}$, $\rho$, $p$ with their perturbed values from Equations~\eqref{3.19}--\eqref{3.21}, we obtain
\begin{equation}\label{3.24}
\Tilde{r_0}\pdv{\left(\delta\Tilde{r}/\Tilde{r}_0\right)}{\Tilde{r}_0}=-\left(3\frac{\delta\Tilde{r}}{\Tilde{r}_0}+\frac{\delta\rho}{\rho_0}\right)
\end{equation}
and
\begin{equation}\label{3.25}
\Tilde{r}_0\rho_0\Ddot{\left(\frac{\delta\Tilde{r}_0}{\Tilde{r}_0}\right)}=-\pdv{p_0}{\Tilde{r}_0}\left(\frac{\delta p}{p_0}+4\frac{\delta\Tilde{r}}{\Tilde{r}_0}+\frac{\delta\Phi}{\Phi_0}\right)-p_0\pdv{\left(\delta p/p_0\right)}{\Tilde{r}_0},
\end{equation}
where $\delta\Phi$ is the Langrangian perturbation in $\Phi$. Let us consider that the perturbations behave as a plane-wave, such that a quantity $f$ can be written as
\begin{equation}\label{3.26}
    \frac{\delta f(t,\Tilde{r}_0)}{f_0}=\frac{\delta f(\Tilde{r}_0)}{f_0}e^{i\sigma t},
\end{equation}
where $\sigma$ is the characteristic frequency. Accordingly, Equations~\eqref{3.24} and~\eqref{3.25} become
\begin{equation}\label{3.27}
\dv{\Tilde{\zeta}}{\Tilde{r}}=-\frac{1}{\Tilde{r}}\left(3\Tilde{\zeta}+\frac{1}{\Gamma}\frac{\delta p}{p}\right)
\end{equation}
and
\begin{equation}\label{3.28}
\dv{\left(\delta p/p\right)}{\Tilde{r}}=-\frac{1}{p}\dv{p}{\Tilde{r}}\left(\frac{\delta p}{p}+4\Tilde{\zeta}+\frac{\delta\Phi}{\Phi}+\frac{\sigma^2\Tilde{r}^3\Phi\Tilde{\zeta}}{GM}\right),
\end{equation}
where
\begin{equation}\label{3.29}
    \frac{\delta\Phi}{\Phi}=\frac{2\kappa^2 c^2\alpha\rho}{\Phi\Gamma}\left(\frac{\delta p}{p}\right).
\end{equation}
Here $\Tilde{\zeta}=\delta\Tilde{r}(\Tilde{r}_0)/\Tilde{r}_0$ and  $\Gamma=(\pdv*{\ln p}{\ln \rho})$. We have introduced ordinary spatial derivatives in Equations~\eqref{3.27} and~\eqref{3.28} because of the $\Tilde{r}$-dependent variables and we have also skipped the subscripts zero for simplicity as all the quantities appearing are in their static configuration. Thus, Equations~\eqref{3.27} and~\eqref{3.28} are the two linear, first-order, time-independent, coupled differential equations governing the radial oscillation in Palatani $f(R)=R+\alpha R^2$ gravity in the Newtonian regime. It is quite evident that the above equations reduce to the Newtonian radial oscillation equations if $\Phi=1$.

We also obtain the corresponding modified equations for radial oscillations in the Jordan frame by following the conformal transformation $\Tilde{r}^2=\Phi r^2$. Since $\Tilde{\zeta}$ is related to $\zeta$ from the Jordan frame by the relation
\begin{equation}\label{3.31}
    \Tilde{\zeta}=\zeta+\frac{1}{2}\frac{\delta\Phi}{\Phi},
\end{equation}
the corresponding radial oscillation equations are 
\begin{align}\label{3.32}
    \dv{\zeta}{r}+\frac{1}{2}\dv{\left(\delta\Phi/\Phi\right)}{r}=-\frac{1}{r}\left(1+\frac{r}{2}\frac{\Phi'}{\Phi}\right)\left[3\left(\zeta+\frac{1}{2}\frac{\delta\Phi}{\Phi}\right) +\frac{1}{\Gamma}\frac{\delta p}{p}\right]
\end{align}
and
\begin{equation}\label{3.33}
\dv{\left(\delta p/p\right)}{r}=-\frac{1}{p}\dv{p}{r}\left[\frac{\delta p}{p}+4\zeta+3\frac{\delta\Phi}{\Phi}+\frac{\sigma^2 r^3\Phi^{\frac{5}{2}}}{GM}\left(\zeta+\frac{1}{2}\frac{\delta\Phi}{\Phi}\right)\right].
\end{equation}
These two equations must be solved simultaneously with the appropriate boundary conditions in order to determine the frequencies of normal modes. Since we are dealing with metric theory, that is, the independent connection is not coupled to the matter fields in Equation~\eqref{1}, the particles are moving along the connection given by $g$; hence the physical variables are given in the Jordan frame only. However, we demonstrate in the next section that the solutions in both frames do not quantitatively differ because of slight modifications introduced by the theory.


\section{Results and discussion}\label{Sec:4}
In this section, we present the results obtained by solving the stellar structure equations together with the modified radial oscillation equations derived in the previous section. Our numerical solutions have a form of the mass--radius relation of the WDs and their corresponding stability analysis. Although the Jordan frame is the physical one, we present our results in both frames, demonstrating similar behavior of the curves.

\subsection{Mass--radius relations in Palatini $f(R)$ gravity}\label{subsec:4,1}
In order to obtain the interior solution of the WD, we numerically solve the stellar structure equations~\eqref{3.12} and~\eqref{3.13} for the Einstein frame, while Equations~\eqref{3.16} and~\eqref{3.17} for the Jordan frame along with the Chandrasekhar EoS, given by Equation~\eqref{Chandrasekhar EoS}. The boundary conditions used at the center of the WD are $M(r=0)=0$ and $\rho(r=0)=\rho_\text{c}$ and on the surface $\rho(r=\mathcal{R})=0$ with $\mathcal{R}$ being the radius of the WD. In order to avoid any violation of the conventional physical laws, $\alpha$ is chosen in such a way that it is well within the bound given by~\cite{2010PhRvD..81j4003N} (see also \cite{2021CQGra..38s4003M}), that is, $|\alpha| \lesssim 5\times 10^{15}\rm\,cm^2$. Let us notice that in the case of neutron stars (since the curvature is higher in high density regime), the bound should be reduced to $\sim 10^{12}\rm\,cm^2$~\cite{2005PhRvL..95z1102O}, while when electric forces taken into account, to $\sim 10^{9}\rm\,cm^2$ in the case of Palatini theories~\cite{2012JCAP...11..022A,2018PhR...727....1B}.

\begin{figure}[htpb]
	\centering
	\includegraphics[scale=0.42]{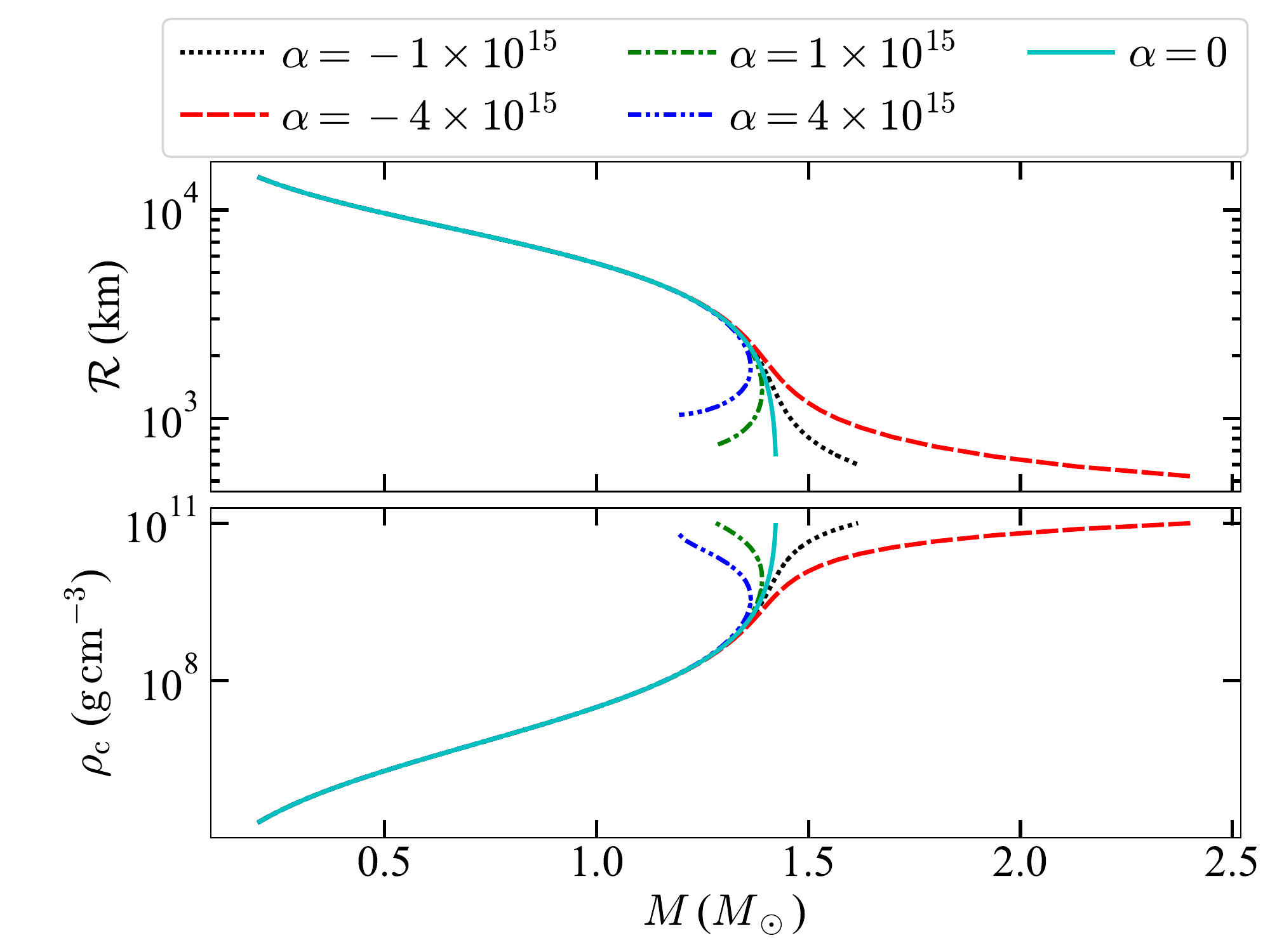}
	\caption{Upper panel: mass--radius relation, Lower panel: variation of mass with respect to $\rho_\text{c}$ for WDs for different values of $\alpha$ in the Jordan frame in Palatini $f(R)$ gravity. In the label, the values of $\alpha$ are shown in cm$^2$ unit.}
	\label{Fig: Palatini_Jordan}
\end{figure}

\begin{figure}[htpb]
	\centering
	\includegraphics[scale=0.42]{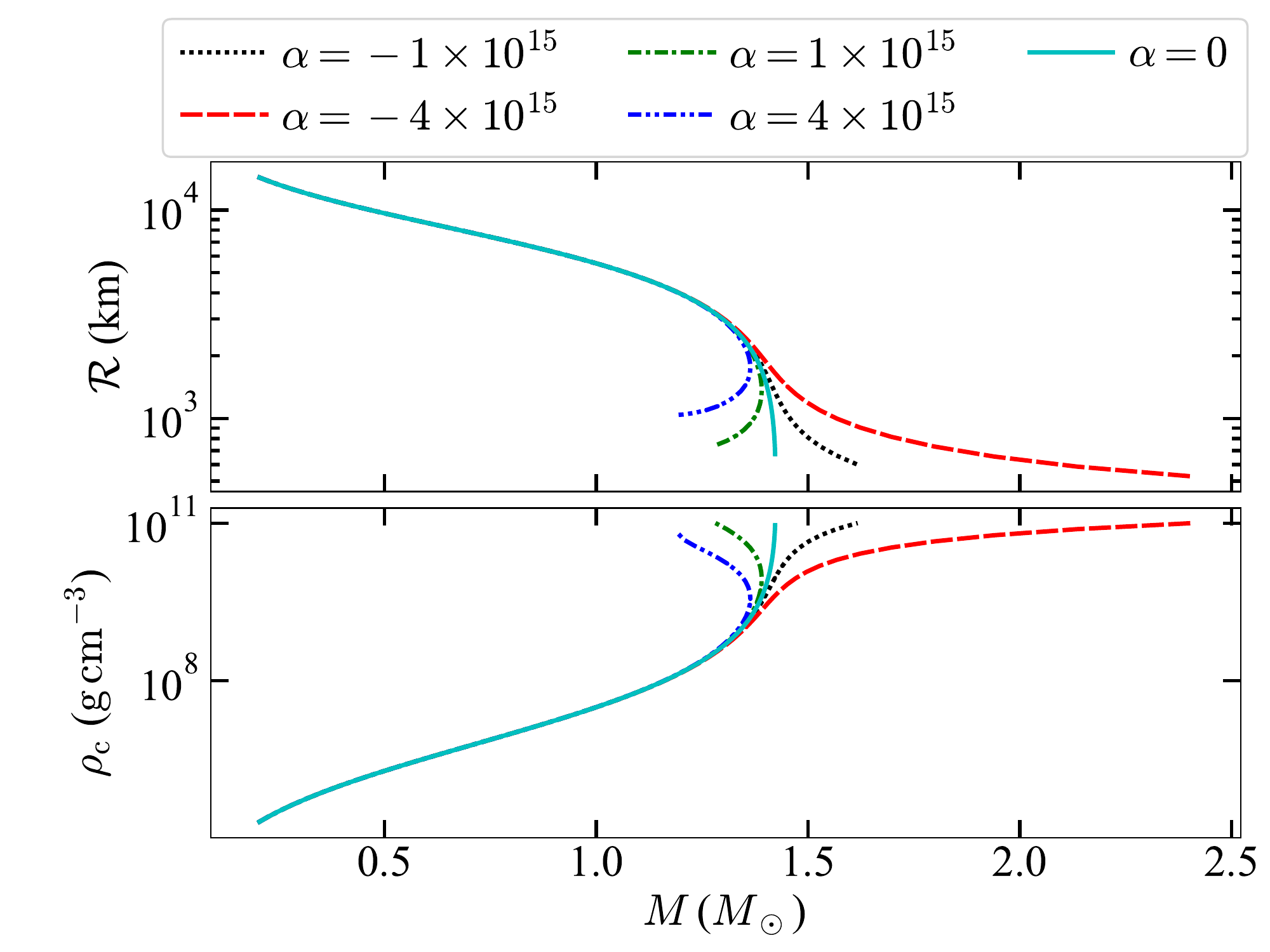}
	\caption{Same as Fig.~\ref{Fig: Palatini_Jordan} except here the Einstein frame is considered.}
	\label{Fig: Palatini_Einstein}
\end{figure}
Figure~\ref{Fig: Palatini_Jordan} illustrates the variation of $\mathcal{R}$ and $\rho_\text{c}$ with respect to $M$ of WDs for different values of $\alpha$ in the Einstein frame. The different mass--radius curves indicate that the interior structure of the WDs with high density gets modified due to the influence of Palatini $f(R)=R+\alpha R^2$ gravity. We know that the Starobinsky model of $f(R)$ gravity reduces to GR on choosing $\alpha=0$. This is also evident from Figure~\ref{Fig: Palatini_Jordan}, where the mass--radius curve corresponding to $\alpha=0$ mimics the Newtonian case, with a mass-limit of about $1.44M_\odot$. Moreover, all the curves merge at low densities, indicating that the effect of modified gravity is negligible in this regime. This is because $R$ is nearly proportional to the density, and hence, in the low-density regime, $R^2$ or any other higher-order corrections do not contribute significantly. However, as $\rho_\text{c}$ increases beyond $10^9\rm\,g\,cm^{-3}$, the curves deviate from the Newtonian case due to the increased contribution of the $\alpha R^2$ term; thereby showing the effect of modified gravity on the mass--radius relation of WDs at the high-density regime. For the case of $\alpha<0$, the curves follow the usual trend of increasing mass with increasing $\rho_\text{c}$ and overshoot the Chandrasekhar mass-limit, thus indicating super-Chandrasekhar WDs. As evident from Figure~\ref{Fig: Palatini_Jordan}, the mass increases with increasing $\alpha$ and can go beyond $2M_\odot$ for high $\alpha$, entering already the mass range reserved usually for neutron stars. On the other hand, for $\alpha>0$, as $\rho_\text{c}$ increases, the mass increases to a maximum value, and then the curve starts turning back, revealing the sub-Chandrasekhar limiting mass WDs. In the next subsection, we show that the portion of the curve corresponds to a decrease in the mass with increasing $\rho_\text{c}$ of the WDs is unstable. Thus, in this case, the limiting mass is the maximum mass attained before the curve starts receding. From Figure~\ref{Fig: Palatini_Jordan}, we see that the mass-limit decreases further from the Chandrasekhar mass-limit for more positive values of $\alpha$.

The mass--radius relations, as well as the variation of mass with $\rho_\text{c}$ in the Einstein frame, are plotted in Figure~\ref{Fig: Palatini_Einstein}. 
Since it is evident from Equation~\eqref{3.15} that for WDs, $\Phi\approx 1$, applying the conformal transformation $\Tilde{r}^2=\Phi r^2$ will not it change too much, and therefore we obtain almost similar curves as in Figure~\ref{Fig: Palatini_Jordan}. Moreover, the mass-radius curves in metric formalism are given by Das \& Mukhopadhyay~\citep{2015JCAP...05..045D}. It is evident that for $f(R) = R+\alpha R^2$ gravity, the results (i.e., the mass-radius curves) are similar in metric or Palatini formalisms. This is because Palatini and metric are just two different formalisms to explain the same phenomena. Hence, even if the modified stellar structure equations look different, they eventually result in similar mass-radius curves.

\subsection{Stability analysis of the modified gravity induced white dwarfs}\label{subsec:4.2}

Let us now study the stable and unstable branches of the mass--radius curves by stability analysis of the WDs in Palatini $f(R)$ gravity. As mentioned in \S\ref{subsec 3.2}, a star in hydrostatic equilibrium may either be stable or unstable against small radial perturbations. According to GR, a branch is considered to be stable if it follows $\pdv*{M}{\rho_\text{c}}>0$, which is also known as the positivity condition -- a necessary condition for stability~\cite{1986bhwd.book.....S,glendenning2010}. It ensures that the stars whose mass increases with the increase in $\rho_\text{c}$ are stable, whereas the stars with decreasing mass with the increase in $\rho_\text{c}$ are unstable~\citep{2000csnp.conf.....G}. We want to examine whether this condition is also valid in the considered theory of gravity with the given equation of state. In other words, whether the WDs on the receding branch of the mass--radius curves in Figure~\ref{Fig: Palatini_Jordan} and~\ref{Fig: Palatini_Einstein} are unstable or not under radial perturbations. To do so, one needs to solve the modified radial oscillation equations derived in \S\ref{subsec 3.2} with appropriate boundary conditions in both the Jordan and the Einstein frames and look for the normal modes of frequencies that are real. The sufficient condition for stability in modified gravity is $\sigma^2>0$, where $\sigma$ is the characteristic frequency of the normal mode (see Equation~\eqref{3.26}). This is so because the perturbations travel as plane waves $(\sim e^{i\sigma t})$ and if $\sigma^2<0$, i.e., $\sigma$ is imaginary, the amplitude of oscillations may grow in time, making the star unstable.

\begin{figure}[htbp]
     \subfloat[$\alpha = 4\times10^{15}\rm\,cm^2$]{%
     \centering
       \includegraphics[scale = 0.42]{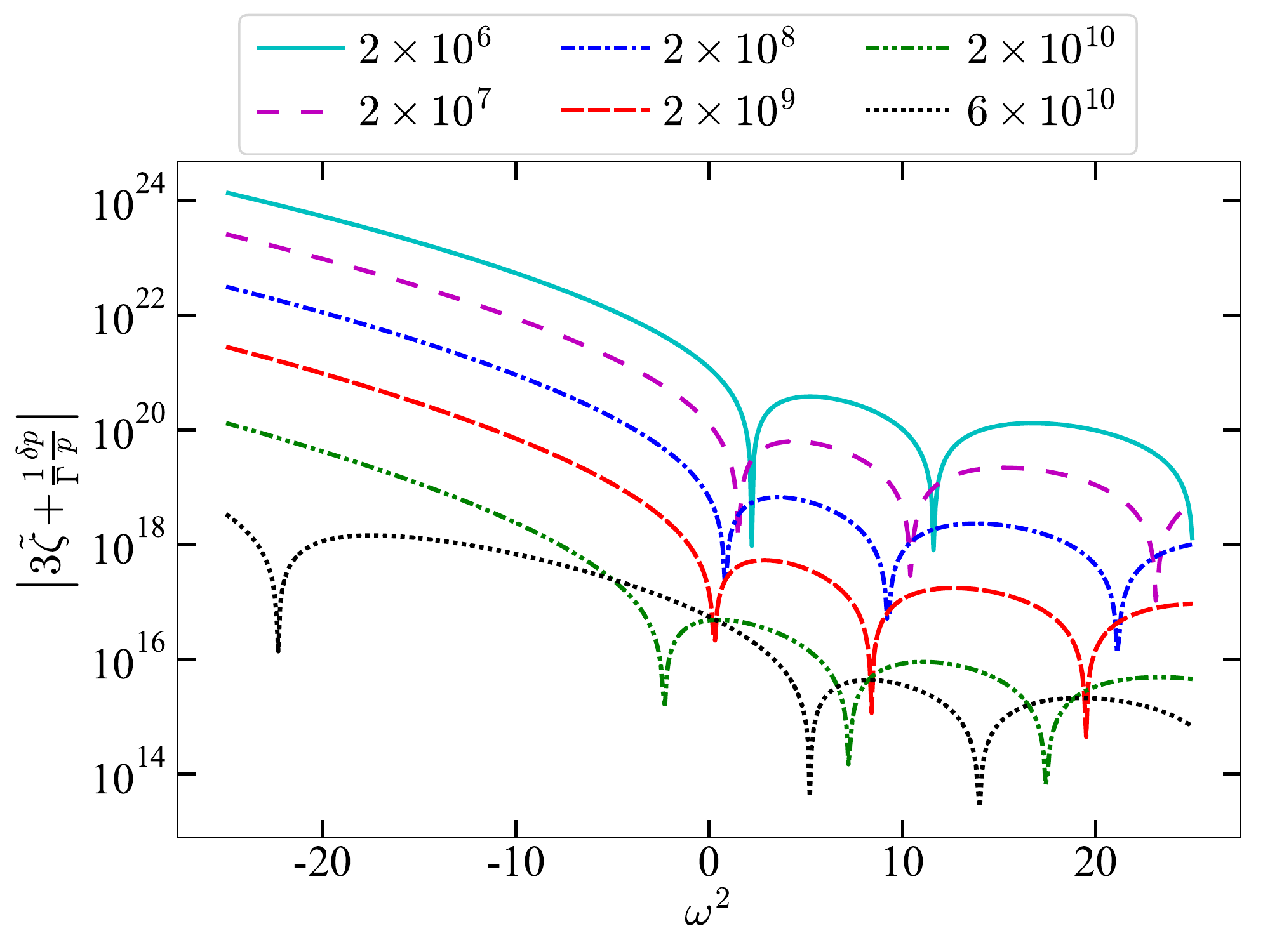}
    }\\
     \subfloat[$\alpha = -4\times10^{15}\rm\,cm^2$]{%
     \centering
       \includegraphics[scale = 0.42]{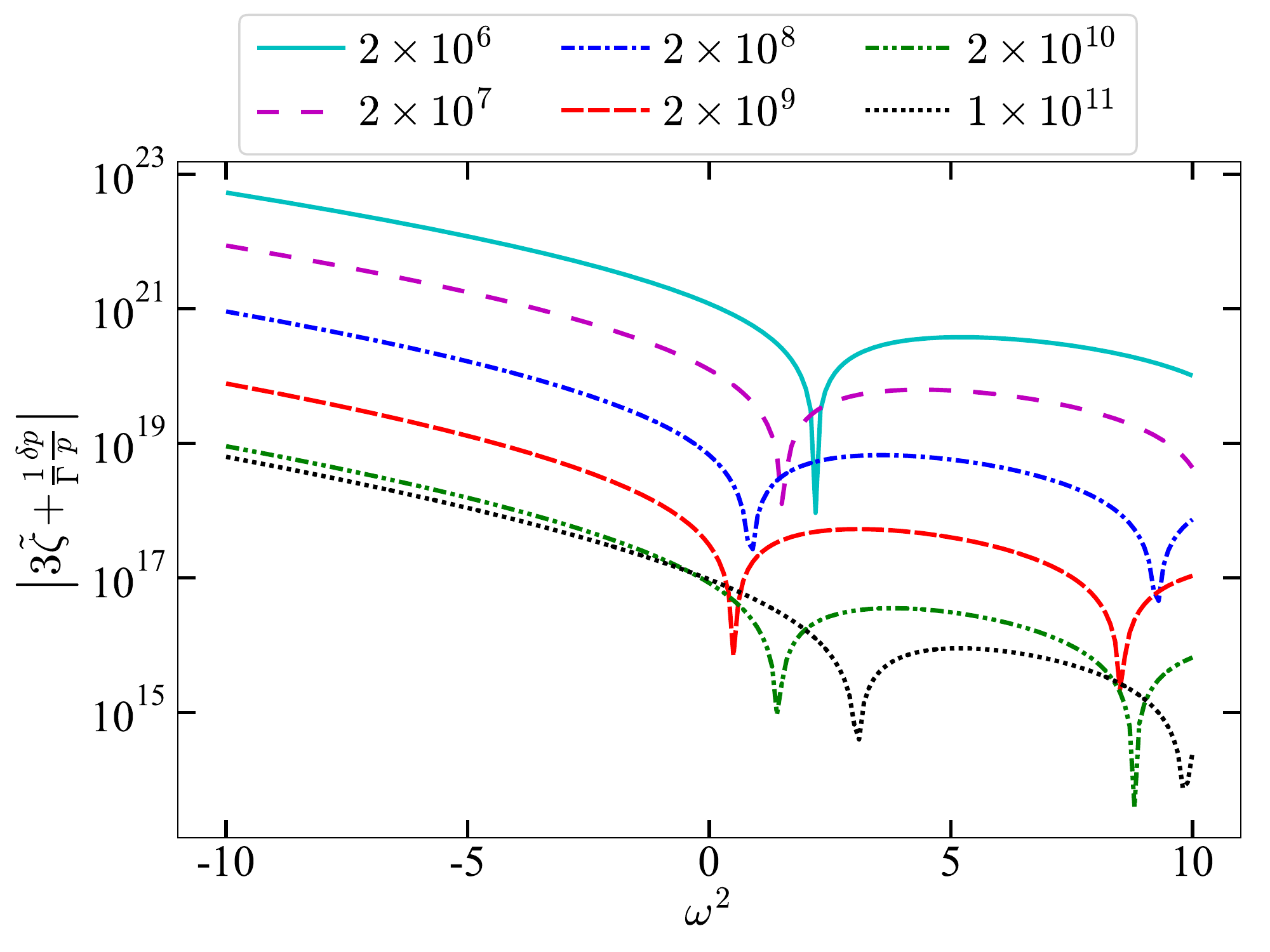}
    }
    \caption{Absolute value of L.H.S. of Equation~\eqref{4.2} at the center of the WD in the Einstein frame for a set of trial values of $\omega^2$. The label shows $\rho_\text{c}$ of the WDs in the unit of $\rm g\,cm^{-3}$.}
    \label{Fig: Stability_Einstein}
\end{figure}

\begin{figure}[htbp]
     \subfloat[$\alpha = 4\times10^{15}\rm\,cm^2$]{%
     \centering
       \includegraphics[scale = 0.42]{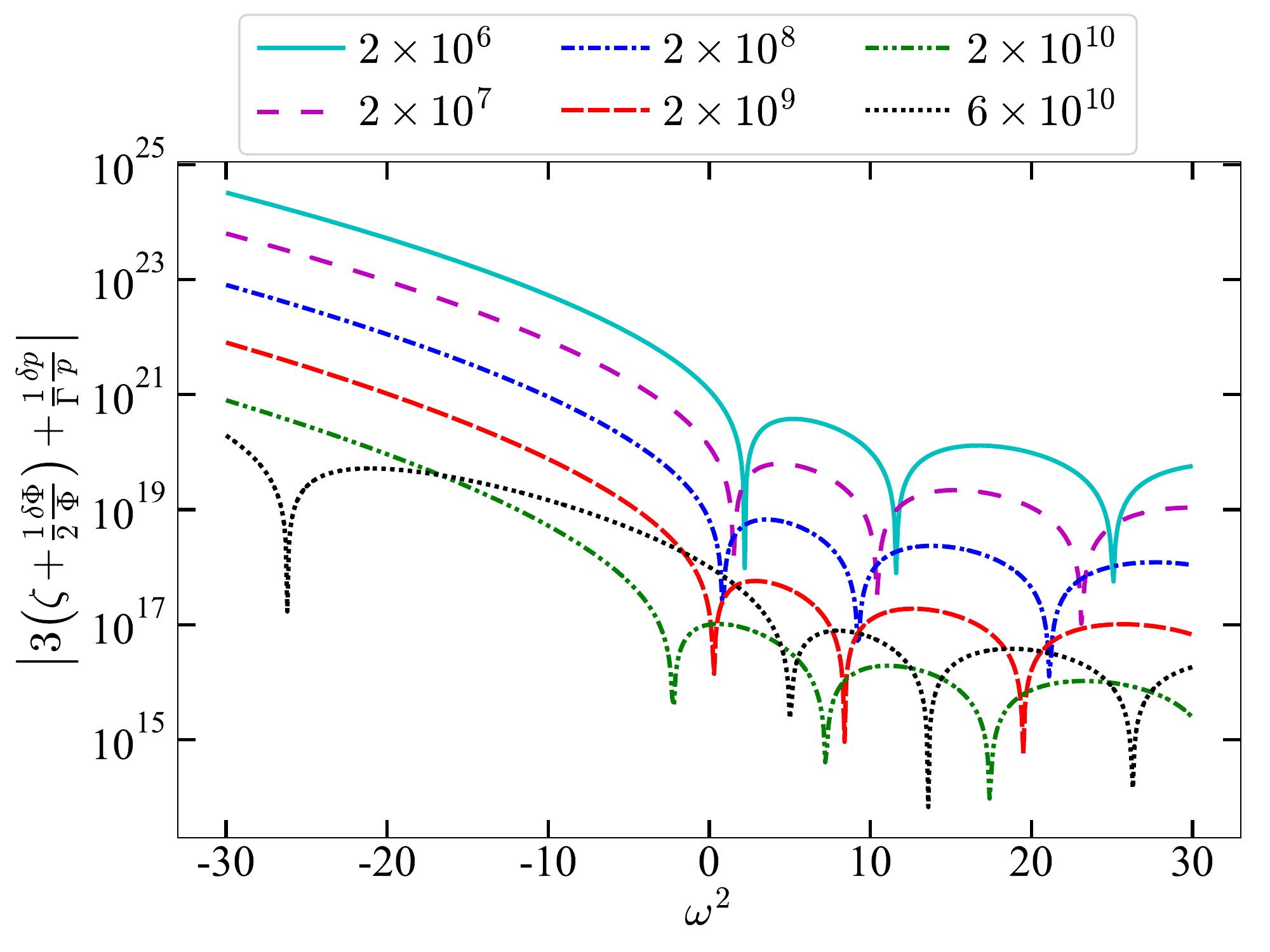}
    }\\
     \subfloat[$\alpha = -4\times10^{15}\rm\,cm^2$]{%
     \centering
       \includegraphics[scale = 0.42]{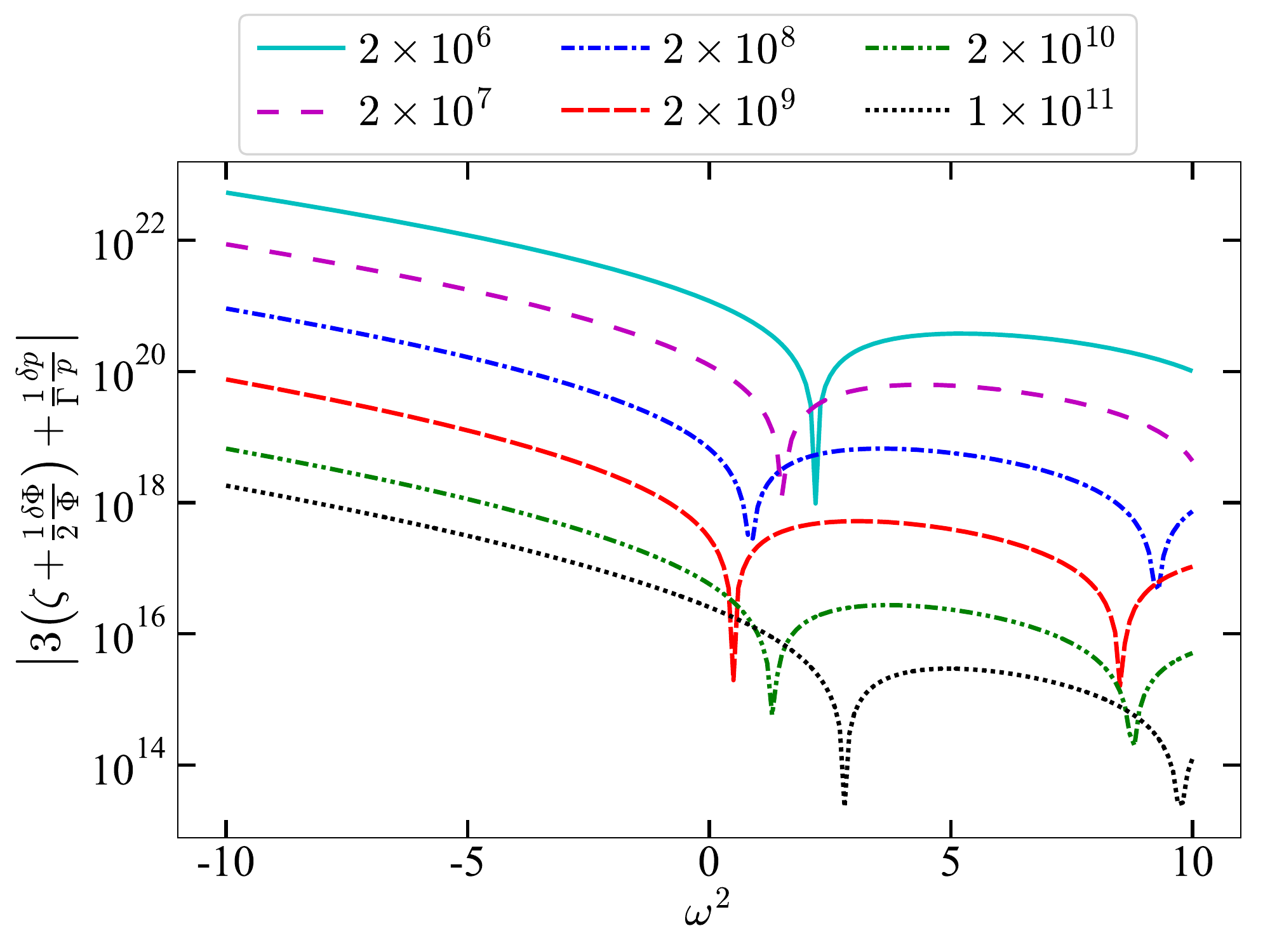}
    }
    \caption{Absolute value of L.H.S. of Equation~\eqref{4.4} at the center of the WD in the Jordan frame for a set of trial values of $\omega^2$. The label shows $\rho_\text{c}$ of the WDs in the unit of $\rm g\,cm^{-3}$.}
    \label{Fig: Stability_Jordan}
\end{figure}
We now numerically solve the modified equations for radial oscillations~\eqref{3.27} and~\eqref{3.28} for the Einstein frame, and Equations~\eqref{3.32} and~\eqref{3.33} for the Jordan frame. In the case of Einstein frame, the boundary conditions on the surface of the WD, i.e., at $\Tilde{r}=\mathcal{R}$, are given by
\begin{equation}\label{4.1}
\begin{aligned}
\Tilde{\zeta}(\Tilde{r}=\mathcal{R})=1,\\
\frac{\delta p}{p}+4\Tilde{\zeta}+\frac{\delta\Phi}{\Phi}+\frac{\sigma^2\mathcal{R}^3\Phi\Tilde{\zeta}}{GM}=0.
\end{aligned}
\end{equation}
The latter condition makes sure that $\dv*{(\delta p/p)}{\Tilde{r}}$ in Equation~\eqref{3.28} is finite everywhere. Moreover, in order to ensure the physical regularity of the solutions, $\Tilde{\zeta}$ and $\dv*{\Tilde{\zeta}}{\Tilde{r}}$ in Equation~\eqref{3.27} must be finite at the center, making the term in the parenthesis of R.H.S. equal to 0. Thus, the boundary condition at the center of the star, i.e., at $r=0$, is given by
\begin{equation}\label{4.2}
3\Tilde{\zeta}+\frac{1}{\Gamma}\frac{\delta p}{p}=0.
\end{equation}
In the same way, we also find the appropriate boundary conditions in the Jordan frame. At the surface, the following conditions need to be satisfied:
\begin{equation}\label{4.3}
\begin{aligned}
\zeta(r=\mathcal{R})=1,\\
\frac{\delta p}{p}+4\zeta+3\frac{\delta\Phi}{\Phi}+\frac{\sigma^2 \mathcal{R}^3\Phi^{\frac{5}{2}}}{GM}\left(\zeta+\frac{1}{2}\frac{\delta\Phi}{\Phi}\right)=0,
\end{aligned}
\end{equation}
whereas at the center, the boundary condition is given by
\begin{equation}\label{4.4}
3\left(\zeta+\frac{1}{2}\frac{\delta\Phi}{\Phi}\right) +\frac{1}{\Gamma}\frac{\delta p}{p}=0.
\end{equation}

We now integrate the radial oscillation equations from the surface to the center of the stars using these boundary conditions for a range of $\omega^2$. Here, $\omega^2$ is the square of the dimensionless frequency, given by $\omega^2=\sigma^2\mathcal{R}^3/GM$. The values of $\omega^2$ which satisfies Equations~\eqref{4.2} and~\eqref{4.4} in the Einstein and Jordan frame, respectively, are the correct normal mode frequencies for the radial oscillations. In Figure~\ref{Fig: Stability_Einstein} and~\ref{Fig: Stability_Jordan}, the normal mode frequencies correspond to the minima in each curve. If the minima occur for $\omega^2<0$, that particular WD is unstable under radial perturbation and usually do not exist in nature. As we have discussed in the previous section, the WD structure is not much different in Einstein and Jordan frames. Hence the normal mode frequencies are also almost the same in both the frames. It is evident from the figures that the WDs up to $\rho_\text{c}=2\times10^9\rm\,g\,cm^{-3}$ are all stable, mimicking the Newtonian case. However, at high enough densities due to the increased contribution of $\alpha R^2$ term, modified gravity may render extra stability to the WD or make the WD unstable under radial perturbations. Considering the case of $\alpha=4\times10^{15}\rm\, cm^2$ in Figures~\ref{Fig: Stability_Einstein}(a) and~\ref{Fig: Stability_Jordan}(a), we see that the WDs with $\rho_\text{c}\gtrsim10^{10}\rm\,g\,cm^{-3}$ are unstable as the first minimum occurs at $\omega^2<0$. From the mass--radius curves in Figures~\ref{Fig: Palatini_Jordan} and~\ref{Fig: Palatini_Einstein}, we notice that these range of WDs lie on the receding branch of the mass--radius curve, for which $\pdv*{M}{\rho_\text{c}}<0$. Thus, the WDs, which are unstable in the Newtonian regime of Palatini $f(R)$ gravity, also violate the positivity condition. On the other hand, for $\alpha=-4\times10^{15}\rm\,cm^2$, modified gravity brings extra stability to the WDs with $\rho_\text{c}\gtrsim10^{10}\rm\,g\,cm^{-3}$ as the first minimum shifts towards more positive values of $\omega^2$, which is evident from Figures~\ref{Fig: Stability_Einstein}(b) and~\ref{Fig: Stability_Jordan}(b). From the mass--radius curve corresponding to positive values of $\alpha$, we also notice that this branch satisfies the positivity condition, i.e., $\pdv*{M}{\rho_\text{c}}>0$. Thus, in this case, the high-density modified gravity induced WDs are more stable than the intermediate-density WDs. In this way, we examine the stability of WDs in Palatani $f(R)$ gravity, thereby establishing that the positivity condition is also a valid condition for stability analysis in Palatini $f(R)$ gravity in the Newtonian regime.

\section{Conclusions}~\label{Sec:5}
The recent observations of several over- and under-luminous SNe\,Ia suggest the violation of the Chandrasekhar mass-limit for WDs. This has led to an extensive study of super- and sub-Chandrasekhar limiting mass WDs, on the grounds of Newtonian and relativistic modified gravity~\citep{2021arXiv210501702S,2018JCAP...09..007K,2015JCAP...05..045D,2021IJGMM..1840006W,2017EPJC...77..871C}. In this work, we have focused on the WDs in the Newtonian limit of Palatini gravity with the quadratic Lagrangian $f(R)=R+\alpha R^2$. 

Solving the modified stellar structure equations for the given gravity model with the Chandrasekhar EoS, the mass--radius relations of the WDs were obtained for positive and negative values of the parameter $\alpha$ within the physically accepted bounds. Positive values of $\alpha$ turn out to provide sub-Chandrasekhar limiting mass WDs, whereas negative values give the super-Chandrasekhar ones. It should be noticed that there is no turn back in the super-Chandrasekhar branch and this is the reason why the maximum mass is only limited by the maximum possible density. At high densities, various nuclear reactions, for instance, pycno-nuclear reaction and inverse $\beta$-decay, may be triggered~\citep{2019ApJ...879...46O}. However, the rate of such reactions are quite uncertain, and hence we have hypothetically extended our curves to approximately $10^{11}\rm\,g\,cm^{-3}$. Analyzing the stellar structure equations or the mass--radius curves, it is clear that the Palatini gravity model reduces to the Newtonian case for $\alpha=0$ with the conventional Chandrasekhar mass-limit. It is also evident that the deviations from the Newtonian case are profound at high densities of the WDs, i.e., about $\rho_\text{c}\gtrsim10^9\rm\,g\,cm^{-3}$ due to the significant contribution of $\alpha R^2$ term, which eventually leads to a possible explanation for the violation of the Chandrasekhar mass-limit. 

Regarding the stability problem, generally, modified gravity models can induce extra stability to the high-density WDs, or it can make them unstable. Let us notice that such an analysis of compact stars in the framework of modified gravity, however with respect to the GR criterion (that is, using the same radial oscillation equations as for the GR case without any modifications) might be confusing~\citep{2020JCAP...11..048P}. Because of that fact, we have examined our model with respect to the appropriate modifications of the radial oscillation equations provided by the Palatini gravity, and also confirmed that the positivity condition $\pdv*{M}{\rho_\text{c}}>0$ still holds in this particular case.

As a concluding remark, let us comment that many theories of gravity modify the Newtonian limit of the hydrostatic equilibrium equations \cite{2020PhR...876....1O}; therefore they also alter other stellar equations, such as, for example, Schwarzschild criterion to determine the energy transport through a given star~\cite{2020PhRvD.102l4045W}, or energy produced in its core~\cite{2015PhRvL.115t1101S,2019PhRvD.100d4020O,2021PhRvD.103d4037W,2019EPJC...79.1030R}. Some parts of stellar evolution, for instance, Hayashi tracks~\cite{2020PhRvD.102l4045W}, Main Sequence \cite{2021JCAP...05..040C}, and cooling models~\cite{2021PhRvD.103f4032B} are also affected by modified gravity such that it can also have an impact on the properties of WDs and their formation processes. Research along these lines is essential to understand those fascinating objects and fully exploit the upcoming observational events. Detections of various compact objects, including WDs, by the use of gravitational wave detectors, such as aLIGO, Einstein Telescope, LISA, TianQin, BBO, DECIGO~\citep{2015CQGra..32a5014M,2020PhRvD.102f3021H} can put constraints on these theories of gravity, or they might shed light on features of the GR extensions~\citep{2021ApJ...909...65K}.

\section*{Acknowledgements}
SK would like to thank Banibrata Mukhopadhyay of IISc for the useful discussion about this work. AW is supported by the EU through the European Regional Development Fund CoE program TK133 “The Dark Side of the Universe.”

\appendix
\section{Relativistic equations for stellar structure in Palatini gravity}\label{appen}

\subsection{Schematic way to get TOV in modified gravity}
Let us notice that field equations of many modified theories of gravity can be written in the following form~\citep{2014PhLB..730..280C,2015JPhCS.600a2047M}:
\begin{equation}\label{3.1}
\sigma(\psi^i)(G_{\mu\nu}-W_{\mu\nu})=\kappa^2T_{\mu\nu},
\end{equation}
where $G_{\mu\nu}=R_{\mu\nu}-g_{\mu\nu}R/2$ is the usual Einstein tensor and $W_{\mu\nu}$ is an additional term including the theory modifications. They can have a geometric origin, for example, as it happens in this work. $\psi^i$ represents some field while $\sigma(\psi^i)$ is a coupling factor to gravity. For such a model, assuming the following spherically symmetric metric
\begin{equation}\label{3.2}
    \dd{s}^2=-B(r)c^2\dd{t}^2+A(r)\dd{r}^2+r^2\dd{\theta}^2+r^2\sin{\theta}^2\dd{\phi}^2,
\end{equation}
the generalized TOV equations are given by~\citep{2016EPJC...76..697W}
\begin{align}\label{3.3}
    \left(\frac{\Pi}{\sigma}\right)'&=-\frac{\frac{GM(r)}{r^2}\left(\frac{Q}{\sigma}+\frac{\Pi}{\sigma}\right)\left(1+\frac{4\pi r^3\frac{\Pi}{\sigma}}{M(r)}\right)}{1-\frac{2GM(r)}{r}} \nonumber\\ &+\frac{2\sigma}{\kappa^2 r}\left(\frac{W_{\theta\theta}}{r^2}-\frac{W_{rr}}{A}\right)
\end{align}
and
\begin{equation}\label{3.4}
    M(r)=\int_{0}^{r} 4\pi r'^2\frac{Q(r')}{\sigma(r')}\dd{r}',
\end{equation}
where $Q$, $\Pi$ are the generalized energy density and pressure, respectively. $M(r)$ stands for the stellar mass within a radius $r$. The physical interpretation of the terms appearing in the TOV equations because of the modifications, can be found in \cite{2016arXiv160103000V,2020EPJC...80..313S}.

\subsection{TOV equations for Palatini $f(R)$ gravity}
Comparing Equation~\eqref{8} with Equation~\eqref{3.1}, we notice that they have the similar form, and thus the modified TOV equations for Palatini $f(R)$ gravity can be written as~\citep{2018EPJC...78..421W,2021IJGMM..1840006W}
\begin{equation}\label{3.5}
    \dv{\Tilde{r}}(\frac{\Pi}{\Phi(\Tilde{r})^2})=-\frac{GAM(\Tilde{r})}{\Tilde{r}^2}\left(\frac{Q+\Pi}{\Phi (\Tilde{r})^2}\right)\left(1+\frac{4\pi\Tilde{r}^3\frac{\Pi}{\Phi (\Tilde{r})^2}}{M(\Tilde{r})}\right)
\end{equation}
and
\begin{equation}\label{3.6}
    M(\Tilde{r})=\int_{0}^{\Tilde{r}} 4\pi \Tilde{x}^2\frac{Q(\Tilde{x})}{\Phi(\Tilde{x})^2}\dd{\Tilde{x}}.
\end{equation}
Here the metric component $A$ (\ref{3.2}) is given by
\begin{equation}\label{3.7}
    A=1-\frac{2GM(\Tilde{r})}{\Tilde{r}}.
\end{equation}
The tilde in Equations~\eqref{3.5} and~\eqref{3.6} denotes the quantities in the Einstein frame, and it is related to the Jordan frame by the conformal transformation $\Tilde{r}^2=\Phi r^2$, where $\Phi$ is the scalar field introduced in \S\ref{Sec: 2}. Moreover, the conformally related energy density $(\Bar{Q})$ and pressure $(\Bar{\Pi})$ can be written as
\begin{align}\label{3.8}
\Bar{Q}&=\Bar{\rho}+\frac{\Bar{U}}{2\kappa^2 c^2}=\frac{\rho}{\Phi^2}+\frac{U}{2\kappa^2 c^2\Phi^2}=\frac{Q}{\Phi^2},\\
\label{3.9}
\Bar{\Pi}&=\Bar{p}-\frac{\Bar{U}}{2\kappa^2}=\frac{p}{\Phi^2}-\frac{U}{2\kappa^2\Phi^2}=\frac{\Pi}{\Phi^2},
\end{align}
where $\rho$ and $p$ denote the density of matter and pressure of the fluid, respectively. 

For the quadratic model $\Phi$ is given by
\begin{equation}\label{3.10}
\Phi=f'(R)=1+2\alpha R=1-2\alpha\kappa^2 T,
\end{equation}
while from the structural equation~\eqref{4} one obtains the second expression in the above formula.


\bibliographystyle{apsrev4-1}
\bibliography{biblio}

\begin{thebibliography}{107}%
\makeatletter
\providecommand \@ifxundefined [1]{%
 \@ifx{#1\undefined}
}%
\providecommand \@ifnum [1]{%
 \ifnum #1\expandafter \@firstoftwo
 \else \expandafter \@secondoftwo
 \fi
}%
\providecommand \@ifx [1]{%
 \ifx #1\expandafter \@firstoftwo
 \else \expandafter \@secondoftwo
 \fi
}%
\providecommand \natexlab [1]{#1}%
\providecommand \enquote  [1]{``#1''}%
\providecommand \bibnamefont  [1]{#1}%
\providecommand \bibfnamefont [1]{#1}%
\providecommand \citenamefont [1]{#1}%
\providecommand \href@noop [0]{\@secondoftwo}%
\providecommand \href [0]{\begingroup \@sanitize@url \@href}%
\providecommand \@href[1]{\@@startlink{#1}\@@href}%
\providecommand \@@href[1]{\endgroup#1\@@endlink}%
\providecommand \@sanitize@url [0]{\catcode `\\12\catcode `\$12\catcode
  `\&12\catcode `\#12\catcode `\^12\catcode `\_12\catcode `\%12\relax}%
\providecommand \@@startlink[1]{}%
\providecommand \@@endlink[0]{}%
\providecommand \url  [0]{\begingroup\@sanitize@url \@url }%
\providecommand \@url [1]{\endgroup\@href {#1}{\urlprefix }}%
\providecommand \urlprefix  [0]{URL }%
\providecommand \Eprint [0]{\href }%
\providecommand \doibase [0]{http://dx.doi.org/}%
\providecommand \selectlanguage [0]{\@gobble}%
\providecommand \bibinfo  [0]{\@secondoftwo}%
\providecommand \bibfield  [0]{\@secondoftwo}%
\providecommand \translation [1]{[#1]}%
\providecommand \BibitemOpen [0]{}%
\providecommand \bibitemStop [0]{}%
\providecommand \bibitemNoStop [0]{.\EOS\space}%
\providecommand \EOS [0]{\spacefactor3000\relax}%
\providecommand \BibitemShut  [1]{\csname bibitem#1\endcsname}%
\let\auto@bib@innerbib\@empty
\bibitem [{\citenamefont {{Carroll}}(2004)}]{2004sgig.book.....C}%
  \BibitemOpen
  \bibfield  {author} {\bibinfo {author} {\bibfnamefont {S.~M.}\ \bibnamefont
  {{Carroll}}},\ }\href@noop {} {\emph {\bibinfo {title} {{Spacetime and
  geometry. An introduction to general relativity}}}}\ (\bibinfo {year}
  {2004})\BibitemShut {NoStop}%
\bibitem [{\citenamefont {{Guth}}(1981)}]{1981PhRvD..23..347G}%
  \BibitemOpen
  \bibfield  {author} {\bibinfo {author} {\bibfnamefont {A.~H.}\ \bibnamefont
  {{Guth}}},\ }\href {\doibase 10.1103/PhysRevD.23.347} {\bibfield  {journal}
  {\bibinfo  {journal} {\prd}\ }\textbf {\bibinfo {volume} {23}},\ \bibinfo
  {pages} {347} (\bibinfo {year} {1981})}\BibitemShut {NoStop}%
\bibitem [{\citenamefont {{Coles}}(2003)}]{2003A&G....44a..32C}%
  \BibitemOpen
  \bibfield  {author} {\bibinfo {author} {\bibfnamefont {P.}~\bibnamefont
  {{Coles}}},\ }\href@noop {} {\bibfield  {journal} {\bibinfo  {journal}
  {Astronomy and Geophysics}\ }\textbf {\bibinfo {volume} {44}},\ \bibinfo
  {pages} {32} (\bibinfo {year} {2003})}\BibitemShut {NoStop}%
\bibitem [{\citenamefont {{Riess}}\ \emph {et~al.}(1998)\citenamefont
  {{Riess}}, \citenamefont {{Filippenko}}, \citenamefont {{Challis}},
  \citenamefont {{Clocchiatti}}, \citenamefont {{Diercks}}, \citenamefont
  {{Garnavich}}, \citenamefont {{Gilliland}}, \citenamefont {{Hogan}},
  \citenamefont {{Jha}}, \citenamefont {{Kirshner}}, \citenamefont
  {{Leibundgut}}, \citenamefont {{Phillips}}, \citenamefont {{Reiss}},
  \citenamefont {{Schmidt}}, \citenamefont {{Schommer}}, \citenamefont
  {{Smith}}, \citenamefont {{Spyromilio}}, \citenamefont {{Stubbs}},
  \citenamefont {{Suntzeff}},\ and\ \citenamefont
  {{Tonry}}}]{1998AJ....116.1009R}%
  \BibitemOpen
  \bibfield  {author} {\bibinfo {author} {\bibfnamefont {A.~G.}\ \bibnamefont
  {{Riess}}}, \bibinfo {author} {\bibfnamefont {A.~V.}\ \bibnamefont
  {{Filippenko}}}, \bibinfo {author} {\bibfnamefont {P.}~\bibnamefont
  {{Challis}}}, \bibinfo {author} {\bibfnamefont {A.}~\bibnamefont
  {{Clocchiatti}}}, \bibinfo {author} {\bibfnamefont {A.}~\bibnamefont
  {{Diercks}}}, \bibinfo {author} {\bibfnamefont {P.~M.}\ \bibnamefont
  {{Garnavich}}}, \bibinfo {author} {\bibfnamefont {R.~L.}\ \bibnamefont
  {{Gilliland}}}, \bibinfo {author} {\bibfnamefont {C.~J.}\ \bibnamefont
  {{Hogan}}}, \bibinfo {author} {\bibfnamefont {S.}~\bibnamefont {{Jha}}},
  \bibinfo {author} {\bibfnamefont {R.~P.}\ \bibnamefont {{Kirshner}}},
  \bibinfo {author} {\bibfnamefont {B.}~\bibnamefont {{Leibundgut}}}, \bibinfo
  {author} {\bibfnamefont {M.~M.}\ \bibnamefont {{Phillips}}}, \bibinfo
  {author} {\bibfnamefont {D.}~\bibnamefont {{Reiss}}}, \bibinfo {author}
  {\bibfnamefont {B.~P.}\ \bibnamefont {{Schmidt}}}, \bibinfo {author}
  {\bibfnamefont {R.~A.}\ \bibnamefont {{Schommer}}}, \bibinfo {author}
  {\bibfnamefont {R.~C.}\ \bibnamefont {{Smith}}}, \bibinfo {author}
  {\bibfnamefont {J.}~\bibnamefont {{Spyromilio}}}, \bibinfo {author}
  {\bibfnamefont {C.}~\bibnamefont {{Stubbs}}}, \bibinfo {author}
  {\bibfnamefont {N.~B.}\ \bibnamefont {{Suntzeff}}}, \ and\ \bibinfo {author}
  {\bibfnamefont {J.}~\bibnamefont {{Tonry}}},\ }\href {\doibase
  10.1086/300499} {\bibfield  {journal} {\bibinfo  {journal} {\aj}\ }\textbf
  {\bibinfo {volume} {116}},\ \bibinfo {pages} {1009} (\bibinfo {year}
  {1998})},\ \Eprint {http://arxiv.org/abs/astro-ph/9805201}
  {arXiv:astro-ph/9805201 [astro-ph]} \BibitemShut {NoStop}%
\bibitem [{\citenamefont {{Perlmutter}}\ \emph {et~al.}(1999)\citenamefont
  {{Perlmutter}}, \citenamefont {{Aldering}}, \citenamefont {{Goldhaber}},
  \citenamefont {{Knop}}, \citenamefont {{Nugent}}, \citenamefont {{Castro}},
  \citenamefont {{Deustua}}, \citenamefont {{Fabbro}}, \citenamefont
  {{Goobar}}, \citenamefont {{Groom}}, \citenamefont {{Hook}}, \citenamefont
  {{Kim}}, \citenamefont {{Kim}}, \citenamefont {{Lee}}, \citenamefont
  {{Nunes}}, \citenamefont {{Pain}}, \citenamefont {{Pennypacker}},
  \citenamefont {{Quimby}}, \citenamefont {{Lidman}}, \citenamefont {{Ellis}},
  \citenamefont {{Irwin}}, \citenamefont {{McMahon}}, \citenamefont
  {{Ruiz-Lapuente}}, \citenamefont {{Walton}}, \citenamefont {{Schaefer}},
  \citenamefont {{Boyle}}, \citenamefont {{Filippenko}}, \citenamefont
  {{Matheson}}, \citenamefont {{Fruchter}}, \citenamefont {{Panagia}},
  \citenamefont {{Newberg}}, \citenamefont {{Couch}},\ and\ \citenamefont
  {{Project}}}]{1999ApJ...517..565P}%
  \BibitemOpen
  \bibfield  {author} {\bibinfo {author} {\bibfnamefont {S.}~\bibnamefont
  {{Perlmutter}}}, \bibinfo {author} {\bibfnamefont {G.}~\bibnamefont
  {{Aldering}}}, \bibinfo {author} {\bibfnamefont {G.}~\bibnamefont
  {{Goldhaber}}}, \bibinfo {author} {\bibfnamefont {R.~A.}\ \bibnamefont
  {{Knop}}}, \bibinfo {author} {\bibfnamefont {P.}~\bibnamefont {{Nugent}}},
  \bibinfo {author} {\bibfnamefont {P.~G.}\ \bibnamefont {{Castro}}}, \bibinfo
  {author} {\bibfnamefont {S.}~\bibnamefont {{Deustua}}}, \bibinfo {author}
  {\bibfnamefont {S.}~\bibnamefont {{Fabbro}}}, \bibinfo {author}
  {\bibfnamefont {A.}~\bibnamefont {{Goobar}}}, \bibinfo {author}
  {\bibfnamefont {D.~E.}\ \bibnamefont {{Groom}}}, \bibinfo {author}
  {\bibfnamefont {I.~M.}\ \bibnamefont {{Hook}}}, \bibinfo {author}
  {\bibfnamefont {A.~G.}\ \bibnamefont {{Kim}}}, \bibinfo {author}
  {\bibfnamefont {M.~Y.}\ \bibnamefont {{Kim}}}, \bibinfo {author}
  {\bibfnamefont {J.~C.}\ \bibnamefont {{Lee}}}, \bibinfo {author}
  {\bibfnamefont {N.~J.}\ \bibnamefont {{Nunes}}}, \bibinfo {author}
  {\bibfnamefont {R.}~\bibnamefont {{Pain}}}, \bibinfo {author} {\bibfnamefont
  {C.~R.}\ \bibnamefont {{Pennypacker}}}, \bibinfo {author} {\bibfnamefont
  {R.}~\bibnamefont {{Quimby}}}, \bibinfo {author} {\bibfnamefont
  {C.}~\bibnamefont {{Lidman}}}, \bibinfo {author} {\bibfnamefont {R.~S.}\
  \bibnamefont {{Ellis}}}, \bibinfo {author} {\bibfnamefont {M.}~\bibnamefont
  {{Irwin}}}, \bibinfo {author} {\bibfnamefont {R.~G.}\ \bibnamefont
  {{McMahon}}}, \bibinfo {author} {\bibfnamefont {P.}~\bibnamefont
  {{Ruiz-Lapuente}}}, \bibinfo {author} {\bibfnamefont {N.}~\bibnamefont
  {{Walton}}}, \bibinfo {author} {\bibfnamefont {B.}~\bibnamefont
  {{Schaefer}}}, \bibinfo {author} {\bibfnamefont {B.~J.}\ \bibnamefont
  {{Boyle}}}, \bibinfo {author} {\bibfnamefont {A.~V.}\ \bibnamefont
  {{Filippenko}}}, \bibinfo {author} {\bibfnamefont {T.}~\bibnamefont
  {{Matheson}}}, \bibinfo {author} {\bibfnamefont {A.~S.}\ \bibnamefont
  {{Fruchter}}}, \bibinfo {author} {\bibfnamefont {N.}~\bibnamefont
  {{Panagia}}}, \bibinfo {author} {\bibfnamefont {H.~J.~M.}\ \bibnamefont
  {{Newberg}}}, \bibinfo {author} {\bibfnamefont {W.~J.}\ \bibnamefont
  {{Couch}}}, \ and\ \bibinfo {author} {\bibfnamefont {T.~S.~C.}\ \bibnamefont
  {{Project}}},\ }\href {\doibase 10.1086/307221} {\bibfield  {journal}
  {\bibinfo  {journal} {\apj}\ }\textbf {\bibinfo {volume} {517}},\ \bibinfo
  {pages} {565} (\bibinfo {year} {1999})},\ \Eprint
  {http://arxiv.org/abs/astro-ph/9812133} {arXiv:astro-ph/9812133 [astro-ph]}
  \BibitemShut {NoStop}%
\bibitem [{\citenamefont {{Huterer}}\ and\ \citenamefont
  {{Turner}}(1999)}]{1999PhRvD..60h1301H}%
  \BibitemOpen
  \bibfield  {author} {\bibinfo {author} {\bibfnamefont {D.}~\bibnamefont
  {{Huterer}}}\ and\ \bibinfo {author} {\bibfnamefont {M.~S.}\ \bibnamefont
  {{Turner}}},\ }\href {\doibase 10.1103/PhysRevD.60.081301} {\bibfield
  {journal} {\bibinfo  {journal} {\prd}\ }\textbf {\bibinfo {volume} {60}},\
  \bibinfo {eid} {081301} (\bibinfo {year} {1999})},\ \Eprint
  {http://arxiv.org/abs/astro-ph/9808133} {arXiv:astro-ph/9808133 [astro-ph]}
  \BibitemShut {NoStop}%
\bibitem [{\citenamefont {{Del Popolo}}\ and\ \citenamefont {{Le
  Delliou}}(2017)}]{2017Galax...5...17D}%
  \BibitemOpen
  \bibfield  {author} {\bibinfo {author} {\bibfnamefont {A.}~\bibnamefont {{Del
  Popolo}}}\ and\ \bibinfo {author} {\bibfnamefont {M.}~\bibnamefont {{Le
  Delliou}}},\ }\href {\doibase 10.3390/galaxies5010017} {\bibfield  {journal}
  {\bibinfo  {journal} {Galaxies}\ }\textbf {\bibinfo {volume} {5}},\ \bibinfo
  {pages} {17} (\bibinfo {year} {2017})},\ \Eprint
  {http://arxiv.org/abs/1606.07790} {arXiv:1606.07790 [astro-ph.CO]}
  \BibitemShut {NoStop}%
\bibitem [{\citenamefont {{Capozziello}}\ and\ \citenamefont {{de
  Laurentis}}(2011)}]{2011PhR...509..167C}%
  \BibitemOpen
  \bibfield  {author} {\bibinfo {author} {\bibfnamefont {S.}~\bibnamefont
  {{Capozziello}}}\ and\ \bibinfo {author} {\bibfnamefont {M.}~\bibnamefont
  {{de Laurentis}}},\ }\href {\doibase 10.1016/j.physrep.2011.09.003}
  {\bibfield  {journal} {\bibinfo  {journal} {\physrep}\ }\textbf {\bibinfo
  {volume} {509}},\ \bibinfo {pages} {167} (\bibinfo {year} {2011})},\ \Eprint
  {http://arxiv.org/abs/1108.6266} {arXiv:1108.6266 [gr-qc]} \BibitemShut
  {NoStop}%
\bibitem [{\citenamefont {{Saridakis}}\ \emph {et~al.}(2021)\citenamefont
  {{Saridakis}}, \citenamefont {{Lazkoz}}, \citenamefont {{Salzano}},
  \citenamefont {{Vargas Moniz}}, \citenamefont {{Capozziello}}, \citenamefont
  {{Beltr{\'a}n Jim{\'e}nez}}, \citenamefont {{De Laurentis}}, \citenamefont
  {{Olmo}}, \citenamefont {{Akrami}}, \citenamefont {{Bahamonde}},
  \citenamefont {{Bl{\'a}zquez-Salcedo}}, \citenamefont {{B{\"o}hmer}},
  \citenamefont {{Bonvin}}, \citenamefont {{Bouhmadi-L{\'o}pez}}, \citenamefont
  {{Brax}}, \citenamefont {{Calcagni}}, \citenamefont {{Casadio}},
  \citenamefont {{Cembranos}}, \citenamefont {{de la Cruz-Dombriz}},
  \citenamefont {{Davis}}, \citenamefont {{Delhom}}, \citenamefont {{Di
  Valentino}}, \citenamefont {{Dialektopoulos}}, \citenamefont {{Elder}},
  \citenamefont {{Mar{\'\i}a Ezquiaga}}, \citenamefont {{Frusciante}},
  \citenamefont {{Garattini}}, \citenamefont {{Gergely}}, \citenamefont
  {{Giusti}}, \citenamefont {{Heisenberg}}, \citenamefont {{Hohmann}},
  \citenamefont {{Iosifidis}}, \citenamefont {{Kazantzidis}}, \citenamefont
  {{Kleihaus}}, \citenamefont {{Koivisto}}, \citenamefont {{Kunz}},
  \citenamefont {{Lobo}}, \citenamefont {{Martinelli}}, \citenamefont
  {{Mart{\'\i}n-Moruno}}, \citenamefont {{Mimoso}}, \citenamefont {{Mota}},
  \citenamefont {{Peirone}}, \citenamefont {{Perivolaropoulos}}, \citenamefont
  {{Pettorino}}, \citenamefont {{Pfeifer}}, \citenamefont {{Pizzuti}},
  \citenamefont {{Rubiera-Garcia}}, \citenamefont {{Levi Said}}, \citenamefont
  {{Sakellariadou}}, \citenamefont {{Saltas}}, \citenamefont {{Spurio
  Mancini}}, \citenamefont {{Voicu}},\ and\ \citenamefont
  {{Wojnar}}}]{2021arXiv210512582S}%
  \BibitemOpen
  \bibfield  {author} {\bibinfo {author} {\bibfnamefont {E.~N.}\ \bibnamefont
  {{Saridakis}}}, \bibinfo {author} {\bibfnamefont {R.}~\bibnamefont
  {{Lazkoz}}}, \bibinfo {author} {\bibfnamefont {V.}~\bibnamefont {{Salzano}}},
  \bibinfo {author} {\bibfnamefont {P.}~\bibnamefont {{Vargas Moniz}}},
  \bibinfo {author} {\bibfnamefont {S.}~\bibnamefont {{Capozziello}}}, \bibinfo
  {author} {\bibfnamefont {J.}~\bibnamefont {{Beltr{\'a}n Jim{\'e}nez}}},
  \bibinfo {author} {\bibfnamefont {M.}~\bibnamefont {{De Laurentis}}},
  \bibinfo {author} {\bibfnamefont {G.~J.}\ \bibnamefont {{Olmo}}}, \bibinfo
  {author} {\bibfnamefont {Y.}~\bibnamefont {{Akrami}}}, \bibinfo {author}
  {\bibfnamefont {S.}~\bibnamefont {{Bahamonde}}}, \bibinfo {author}
  {\bibfnamefont {J.~L.}\ \bibnamefont {{Bl{\'a}zquez-Salcedo}}}, \bibinfo
  {author} {\bibfnamefont {C.~G.}\ \bibnamefont {{B{\"o}hmer}}}, \bibinfo
  {author} {\bibfnamefont {C.}~\bibnamefont {{Bonvin}}}, \bibinfo {author}
  {\bibfnamefont {M.}~\bibnamefont {{Bouhmadi-L{\'o}pez}}}, \bibinfo {author}
  {\bibfnamefont {P.}~\bibnamefont {{Brax}}}, \bibinfo {author} {\bibfnamefont
  {G.}~\bibnamefont {{Calcagni}}}, \bibinfo {author} {\bibfnamefont
  {R.}~\bibnamefont {{Casadio}}}, \bibinfo {author} {\bibfnamefont {J.~A.~R.}\
  \bibnamefont {{Cembranos}}}, \bibinfo {author} {\bibfnamefont
  {{\'A}.}~\bibnamefont {{de la Cruz-Dombriz}}}, \bibinfo {author}
  {\bibfnamefont {A.-C.}\ \bibnamefont {{Davis}}}, \bibinfo {author}
  {\bibfnamefont {A.}~\bibnamefont {{Delhom}}}, \bibinfo {author}
  {\bibfnamefont {E.}~\bibnamefont {{Di Valentino}}}, \bibinfo {author}
  {\bibfnamefont {K.~F.}\ \bibnamefont {{Dialektopoulos}}}, \bibinfo {author}
  {\bibfnamefont {B.}~\bibnamefont {{Elder}}}, \bibinfo {author} {\bibfnamefont
  {J.}~\bibnamefont {{Mar{\'\i}a Ezquiaga}}}, \bibinfo {author} {\bibfnamefont
  {N.}~\bibnamefont {{Frusciante}}}, \bibinfo {author} {\bibfnamefont
  {R.}~\bibnamefont {{Garattini}}}, \bibinfo {author} {\bibfnamefont
  {L.~{\'A}.}\ \bibnamefont {{Gergely}}}, \bibinfo {author} {\bibfnamefont
  {A.}~\bibnamefont {{Giusti}}}, \bibinfo {author} {\bibfnamefont
  {L.}~\bibnamefont {{Heisenberg}}}, \bibinfo {author} {\bibfnamefont
  {M.}~\bibnamefont {{Hohmann}}}, \bibinfo {author} {\bibfnamefont
  {D.}~\bibnamefont {{Iosifidis}}}, \bibinfo {author} {\bibfnamefont
  {L.}~\bibnamefont {{Kazantzidis}}}, \bibinfo {author} {\bibfnamefont
  {B.}~\bibnamefont {{Kleihaus}}}, \bibinfo {author} {\bibfnamefont {T.~S.}\
  \bibnamefont {{Koivisto}}}, \bibinfo {author} {\bibfnamefont
  {J.}~\bibnamefont {{Kunz}}}, \bibinfo {author} {\bibfnamefont {F.~S.~N.}\
  \bibnamefont {{Lobo}}}, \bibinfo {author} {\bibfnamefont {M.}~\bibnamefont
  {{Martinelli}}}, \bibinfo {author} {\bibfnamefont {P.}~\bibnamefont
  {{Mart{\'\i}n-Moruno}}}, \bibinfo {author} {\bibfnamefont {J.~P.}\
  \bibnamefont {{Mimoso}}}, \bibinfo {author} {\bibfnamefont {D.~F.}\
  \bibnamefont {{Mota}}}, \bibinfo {author} {\bibfnamefont {S.}~\bibnamefont
  {{Peirone}}}, \bibinfo {author} {\bibfnamefont {L.}~\bibnamefont
  {{Perivolaropoulos}}}, \bibinfo {author} {\bibfnamefont {V.}~\bibnamefont
  {{Pettorino}}}, \bibinfo {author} {\bibfnamefont {C.}~\bibnamefont
  {{Pfeifer}}}, \bibinfo {author} {\bibfnamefont {L.}~\bibnamefont
  {{Pizzuti}}}, \bibinfo {author} {\bibfnamefont {D.}~\bibnamefont
  {{Rubiera-Garcia}}}, \bibinfo {author} {\bibfnamefont {J.}~\bibnamefont
  {{Levi Said}}}, \bibinfo {author} {\bibfnamefont {M.}~\bibnamefont
  {{Sakellariadou}}}, \bibinfo {author} {\bibfnamefont {I.~D.}\ \bibnamefont
  {{Saltas}}}, \bibinfo {author} {\bibfnamefont {A.}~\bibnamefont {{Spurio
  Mancini}}}, \bibinfo {author} {\bibfnamefont {N.}~\bibnamefont {{Voicu}}}, \
  and\ \bibinfo {author} {\bibfnamefont {A.}~\bibnamefont {{Wojnar}}},\
  }\href@noop {} {\bibfield  {journal} {\bibinfo  {journal} {arXiv e-prints}\
  ,\ \bibinfo {eid} {arXiv:2105.12582}} (\bibinfo {year} {2021})},\ \Eprint
  {http://arxiv.org/abs/2105.12582} {arXiv:2105.12582 [gr-qc]} \BibitemShut
  {NoStop}%
\bibitem [{\citenamefont {{Faraoni}}(2004)}]{2004cstg.book.....F}%
  \BibitemOpen
  \bibfield  {author} {\bibinfo {author} {\bibfnamefont {V.}~\bibnamefont
  {{Faraoni}}},\ }\href@noop {} {\emph {\bibinfo {title} {{Cosmology in
  Scalar-Tensor Gravity}}}}\ (\bibinfo {year} {2004})\BibitemShut {NoStop}%
\bibitem [{\citenamefont {{Naruko}}\ \emph {et~al.}(2016)\citenamefont
  {{Naruko}}, \citenamefont {{Yoshida}},\ and\ \citenamefont
  {{Mukohyama}}}]{2016CQGra..33iLT01N}%
  \BibitemOpen
  \bibfield  {author} {\bibinfo {author} {\bibfnamefont {A.}~\bibnamefont
  {{Naruko}}}, \bibinfo {author} {\bibfnamefont {D.}~\bibnamefont {{Yoshida}}},
  \ and\ \bibinfo {author} {\bibfnamefont {S.}~\bibnamefont {{Mukohyama}}},\
  }\href {\doibase 10.1088/0264-9381/33/9/09LT01} {\bibfield  {journal}
  {\bibinfo  {journal} {Classical and Quantum Gravity}\ }\textbf {\bibinfo
  {volume} {33}},\ \bibinfo {eid} {09LT01} (\bibinfo {year} {2016})},\ \Eprint
  {http://arxiv.org/abs/1512.06977} {arXiv:1512.06977 [gr-qc]} \BibitemShut
  {NoStop}%
\bibitem [{\citenamefont {{Buchdahl}}(1970)}]{1970MNRAS.150....1B}%
  \BibitemOpen
  \bibfield  {author} {\bibinfo {author} {\bibfnamefont {H.~A.}\ \bibnamefont
  {{Buchdahl}}},\ }\href {\doibase 10.1093/mnras/150.1.1} {\bibfield  {journal}
  {\bibinfo  {journal} {\mnras}\ }\textbf {\bibinfo {volume} {150}},\ \bibinfo
  {pages} {1} (\bibinfo {year} {1970})}\BibitemShut {NoStop}%
\bibitem [{\citenamefont {{Bhattacharjee}}\ \emph {et~al.}(2020)\citenamefont
  {{Bhattacharjee}}, \citenamefont {{Santos}}, \citenamefont {{Moraes}},\ and\
  \citenamefont {{Sahoo}}}]{2020EPJP..135..576B}%
  \BibitemOpen
  \bibfield  {author} {\bibinfo {author} {\bibfnamefont {S.}~\bibnamefont
  {{Bhattacharjee}}}, \bibinfo {author} {\bibfnamefont {J.~R.~L.}\ \bibnamefont
  {{Santos}}}, \bibinfo {author} {\bibfnamefont {P.~H.~R.~S.}\ \bibnamefont
  {{Moraes}}}, \ and\ \bibinfo {author} {\bibfnamefont {P.~K.}\ \bibnamefont
  {{Sahoo}}},\ }\href {\doibase 10.1140/epjp/s13360-020-00583-6} {\bibfield
  {journal} {\bibinfo  {journal} {European Physical Journal Plus}\ }\textbf
  {\bibinfo {volume} {135}},\ \bibinfo {eid} {576} (\bibinfo {year} {2020})},\
  \Eprint {http://arxiv.org/abs/2006.04336} {arXiv:2006.04336 [gr-qc]}
  \BibitemShut {NoStop}%
\bibitem [{\citenamefont {{Oikonomou}}(2018)}]{2018PhRvD..97f4001O}%
  \BibitemOpen
  \bibfield  {author} {\bibinfo {author} {\bibfnamefont {V.~K.}\ \bibnamefont
  {{Oikonomou}}},\ }\href {\doibase 10.1103/PhysRevD.97.064001} {\bibfield
  {journal} {\bibinfo  {journal} {\prd}\ }\textbf {\bibinfo {volume} {97}},\
  \bibinfo {eid} {064001} (\bibinfo {year} {2018})},\ \Eprint
  {http://arxiv.org/abs/1801.03426} {arXiv:1801.03426 [gr-qc]} \BibitemShut
  {NoStop}%
\bibitem [{\citenamefont {{Odintsov}}\ and\ \citenamefont
  {{Oikonomou}}(2019)}]{2019PhRvD..99f4049O}%
  \BibitemOpen
  \bibfield  {author} {\bibinfo {author} {\bibfnamefont {S.~D.}\ \bibnamefont
  {{Odintsov}}}\ and\ \bibinfo {author} {\bibfnamefont {V.~K.}\ \bibnamefont
  {{Oikonomou}}},\ }\href {\doibase 10.1103/PhysRevD.99.064049} {\bibfield
  {journal} {\bibinfo  {journal} {\prd}\ }\textbf {\bibinfo {volume} {99}},\
  \bibinfo {eid} {064049} (\bibinfo {year} {2019})},\ \Eprint
  {http://arxiv.org/abs/1901.05363} {arXiv:1901.05363 [gr-qc]} \BibitemShut
  {NoStop}%
\bibitem [{\citenamefont {{Capozziello}}\ and\ \citenamefont {{De
  Laurentis}}(2012)}]{2012AnP...524..545C}%
  \BibitemOpen
  \bibfield  {author} {\bibinfo {author} {\bibfnamefont {S.}~\bibnamefont
  {{Capozziello}}}\ and\ \bibinfo {author} {\bibfnamefont {M.}~\bibnamefont
  {{De Laurentis}}},\ }\href {\doibase 10.1002/andp.201200109} {\bibfield
  {journal} {\bibinfo  {journal} {Annalen der Physik}\ }\textbf {\bibinfo
  {volume} {524}},\ \bibinfo {pages} {545} (\bibinfo {year}
  {2012})}\BibitemShut {NoStop}%
\bibitem [{\citenamefont {{Nojiri}}\ and\ \citenamefont
  {{Odintsov}}(2007)}]{2007IJGMM..04..115N}%
  \BibitemOpen
  \bibfield  {author} {\bibinfo {author} {\bibfnamefont {S.}~\bibnamefont
  {{Nojiri}}}\ and\ \bibinfo {author} {\bibfnamefont {S.~D.}\ \bibnamefont
  {{Odintsov}}},\ }\href {\doibase 10.1142/S0219887807001928} {\bibfield
  {journal} {\bibinfo  {journal} {International Journal of Geometric Methods in
  Modern Physics}\ }\textbf {\bibinfo {volume} {04}},\ \bibinfo {pages} {115}
  (\bibinfo {year} {2007})},\ \Eprint {http://arxiv.org/abs/hep-th/0601213}
  {arXiv:hep-th/0601213 [hep-th]} \BibitemShut {NoStop}%
\bibitem [{\citenamefont {{Katsuragawa}}\ \emph {et~al.}(2019)\citenamefont
  {{Katsuragawa}}, \citenamefont {{Nakamura}}, \citenamefont {{Ikeda}},\ and\
  \citenamefont {{Capozziello}}}]{2019PhRvD..99l4050K}%
  \BibitemOpen
  \bibfield  {author} {\bibinfo {author} {\bibfnamefont {T.}~\bibnamefont
  {{Katsuragawa}}}, \bibinfo {author} {\bibfnamefont {T.}~\bibnamefont
  {{Nakamura}}}, \bibinfo {author} {\bibfnamefont {T.}~\bibnamefont {{Ikeda}}},
  \ and\ \bibinfo {author} {\bibfnamefont {S.}~\bibnamefont {{Capozziello}}},\
  }\href {\doibase 10.1103/PhysRevD.99.124050} {\bibfield  {journal} {\bibinfo
  {journal} {\prd}\ }\textbf {\bibinfo {volume} {99}},\ \bibinfo {eid} {124050}
  (\bibinfo {year} {2019})},\ \Eprint {http://arxiv.org/abs/1902.02494}
  {arXiv:1902.02494 [gr-qc]} \BibitemShut {NoStop}%
\bibitem [{\citenamefont {{Kalita}}\ and\ \citenamefont
  {{Mukhopadhyay}}(2021)}]{2021ApJ...909...65K}%
  \BibitemOpen
  \bibfield  {author} {\bibinfo {author} {\bibfnamefont {S.}~\bibnamefont
  {{Kalita}}}\ and\ \bibinfo {author} {\bibfnamefont {B.}~\bibnamefont
  {{Mukhopadhyay}}},\ }\href {\doibase 10.3847/1538-4357/abddb8} {\bibfield
  {journal} {\bibinfo  {journal} {\apj}\ }\textbf {\bibinfo {volume} {909}},\
  \bibinfo {eid} {65} (\bibinfo {year} {2021})},\ \Eprint
  {http://arxiv.org/abs/2101.07278} {arXiv:2101.07278 [astro-ph.HE]}
  \BibitemShut {NoStop}%
\bibitem [{\citenamefont {{Astashenok}}\ \emph {et~al.}(2013)\citenamefont
  {{Astashenok}}, \citenamefont {{Capozziello}},\ and\ \citenamefont
  {{Odintsov}}}]{2013JCAP...12..040A}%
  \BibitemOpen
  \bibfield  {author} {\bibinfo {author} {\bibfnamefont {A.~V.}\ \bibnamefont
  {{Astashenok}}}, \bibinfo {author} {\bibfnamefont {S.}~\bibnamefont
  {{Capozziello}}}, \ and\ \bibinfo {author} {\bibfnamefont {S.~D.}\
  \bibnamefont {{Odintsov}}},\ }\href {\doibase 10.1088/1475-7516/2013/12/040}
  {\bibfield  {journal} {\bibinfo  {journal} {\jcap}\ }\textbf {\bibinfo
  {volume} {2013}},\ \bibinfo {eid} {040} (\bibinfo {year} {2013})},\ \Eprint
  {http://arxiv.org/abs/1309.1978} {arXiv:1309.1978 [gr-qc]} \BibitemShut
  {NoStop}%
\bibitem [{\citenamefont {{Astashenok}}\ \emph {et~al.}(2014)\citenamefont
  {{Astashenok}}, \citenamefont {{Capozziello}},\ and\ \citenamefont
  {{Odintsov}}}]{2014PhRvD..89j3509A}%
  \BibitemOpen
  \bibfield  {author} {\bibinfo {author} {\bibfnamefont {A.~V.}\ \bibnamefont
  {{Astashenok}}}, \bibinfo {author} {\bibfnamefont {S.}~\bibnamefont
  {{Capozziello}}}, \ and\ \bibinfo {author} {\bibfnamefont {S.~D.}\
  \bibnamefont {{Odintsov}}},\ }\href {\doibase 10.1103/PhysRevD.89.103509}
  {\bibfield  {journal} {\bibinfo  {journal} {\prd}\ }\textbf {\bibinfo
  {volume} {89}},\ \bibinfo {eid} {103509} (\bibinfo {year} {2014})},\ \Eprint
  {http://arxiv.org/abs/1401.4546} {arXiv:1401.4546 [gr-qc]} \BibitemShut
  {NoStop}%
\bibitem [{\citenamefont {{Ganguly}}\ \emph {et~al.}(2014)\citenamefont
  {{Ganguly}}, \citenamefont {{Gannouji}}, \citenamefont {{Goswami}},\ and\
  \citenamefont {{Ray}}}]{2014PhRvD..89f4019G}%
  \BibitemOpen
  \bibfield  {author} {\bibinfo {author} {\bibfnamefont {A.}~\bibnamefont
  {{Ganguly}}}, \bibinfo {author} {\bibfnamefont {R.}~\bibnamefont
  {{Gannouji}}}, \bibinfo {author} {\bibfnamefont {R.}~\bibnamefont
  {{Goswami}}}, \ and\ \bibinfo {author} {\bibfnamefont {S.}~\bibnamefont
  {{Ray}}},\ }\href {\doibase 10.1103/PhysRevD.89.064019} {\bibfield  {journal}
  {\bibinfo  {journal} {\prd}\ }\textbf {\bibinfo {volume} {89}},\ \bibinfo
  {eid} {064019} (\bibinfo {year} {2014})},\ \Eprint
  {http://arxiv.org/abs/1309.3279} {arXiv:1309.3279 [gr-qc]} \BibitemShut
  {NoStop}%
\bibitem [{\citenamefont {{Capozziello}}\ \emph {et~al.}(2016)\citenamefont
  {{Capozziello}}, \citenamefont {{De Laurentis}}, \citenamefont
  {{Farinelli}},\ and\ \citenamefont {{Odintsov}}}]{2016PhRvD..93b3501C}%
  \BibitemOpen
  \bibfield  {author} {\bibinfo {author} {\bibfnamefont {S.}~\bibnamefont
  {{Capozziello}}}, \bibinfo {author} {\bibfnamefont {M.}~\bibnamefont {{De
  Laurentis}}}, \bibinfo {author} {\bibfnamefont {R.}~\bibnamefont
  {{Farinelli}}}, \ and\ \bibinfo {author} {\bibfnamefont {S.~D.}\ \bibnamefont
  {{Odintsov}}},\ }\href {\doibase 10.1103/PhysRevD.93.023501} {\bibfield
  {journal} {\bibinfo  {journal} {\prd}\ }\textbf {\bibinfo {volume} {93}},\
  \bibinfo {eid} {023501} (\bibinfo {year} {2016})},\ \Eprint
  {http://arxiv.org/abs/1509.04163} {arXiv:1509.04163 [gr-qc]} \BibitemShut
  {NoStop}%
\bibitem [{\citenamefont {{Astashenok}}\ and\ \citenamefont
  {{Odintsov}}(2016)}]{2016PhRvD..94f3008A}%
  \BibitemOpen
  \bibfield  {author} {\bibinfo {author} {\bibfnamefont {A.~V.}\ \bibnamefont
  {{Astashenok}}}\ and\ \bibinfo {author} {\bibfnamefont {S.~D.}\ \bibnamefont
  {{Odintsov}}},\ }\href {\doibase 10.1103/PhysRevD.94.063008} {\bibfield
  {journal} {\bibinfo  {journal} {\prd}\ }\textbf {\bibinfo {volume} {94}},\
  \bibinfo {eid} {063008} (\bibinfo {year} {2016})},\ \Eprint
  {http://arxiv.org/abs/1512.07279} {arXiv:1512.07279 [gr-qc]} \BibitemShut
  {NoStop}%
\bibitem [{\citenamefont {{Kalita}}\ and\ \citenamefont
  {{Mukhopadhyay}}(2018)}]{2018JCAP...09..007K}%
  \BibitemOpen
  \bibfield  {author} {\bibinfo {author} {\bibfnamefont {S.}~\bibnamefont
  {{Kalita}}}\ and\ \bibinfo {author} {\bibfnamefont {B.}~\bibnamefont
  {{Mukhopadhyay}}},\ }\href {\doibase 10.1088/1475-7516/2018/09/007}
  {\bibfield  {journal} {\bibinfo  {journal} {\jcap}\ }\textbf {\bibinfo
  {volume} {2018}},\ \bibinfo {eid} {007} (\bibinfo {year} {2018})},\ \Eprint
  {http://arxiv.org/abs/1805.12550} {arXiv:1805.12550 [gr-qc]} \BibitemShut
  {NoStop}%
\bibitem [{\citenamefont {{Das}}\ and\ \citenamefont
  {{Mukhopadhyay}}(2015{\natexlab{a}})}]{2015JCAP...05..045D}%
  \BibitemOpen
  \bibfield  {author} {\bibinfo {author} {\bibfnamefont {U.}~\bibnamefont
  {{Das}}}\ and\ \bibinfo {author} {\bibfnamefont {B.}~\bibnamefont
  {{Mukhopadhyay}}},\ }\href {\doibase 10.1088/1475-7516/2015/05/045}
  {\bibfield  {journal} {\bibinfo  {journal} {\jcap}\ }\textbf {\bibinfo
  {volume} {2015}},\ \bibinfo {eid} {045} (\bibinfo {year}
  {2015}{\natexlab{a}})},\ \Eprint {http://arxiv.org/abs/1411.1515}
  {arXiv:1411.1515 [astro-ph.SR]} \BibitemShut {NoStop}%
\bibitem [{\citenamefont {{Astashenok}}\ \emph {et~al.}(2015)\citenamefont
  {{Astashenok}}, \citenamefont {{Capozziello}},\ and\ \citenamefont
  {{Odintsov}}}]{2015JCAP...01..001A}%
  \BibitemOpen
  \bibfield  {author} {\bibinfo {author} {\bibfnamefont {A.~V.}\ \bibnamefont
  {{Astashenok}}}, \bibinfo {author} {\bibfnamefont {S.}~\bibnamefont
  {{Capozziello}}}, \ and\ \bibinfo {author} {\bibfnamefont {S.~D.}\
  \bibnamefont {{Odintsov}}},\ }\href {\doibase 10.1088/1475-7516/2015/01/001}
  {\bibfield  {journal} {\bibinfo  {journal} {\jcap}\ }\textbf {\bibinfo
  {volume} {2015}},\ \bibinfo {eid} {001} (\bibinfo {year} {2015})},\ \Eprint
  {http://arxiv.org/abs/1408.3856} {arXiv:1408.3856 [gr-qc]} \BibitemShut
  {NoStop}%
\bibitem [{\citenamefont {{Astashenok}}(2016)}]{2016IJMPS..4160130A}%
  \BibitemOpen
  \bibfield  {author} {\bibinfo {author} {\bibfnamefont {A.~V.}\ \bibnamefont
  {{Astashenok}}},\ }in\ \href {\doibase 10.1142/S2010194516601307} {\emph
  {\bibinfo {booktitle} {International Journal of Modern Physics Conference
  Series}}},\ \bibinfo {series} {International Journal of Modern Physics
  Conference Series}, Vol.~\bibinfo {volume} {41}\ (\bibinfo {year} {2016})\
  p.\ \bibinfo {pages} {1660130}\BibitemShut {NoStop}%
\bibitem [{\citenamefont {{Chiba}}(2003)}]{2003PhLB..575....1C}%
  \BibitemOpen
  \bibfield  {author} {\bibinfo {author} {\bibfnamefont {T.}~\bibnamefont
  {{Chiba}}},\ }\href {\doibase 10.1016/j.physletb.2003.09.033} {\bibfield
  {journal} {\bibinfo  {journal} {Physics Letters B}\ }\textbf {\bibinfo
  {volume} {575}},\ \bibinfo {pages} {1} (\bibinfo {year} {2003})},\ \Eprint
  {http://arxiv.org/abs/astro-ph/0307338} {arXiv:astro-ph/0307338 [astro-ph]}
  \BibitemShut {NoStop}%
\bibitem [{\citenamefont {{Dolgov}}\ and\ \citenamefont
  {{Kawasaki}}(2003)}]{2003PhLB..573....1D}%
  \BibitemOpen
  \bibfield  {author} {\bibinfo {author} {\bibfnamefont {A.~D.}\ \bibnamefont
  {{Dolgov}}}\ and\ \bibinfo {author} {\bibfnamefont {M.}~\bibnamefont
  {{Kawasaki}}},\ }\href {\doibase 10.1016/j.physletb.2003.08.039} {\bibfield
  {journal} {\bibinfo  {journal} {Physics Letters B}\ }\textbf {\bibinfo
  {volume} {573}},\ \bibinfo {pages} {1} (\bibinfo {year} {2003})},\ \Eprint
  {http://arxiv.org/abs/astro-ph/0307285} {arXiv:astro-ph/0307285 [astro-ph]}
  \BibitemShut {NoStop}%
\bibitem [{\citenamefont {{Fay}}\ \emph {et~al.}(2007)\citenamefont {{Fay}},
  \citenamefont {{Tavakol}},\ and\ \citenamefont
  {{Tsujikawa}}}]{2007PhRvD..75f3509F}%
  \BibitemOpen
  \bibfield  {author} {\bibinfo {author} {\bibfnamefont {S.}~\bibnamefont
  {{Fay}}}, \bibinfo {author} {\bibfnamefont {R.}~\bibnamefont {{Tavakol}}}, \
  and\ \bibinfo {author} {\bibfnamefont {S.}~\bibnamefont {{Tsujikawa}}},\
  }\href {\doibase 10.1103/PhysRevD.75.063509} {\bibfield  {journal} {\bibinfo
  {journal} {\prd}\ }\textbf {\bibinfo {volume} {75}},\ \bibinfo {eid} {063509}
  (\bibinfo {year} {2007})},\ \Eprint {http://arxiv.org/abs/astro-ph/0701479}
  {arXiv:astro-ph/0701479 [astro-ph]} \BibitemShut {NoStop}%
\bibitem [{\citenamefont
  {{Sotiriou}}(2006{\natexlab{a}})}]{2006CQGra..23.1253S}%
  \BibitemOpen
  \bibfield  {author} {\bibinfo {author} {\bibfnamefont {T.~P.}\ \bibnamefont
  {{Sotiriou}}},\ }\href {\doibase 10.1088/0264-9381/23/4/012} {\bibfield
  {journal} {\bibinfo  {journal} {Classical and Quantum Gravity}\ }\textbf
  {\bibinfo {volume} {23}},\ \bibinfo {pages} {1253} (\bibinfo {year}
  {2006}{\natexlab{a}})},\ \Eprint {http://arxiv.org/abs/gr-qc/0512017}
  {arXiv:gr-qc/0512017 [gr-qc]} \BibitemShut {NoStop}%
\bibitem [{\citenamefont {{Toniato}}\ \emph {et~al.}(2020)\citenamefont
  {{Toniato}}, \citenamefont {{Rodrigues}},\ and\ \citenamefont
  {{Wojnar}}}]{toniato2020palatini}%
  \BibitemOpen
  \bibfield  {author} {\bibinfo {author} {\bibfnamefont {J.~D.}\ \bibnamefont
  {{Toniato}}}, \bibinfo {author} {\bibfnamefont {D.~C.}\ \bibnamefont
  {{Rodrigues}}}, \ and\ \bibinfo {author} {\bibfnamefont {A.}~\bibnamefont
  {{Wojnar}}},\ }\href {\doibase 10.1103/PhysRevD.101.064050} {\bibfield
  {journal} {\bibinfo  {journal} {\prd}\ }\textbf {\bibinfo {volume} {101}},\
  \bibinfo {eid} {064050} (\bibinfo {year} {2020})},\ \Eprint
  {http://arxiv.org/abs/1912.12234} {arXiv:1912.12234 [gr-qc]} \BibitemShut
  {NoStop}%
\bibitem [{\citenamefont {{Ferraris}}\ \emph {et~al.}(1994)\citenamefont
  {{Ferraris}}, \citenamefont {{Francaviglia}},\ and\ \citenamefont
  {{Volovich}}}]{1994CQGra..11.1505F}%
  \BibitemOpen
  \bibfield  {author} {\bibinfo {author} {\bibfnamefont {M.}~\bibnamefont
  {{Ferraris}}}, \bibinfo {author} {\bibfnamefont {M.}~\bibnamefont
  {{Francaviglia}}}, \ and\ \bibinfo {author} {\bibfnamefont {I.}~\bibnamefont
  {{Volovich}}},\ }\href {\doibase 10.1088/0264-9381/11/6/015} {\bibfield
  {journal} {\bibinfo  {journal} {Classical and Quantum Gravity}\ }\textbf
  {\bibinfo {volume} {11}},\ \bibinfo {pages} {1505} (\bibinfo {year}
  {1994})},\ \Eprint {http://arxiv.org/abs/gr-qc/9303007} {arXiv:gr-qc/9303007
  [gr-qc]} \BibitemShut {NoStop}%
\bibitem [{\citenamefont {{Sotiriou}}(2007)}]{2007PhLB..645..389S}%
  \BibitemOpen
  \bibfield  {author} {\bibinfo {author} {\bibfnamefont {T.~P.}\ \bibnamefont
  {{Sotiriou}}},\ }\href {\doibase 10.1016/j.physletb.2007.01.003} {\bibfield
  {journal} {\bibinfo  {journal} {Physics Letters B}\ }\textbf {\bibinfo
  {volume} {645}},\ \bibinfo {pages} {389} (\bibinfo {year} {2007})},\ \Eprint
  {http://arxiv.org/abs/gr-qc/0611107} {arXiv:gr-qc/0611107 [gr-qc]}
  \BibitemShut {NoStop}%
\bibitem [{\citenamefont {{Kausar}}\ \emph {et~al.}(2016)\citenamefont
  {{Kausar}}, \citenamefont {{Philippoz}},\ and\ \citenamefont
  {{Jetzer}}}]{2016PhRvD..93l4071K}%
  \BibitemOpen
  \bibfield  {author} {\bibinfo {author} {\bibfnamefont {H.~R.}\ \bibnamefont
  {{Kausar}}}, \bibinfo {author} {\bibfnamefont {L.}~\bibnamefont
  {{Philippoz}}}, \ and\ \bibinfo {author} {\bibfnamefont {P.}~\bibnamefont
  {{Jetzer}}},\ }\href {\doibase 10.1103/PhysRevD.93.124071} {\bibfield
  {journal} {\bibinfo  {journal} {\prd}\ }\textbf {\bibinfo {volume} {93}},\
  \bibinfo {eid} {124071} (\bibinfo {year} {2016})},\ \Eprint
  {http://arxiv.org/abs/1606.07000} {arXiv:1606.07000 [gr-qc]} \BibitemShut
  {NoStop}%
\bibitem [{\citenamefont {{Meng}}\ and\ \citenamefont
  {{Wang}}(2004)}]{2004GReGr..36.1947M}%
  \BibitemOpen
  \bibfield  {author} {\bibinfo {author} {\bibfnamefont {X.-H.}\ \bibnamefont
  {{Meng}}}\ and\ \bibinfo {author} {\bibfnamefont {P.}~\bibnamefont
  {{Wang}}},\ }\href {\doibase 10.1023/B:GERG.0000036052.81522.fe} {\bibfield
  {journal} {\bibinfo  {journal} {General Relativity and Gravitation}\ }\textbf
  {\bibinfo {volume} {36}},\ \bibinfo {pages} {1947} (\bibinfo {year}
  {2004})},\ \Eprint {http://arxiv.org/abs/gr-qc/0311019} {arXiv:gr-qc/0311019
  [gr-qc]} \BibitemShut {NoStop}%
\bibitem [{\citenamefont {{Dom{\'\i}nguez}}\ and\ \citenamefont
  {{Barraco}}(2004)}]{2004PhRvD..70d3505D}%
  \BibitemOpen
  \bibfield  {author} {\bibinfo {author} {\bibfnamefont {A.~E.}\ \bibnamefont
  {{Dom{\'\i}nguez}}}\ and\ \bibinfo {author} {\bibfnamefont {D.~E.}\
  \bibnamefont {{Barraco}}},\ }\href {\doibase 10.1103/PhysRevD.70.043505}
  {\bibfield  {journal} {\bibinfo  {journal} {\prd}\ }\textbf {\bibinfo
  {volume} {70}},\ \bibinfo {eid} {043505} (\bibinfo {year} {2004})},\ \Eprint
  {http://arxiv.org/abs/gr-qc/0408069} {arXiv:gr-qc/0408069 [astro-ph]}
  \BibitemShut {NoStop}%
\bibitem [{\citenamefont
  {{Sotiriou}}(2006{\natexlab{b}})}]{2006GReGr..38.1407S}%
  \BibitemOpen
  \bibfield  {author} {\bibinfo {author} {\bibfnamefont {T.~P.}\ \bibnamefont
  {{Sotiriou}}},\ }\href {\doibase 10.1007/s10714-006-0328-8} {\bibfield
  {journal} {\bibinfo  {journal} {General Relativity and Gravitation}\ }\textbf
  {\bibinfo {volume} {38}},\ \bibinfo {pages} {1407} (\bibinfo {year}
  {2006}{\natexlab{b}})},\ \Eprint {http://arxiv.org/abs/gr-qc/0507027}
  {arXiv:gr-qc/0507027 [gr-qc]} \BibitemShut {NoStop}%
\bibitem [{\citenamefont
  {{Sotiriou}}(2006{\natexlab{c}})}]{2006PhRvD..73f3515S}%
  \BibitemOpen
  \bibfield  {author} {\bibinfo {author} {\bibfnamefont {T.~P.}\ \bibnamefont
  {{Sotiriou}}},\ }\href {\doibase 10.1103/PhysRevD.73.063515} {\bibfield
  {journal} {\bibinfo  {journal} {\prd}\ }\textbf {\bibinfo {volume} {73}},\
  \bibinfo {eid} {063515} (\bibinfo {year} {2006}{\natexlab{c}})},\ \Eprint
  {http://arxiv.org/abs/gr-qc/0509029} {arXiv:gr-qc/0509029 [gr-qc]}
  \BibitemShut {NoStop}%
\bibitem [{\citenamefont {{Nojiri}}\ and\ \citenamefont
  {{Odintsov}}(2004)}]{2004GReGr..36.1765N}%
  \BibitemOpen
  \bibfield  {author} {\bibinfo {author} {\bibfnamefont {S.}~\bibnamefont
  {{Nojiri}}}\ and\ \bibinfo {author} {\bibfnamefont {S.~D.}\ \bibnamefont
  {{Odintsov}}},\ }\href {\doibase 10.1023/B:GERG.0000035950.40718.48}
  {\bibfield  {journal} {\bibinfo  {journal} {General Relativity and
  Gravitation}\ }\textbf {\bibinfo {volume} {36}},\ \bibinfo {pages} {1765}
  (\bibinfo {year} {2004})},\ \Eprint {http://arxiv.org/abs/hep-th/0308176}
  {arXiv:hep-th/0308176 [hep-th]} \BibitemShut {NoStop}%
\bibitem [{\citenamefont {{Amarzguioui}}\ \emph {et~al.}(2006)\citenamefont
  {{Amarzguioui}}, \citenamefont {{Elgar{\o}y}}, \citenamefont {{Mota}},\ and\
  \citenamefont {{Multam{\"a}ki}}}]{2006A&A...454..707A}%
  \BibitemOpen
  \bibfield  {author} {\bibinfo {author} {\bibfnamefont {M.}~\bibnamefont
  {{Amarzguioui}}}, \bibinfo {author} {\bibfnamefont {{\O}.}~\bibnamefont
  {{Elgar{\o}y}}}, \bibinfo {author} {\bibfnamefont {D.~F.}\ \bibnamefont
  {{Mota}}}, \ and\ \bibinfo {author} {\bibfnamefont {T.}~\bibnamefont
  {{Multam{\"a}ki}}},\ }\href {\doibase 10.1051/0004-6361:20064994} {\bibfield
  {journal} {\bibinfo  {journal} {\aap}\ }\textbf {\bibinfo {volume} {454}},\
  \bibinfo {pages} {707} (\bibinfo {year} {2006})},\ \Eprint
  {http://arxiv.org/abs/astro-ph/0510519} {arXiv:astro-ph/0510519 [astro-ph]}
  \BibitemShut {NoStop}%
\bibitem [{\citenamefont {{Szyd{\l}owski}}\ \emph {et~al.}(2016)\citenamefont
  {{Szyd{\l}owski}}, \citenamefont {{Stachowski}}, \citenamefont {{Borowiec}},\
  and\ \citenamefont {{Wojnar}}}]{2016EPJC...76..567S}%
  \BibitemOpen
  \bibfield  {author} {\bibinfo {author} {\bibfnamefont {M.}~\bibnamefont
  {{Szyd{\l}owski}}}, \bibinfo {author} {\bibfnamefont {A.}~\bibnamefont
  {{Stachowski}}}, \bibinfo {author} {\bibfnamefont {A.}~\bibnamefont
  {{Borowiec}}}, \ and\ \bibinfo {author} {\bibfnamefont {A.}~\bibnamefont
  {{Wojnar}}},\ }\href {\doibase 10.1140/epjc/s10052-016-4426-9} {\bibfield
  {journal} {\bibinfo  {journal} {European Physical Journal C}\ }\textbf
  {\bibinfo {volume} {76}},\ \bibinfo {eid} {567} (\bibinfo {year} {2016})},\
  \Eprint {http://arxiv.org/abs/1512.04580} {arXiv:1512.04580 [gr-qc]}
  \BibitemShut {NoStop}%
\bibitem [{\citenamefont {{Borowiec}}\ \emph {et~al.}(2016)\citenamefont
  {{Borowiec}}, \citenamefont {{Stachowski}}, \citenamefont {{Szyd{\l}owski}},\
  and\ \citenamefont {{Wojnar}}}]{2016JCAP...01..040B}%
  \BibitemOpen
  \bibfield  {author} {\bibinfo {author} {\bibfnamefont {A.}~\bibnamefont
  {{Borowiec}}}, \bibinfo {author} {\bibfnamefont {A.}~\bibnamefont
  {{Stachowski}}}, \bibinfo {author} {\bibfnamefont {M.}~\bibnamefont
  {{Szyd{\l}owski}}}, \ and\ \bibinfo {author} {\bibfnamefont {A.}~\bibnamefont
  {{Wojnar}}},\ }\href {\doibase 10.1088/1475-7516/2016/01/040} {\bibfield
  {journal} {\bibinfo  {journal} {\jcap}\ }\textbf {\bibinfo {volume} {2016}},\
  \bibinfo {eid} {040} (\bibinfo {year} {2016})},\ \Eprint
  {http://arxiv.org/abs/1512.01199} {arXiv:1512.01199 [gr-qc]} \BibitemShut
  {NoStop}%
\bibitem [{\citenamefont {{Borowiec}}\ and\ \citenamefont
  {{Kozak}}(2020)}]{2020JCAP...07..003B}%
  \BibitemOpen
  \bibfield  {author} {\bibinfo {author} {\bibfnamefont {A.}~\bibnamefont
  {{Borowiec}}}\ and\ \bibinfo {author} {\bibfnamefont {A.}~\bibnamefont
  {{Kozak}}},\ }\href {\doibase 10.1088/1475-7516/2020/07/003} {\bibfield
  {journal} {\bibinfo  {journal} {\jcap}\ }\textbf {\bibinfo {volume} {2020}},\
  \bibinfo {eid} {003} (\bibinfo {year} {2020})},\ \Eprint
  {http://arxiv.org/abs/2003.02741} {arXiv:2003.02741 [gr-qc]} \BibitemShut
  {NoStop}%
\bibitem [{\citenamefont {{J{\"a}rv}}\ \emph {et~al.}(2020)\citenamefont
  {{J{\"a}rv}}, \citenamefont {{Karam}}, \citenamefont {{Kozak}}, \citenamefont
  {{Lykkas}}, \citenamefont {{Racioppi}},\ and\ \citenamefont
  {{Saal}}}]{2020PhRvD.102d4029J}%
  \BibitemOpen
  \bibfield  {author} {\bibinfo {author} {\bibfnamefont {L.}~\bibnamefont
  {{J{\"a}rv}}}, \bibinfo {author} {\bibfnamefont {A.}~\bibnamefont {{Karam}}},
  \bibinfo {author} {\bibfnamefont {A.}~\bibnamefont {{Kozak}}}, \bibinfo
  {author} {\bibfnamefont {A.}~\bibnamefont {{Lykkas}}}, \bibinfo {author}
  {\bibfnamefont {A.}~\bibnamefont {{Racioppi}}}, \ and\ \bibinfo {author}
  {\bibfnamefont {M.}~\bibnamefont {{Saal}}},\ }\href {\doibase
  10.1103/PhysRevD.102.044029} {\bibfield  {journal} {\bibinfo  {journal}
  {\prd}\ }\textbf {\bibinfo {volume} {102}},\ \bibinfo {eid} {044029}
  (\bibinfo {year} {2020})},\ \Eprint {http://arxiv.org/abs/2005.14571}
  {arXiv:2005.14571 [gr-qc]} \BibitemShut {NoStop}%
\bibitem [{\citenamefont {{Gialamas}}\ \emph {et~al.}(2021)\citenamefont
  {{Gialamas}}, \citenamefont {{Karam}}, \citenamefont {{Pappas}},\ and\
  \citenamefont {{Spanos}}}]{2021PhRvD.104b3521G}%
  \BibitemOpen
  \bibfield  {author} {\bibinfo {author} {\bibfnamefont {I.~D.}\ \bibnamefont
  {{Gialamas}}}, \bibinfo {author} {\bibfnamefont {A.}~\bibnamefont {{Karam}}},
  \bibinfo {author} {\bibfnamefont {T.~D.}\ \bibnamefont {{Pappas}}}, \ and\
  \bibinfo {author} {\bibfnamefont {V.~C.}\ \bibnamefont {{Spanos}}},\ }\href
  {\doibase 10.1103/PhysRevD.104.023521} {\bibfield  {journal} {\bibinfo
  {journal} {\prd}\ }\textbf {\bibinfo {volume} {104}},\ \bibinfo {eid}
  {023521} (\bibinfo {year} {2021})},\ \Eprint
  {http://arxiv.org/abs/2104.04550} {arXiv:2104.04550 [astro-ph.CO]}
  \BibitemShut {NoStop}%
\bibitem [{\citenamefont {{Herzog}}\ and\ \citenamefont
  {{Sanchis-Alepuz}}(2021)}]{2021EPJC...81..888H}%
  \BibitemOpen
  \bibfield  {author} {\bibinfo {author} {\bibfnamefont {G.}~\bibnamefont
  {{Herzog}}}\ and\ \bibinfo {author} {\bibfnamefont {H.}~\bibnamefont
  {{Sanchis-Alepuz}}},\ }\href {\doibase 10.1140/epjc/s10052-021-09662-z}
  {\bibfield  {journal} {\bibinfo  {journal} {European Physical Journal C}\
  }\textbf {\bibinfo {volume} {81}},\ \bibinfo {eid} {888} (\bibinfo {year}
  {2021})},\ \Eprint {http://arxiv.org/abs/2102.05722} {arXiv:2102.05722
  [gr-qc]} \BibitemShut {NoStop}%
\bibitem [{\citenamefont {{Wojnar}}(2021{\natexlab{a}})}]{2021IJGMM..1840006W}%
  \BibitemOpen
  \bibfield  {author} {\bibinfo {author} {\bibfnamefont {A.}~\bibnamefont
  {{Wojnar}}},\ }\href {\doibase 10.1142/S0219887821400065} {\bibfield
  {journal} {\bibinfo  {journal} {International Journal of Geometric Methods in
  Modern Physics}\ }\textbf {\bibinfo {volume} {18}},\ \bibinfo {eid}
  {2140006-60} (\bibinfo {year} {2021}{\natexlab{a}})},\ \Eprint
  {http://arxiv.org/abs/2012.13927} {arXiv:2012.13927 [gr-qc]} \BibitemShut
  {NoStop}%
\bibitem [{\citenamefont {{Banerjee}}\ \emph {et~al.}(2017)\citenamefont
  {{Banerjee}}, \citenamefont {{Shankar}},\ and\ \citenamefont
  {{Singh}}}]{2017JCAP...10..004B}%
  \BibitemOpen
  \bibfield  {author} {\bibinfo {author} {\bibfnamefont {S.}~\bibnamefont
  {{Banerjee}}}, \bibinfo {author} {\bibfnamefont {S.}~\bibnamefont
  {{Shankar}}}, \ and\ \bibinfo {author} {\bibfnamefont {T.~P.}\ \bibnamefont
  {{Singh}}},\ }\href {\doibase 10.1088/1475-7516/2017/10/004} {\bibfield
  {journal} {\bibinfo  {journal} {\jcap}\ }\textbf {\bibinfo {volume} {2017}},\
  \bibinfo {eid} {004} (\bibinfo {year} {2017})},\ \Eprint
  {http://arxiv.org/abs/1705.01048} {arXiv:1705.01048 [gr-qc]} \BibitemShut
  {NoStop}%
\bibitem [{\citenamefont {{Shapiro}}\ and\ \citenamefont
  {{Teukolsky}}(1986)}]{1986bhwd.book.....S}%
  \BibitemOpen
  \bibfield  {author} {\bibinfo {author} {\bibfnamefont {S.~L.}\ \bibnamefont
  {{Shapiro}}}\ and\ \bibinfo {author} {\bibfnamefont {S.~A.}\ \bibnamefont
  {{Teukolsky}}},\ }\href@noop {} {\emph {\bibinfo {title} {{Black Holes, White
  Dwarfs and Neutron Stars: The Physics of Compact Objects}}}}\ (\bibinfo
  {year} {1986})\BibitemShut {NoStop}%
\bibitem [{\citenamefont {{Lauffer}}\ \emph {et~al.}(2018)\citenamefont
  {{Lauffer}}, \citenamefont {{Romero}},\ and\ \citenamefont
  {{Kepler}}}]{2018MNRAS.480.1547L}%
  \BibitemOpen
  \bibfield  {author} {\bibinfo {author} {\bibfnamefont {G.~R.}\ \bibnamefont
  {{Lauffer}}}, \bibinfo {author} {\bibfnamefont {A.~D.}\ \bibnamefont
  {{Romero}}}, \ and\ \bibinfo {author} {\bibfnamefont {S.~O.}\ \bibnamefont
  {{Kepler}}},\ }\href {\doibase 10.1093/mnras/sty1925} {\bibfield  {journal}
  {\bibinfo  {journal} {\mnras}\ }\textbf {\bibinfo {volume} {480}},\ \bibinfo
  {pages} {1547} (\bibinfo {year} {2018})},\ \Eprint
  {http://arxiv.org/abs/1807.04774} {arXiv:1807.04774 [astro-ph.SR]}
  \BibitemShut {NoStop}%
\bibitem [{\citenamefont {{Chandrasekhar}}(1935)}]{1935MNRAS..95..207C}%
  \BibitemOpen
  \bibfield  {author} {\bibinfo {author} {\bibfnamefont {S.}~\bibnamefont
  {{Chandrasekhar}}},\ }\href {\doibase 10.1093/mnras/95.3.207} {\bibfield
  {journal} {\bibinfo  {journal} {\mnras}\ }\textbf {\bibinfo {volume} {95}},\
  \bibinfo {pages} {207} (\bibinfo {year} {1935})}\BibitemShut {NoStop}%
\bibitem [{\citenamefont {{Nomoto}}\ \emph {et~al.}(1997)\citenamefont
  {{Nomoto}}, \citenamefont {{Iwamoto}},\ and\ \citenamefont
  {{Kishimoto}}}]{1997Sci...276.1378N}%
  \BibitemOpen
  \bibfield  {author} {\bibinfo {author} {\bibfnamefont {K.}~\bibnamefont
  {{Nomoto}}}, \bibinfo {author} {\bibfnamefont {K.}~\bibnamefont {{Iwamoto}}},
  \ and\ \bibinfo {author} {\bibfnamefont {N.}~\bibnamefont {{Kishimoto}}},\
  }\href {\doibase 10.1126/science.276.5317.1378} {\bibfield  {journal}
  {\bibinfo  {journal} {Science}\ }\textbf {\bibinfo {volume} {276}},\ \bibinfo
  {pages} {1378} (\bibinfo {year} {1997})},\ \Eprint
  {http://arxiv.org/abs/astro-ph/9706007} {arXiv:astro-ph/9706007 [astro-ph]}
  \BibitemShut {NoStop}%
\bibitem [{\citenamefont {{Wright}}\ and\ \citenamefont
  {{Li}}(2018)}]{2018PhRvD..97h3505W}%
  \BibitemOpen
  \bibfield  {author} {\bibinfo {author} {\bibfnamefont {B.~S.}\ \bibnamefont
  {{Wright}}}\ and\ \bibinfo {author} {\bibfnamefont {B.}~\bibnamefont
  {{Li}}},\ }\href {\doibase 10.1103/PhysRevD.97.083505} {\bibfield  {journal}
  {\bibinfo  {journal} {\prd}\ }\textbf {\bibinfo {volume} {97}},\ \bibinfo
  {eid} {083505} (\bibinfo {year} {2018})},\ \Eprint
  {http://arxiv.org/abs/1710.07018} {arXiv:1710.07018 [astro-ph.CO]}
  \BibitemShut {NoStop}%
\bibitem [{\citenamefont {{Howell}}\ \emph {et~al.}(2006)\citenamefont
  {{Howell}}, \citenamefont {{Sullivan}}, \citenamefont {{Nugent}},
  \citenamefont {{Ellis}}, \citenamefont {{Conley}}, \citenamefont {{Le
  Borgne}}, \citenamefont {{Carlberg}}, \citenamefont {{Guy}}, \citenamefont
  {{Balam}}, \citenamefont {{Basa}}, \citenamefont {{Fouchez}}, \citenamefont
  {{Hook}}, \citenamefont {{Hsiao}}, \citenamefont {{Neill}}, \citenamefont
  {{Pain}}, \citenamefont {{Perrett}},\ and\ \citenamefont
  {{Pritchet}}}]{2006Natur.443..308H}%
  \BibitemOpen
  \bibfield  {author} {\bibinfo {author} {\bibfnamefont {D.~A.}\ \bibnamefont
  {{Howell}}}, \bibinfo {author} {\bibfnamefont {M.}~\bibnamefont
  {{Sullivan}}}, \bibinfo {author} {\bibfnamefont {P.~E.}\ \bibnamefont
  {{Nugent}}}, \bibinfo {author} {\bibfnamefont {R.~S.}\ \bibnamefont
  {{Ellis}}}, \bibinfo {author} {\bibfnamefont {A.~J.}\ \bibnamefont
  {{Conley}}}, \bibinfo {author} {\bibfnamefont {D.}~\bibnamefont {{Le
  Borgne}}}, \bibinfo {author} {\bibfnamefont {R.~G.}\ \bibnamefont
  {{Carlberg}}}, \bibinfo {author} {\bibfnamefont {J.}~\bibnamefont {{Guy}}},
  \bibinfo {author} {\bibfnamefont {D.}~\bibnamefont {{Balam}}}, \bibinfo
  {author} {\bibfnamefont {S.}~\bibnamefont {{Basa}}}, \bibinfo {author}
  {\bibfnamefont {D.}~\bibnamefont {{Fouchez}}}, \bibinfo {author}
  {\bibfnamefont {I.~M.}\ \bibnamefont {{Hook}}}, \bibinfo {author}
  {\bibfnamefont {E.~Y.}\ \bibnamefont {{Hsiao}}}, \bibinfo {author}
  {\bibfnamefont {J.~D.}\ \bibnamefont {{Neill}}}, \bibinfo {author}
  {\bibfnamefont {R.}~\bibnamefont {{Pain}}}, \bibinfo {author} {\bibfnamefont
  {K.~M.}\ \bibnamefont {{Perrett}}}, \ and\ \bibinfo {author} {\bibfnamefont
  {C.~J.}\ \bibnamefont {{Pritchet}}},\ }\href {\doibase 10.1038/nature05103}
  {\bibfield  {journal} {\bibinfo  {journal} {\nat}\ }\textbf {\bibinfo
  {volume} {443}},\ \bibinfo {pages} {308} (\bibinfo {year} {2006})},\ \Eprint
  {http://arxiv.org/abs/astro-ph/0609616} {arXiv:astro-ph/0609616 [astro-ph]}
  \BibitemShut {NoStop}%
\bibitem [{\citenamefont {{Scalzo}}\ \emph {et~al.}(2010)\citenamefont
  {{Scalzo}}, \citenamefont {{Aldering}}, \citenamefont {{Antilogus}},
  \citenamefont {{Aragon}}, \citenamefont {{Bailey}}, \citenamefont {{Baltay}},
  \citenamefont {{Bongard}}, \citenamefont {{Buton}}, \citenamefont
  {{Childress}}, \citenamefont {{Chotard}}, \citenamefont {{Copin}},
  \citenamefont {{Fakhouri}}, \citenamefont {{Gal-Yam}}, \citenamefont
  {{Gangler}}, \citenamefont {{Hoyer}}, \citenamefont {{Kasliwal}},
  \citenamefont {{Loken}}, \citenamefont {{Nugent}}, \citenamefont {{Pain}},
  \citenamefont {{P{\'e}contal}}, \citenamefont {{Pereira}}, \citenamefont
  {{Perlmutter}}, \citenamefont {{Rabinowitz}}, \citenamefont {{Rau}},
  \citenamefont {{Rigaudier}}, \citenamefont {{Runge}}, \citenamefont
  {{Smadja}}, \citenamefont {{Tao}}, \citenamefont {{Thomas}}, \citenamefont
  {{Weaver}},\ and\ \citenamefont {{Wu}}}]{2010ApJ...713.1073S}%
  \BibitemOpen
  \bibfield  {author} {\bibinfo {author} {\bibfnamefont {R.~A.}\ \bibnamefont
  {{Scalzo}}}, \bibinfo {author} {\bibfnamefont {G.}~\bibnamefont
  {{Aldering}}}, \bibinfo {author} {\bibfnamefont {P.}~\bibnamefont
  {{Antilogus}}}, \bibinfo {author} {\bibfnamefont {C.}~\bibnamefont
  {{Aragon}}}, \bibinfo {author} {\bibfnamefont {S.}~\bibnamefont {{Bailey}}},
  \bibinfo {author} {\bibfnamefont {C.}~\bibnamefont {{Baltay}}}, \bibinfo
  {author} {\bibfnamefont {S.}~\bibnamefont {{Bongard}}}, \bibinfo {author}
  {\bibfnamefont {C.}~\bibnamefont {{Buton}}}, \bibinfo {author} {\bibfnamefont
  {M.}~\bibnamefont {{Childress}}}, \bibinfo {author} {\bibfnamefont
  {N.}~\bibnamefont {{Chotard}}}, \bibinfo {author} {\bibfnamefont
  {Y.}~\bibnamefont {{Copin}}}, \bibinfo {author} {\bibfnamefont {H.~K.}\
  \bibnamefont {{Fakhouri}}}, \bibinfo {author} {\bibfnamefont
  {A.}~\bibnamefont {{Gal-Yam}}}, \bibinfo {author} {\bibfnamefont
  {E.}~\bibnamefont {{Gangler}}}, \bibinfo {author} {\bibfnamefont
  {S.}~\bibnamefont {{Hoyer}}}, \bibinfo {author} {\bibfnamefont
  {M.}~\bibnamefont {{Kasliwal}}}, \bibinfo {author} {\bibfnamefont
  {S.}~\bibnamefont {{Loken}}}, \bibinfo {author} {\bibfnamefont
  {P.}~\bibnamefont {{Nugent}}}, \bibinfo {author} {\bibfnamefont
  {R.}~\bibnamefont {{Pain}}}, \bibinfo {author} {\bibfnamefont
  {E.}~\bibnamefont {{P{\'e}contal}}}, \bibinfo {author} {\bibfnamefont
  {R.}~\bibnamefont {{Pereira}}}, \bibinfo {author} {\bibfnamefont
  {S.}~\bibnamefont {{Perlmutter}}}, \bibinfo {author} {\bibfnamefont
  {D.}~\bibnamefont {{Rabinowitz}}}, \bibinfo {author} {\bibfnamefont
  {A.}~\bibnamefont {{Rau}}}, \bibinfo {author} {\bibfnamefont
  {G.}~\bibnamefont {{Rigaudier}}}, \bibinfo {author} {\bibfnamefont
  {K.}~\bibnamefont {{Runge}}}, \bibinfo {author} {\bibfnamefont
  {G.}~\bibnamefont {{Smadja}}}, \bibinfo {author} {\bibfnamefont
  {C.}~\bibnamefont {{Tao}}}, \bibinfo {author} {\bibfnamefont {R.~C.}\
  \bibnamefont {{Thomas}}}, \bibinfo {author} {\bibfnamefont {B.}~\bibnamefont
  {{Weaver}}}, \ and\ \bibinfo {author} {\bibfnamefont {C.}~\bibnamefont
  {{Wu}}},\ }\href {\doibase 10.1088/0004-637X/713/2/1073} {\bibfield
  {journal} {\bibinfo  {journal} {\apj}\ }\textbf {\bibinfo {volume} {713}},\
  \bibinfo {pages} {1073} (\bibinfo {year} {2010})},\ \Eprint
  {http://arxiv.org/abs/1003.2217} {arXiv:1003.2217 [astro-ph.CO]} \BibitemShut
  {NoStop}%
\bibitem [{\citenamefont {{Yamanaka}}\ \emph {et~al.}(2009)\citenamefont
  {{Yamanaka}}, \citenamefont {{Kawabata}}, \citenamefont {{Kinugasa}},
  \citenamefont {{Tanaka}}, \citenamefont {{Imada}}, \citenamefont {{Maeda}},
  \citenamefont {{Nomoto}}, \citenamefont {{Arai}}, \citenamefont
  {{Chiyonobu}}, \citenamefont {{Fukazawa}}, \citenamefont {{Hashimoto}},
  \citenamefont {{Honda}}, \citenamefont {{Ikejiri}}, \citenamefont {{Itoh}},
  \citenamefont {{Kamata}}, \citenamefont {{Kawai}}, \citenamefont {{Komatsu}},
  \citenamefont {{Konishi}}, \citenamefont {{Kuroda}}, \citenamefont
  {{Miyamoto}}, \citenamefont {{Miyazaki}}, \citenamefont {{Nagae}},
  \citenamefont {{Nakaya}}, \citenamefont {{Ohsugi}}, \citenamefont
  {{Omodaka}}, \citenamefont {{Sakai}}, \citenamefont {{Sasada}}, \citenamefont
  {{Suzuki}}, \citenamefont {{Taguchi}}, \citenamefont {{Takahashi}},
  \citenamefont {{Tanaka}}, \citenamefont {{Uemura}}, \citenamefont
  {{Yamashita}}, \citenamefont {{Yanagisawa}},\ and\ \citenamefont
  {{Yoshida}}}]{2009ApJ...707L.118Y}%
  \BibitemOpen
  \bibfield  {author} {\bibinfo {author} {\bibfnamefont {M.}~\bibnamefont
  {{Yamanaka}}}, \bibinfo {author} {\bibfnamefont {K.~S.}\ \bibnamefont
  {{Kawabata}}}, \bibinfo {author} {\bibfnamefont {K.}~\bibnamefont
  {{Kinugasa}}}, \bibinfo {author} {\bibfnamefont {M.}~\bibnamefont
  {{Tanaka}}}, \bibinfo {author} {\bibfnamefont {A.}~\bibnamefont {{Imada}}},
  \bibinfo {author} {\bibfnamefont {K.}~\bibnamefont {{Maeda}}}, \bibinfo
  {author} {\bibfnamefont {K.}~\bibnamefont {{Nomoto}}}, \bibinfo {author}
  {\bibfnamefont {A.}~\bibnamefont {{Arai}}}, \bibinfo {author} {\bibfnamefont
  {S.}~\bibnamefont {{Chiyonobu}}}, \bibinfo {author} {\bibfnamefont
  {Y.}~\bibnamefont {{Fukazawa}}}, \bibinfo {author} {\bibfnamefont
  {O.}~\bibnamefont {{Hashimoto}}}, \bibinfo {author} {\bibfnamefont
  {S.}~\bibnamefont {{Honda}}}, \bibinfo {author} {\bibfnamefont
  {Y.}~\bibnamefont {{Ikejiri}}}, \bibinfo {author} {\bibfnamefont
  {R.}~\bibnamefont {{Itoh}}}, \bibinfo {author} {\bibfnamefont
  {Y.}~\bibnamefont {{Kamata}}}, \bibinfo {author} {\bibfnamefont
  {N.}~\bibnamefont {{Kawai}}}, \bibinfo {author} {\bibfnamefont
  {T.}~\bibnamefont {{Komatsu}}}, \bibinfo {author} {\bibfnamefont
  {K.}~\bibnamefont {{Konishi}}}, \bibinfo {author} {\bibfnamefont
  {D.}~\bibnamefont {{Kuroda}}}, \bibinfo {author} {\bibfnamefont
  {H.}~\bibnamefont {{Miyamoto}}}, \bibinfo {author} {\bibfnamefont
  {S.}~\bibnamefont {{Miyazaki}}}, \bibinfo {author} {\bibfnamefont
  {O.}~\bibnamefont {{Nagae}}}, \bibinfo {author} {\bibfnamefont
  {H.}~\bibnamefont {{Nakaya}}}, \bibinfo {author} {\bibfnamefont
  {T.}~\bibnamefont {{Ohsugi}}}, \bibinfo {author} {\bibfnamefont
  {T.}~\bibnamefont {{Omodaka}}}, \bibinfo {author} {\bibfnamefont
  {N.}~\bibnamefont {{Sakai}}}, \bibinfo {author} {\bibfnamefont
  {M.}~\bibnamefont {{Sasada}}}, \bibinfo {author} {\bibfnamefont
  {M.}~\bibnamefont {{Suzuki}}}, \bibinfo {author} {\bibfnamefont
  {H.}~\bibnamefont {{Taguchi}}}, \bibinfo {author} {\bibfnamefont
  {H.}~\bibnamefont {{Takahashi}}}, \bibinfo {author} {\bibfnamefont
  {H.}~\bibnamefont {{Tanaka}}}, \bibinfo {author} {\bibfnamefont
  {M.}~\bibnamefont {{Uemura}}}, \bibinfo {author} {\bibfnamefont
  {T.}~\bibnamefont {{Yamashita}}}, \bibinfo {author} {\bibfnamefont
  {K.}~\bibnamefont {{Yanagisawa}}}, \ and\ \bibinfo {author} {\bibfnamefont
  {M.}~\bibnamefont {{Yoshida}}},\ }\href {\doibase
  10.1088/0004-637X/707/2/L118} {\bibfield  {journal} {\bibinfo  {journal}
  {\apjl}\ }\textbf {\bibinfo {volume} {707}},\ \bibinfo {pages} {L118}
  (\bibinfo {year} {2009})},\ \Eprint {http://arxiv.org/abs/0908.2059}
  {arXiv:0908.2059 [astro-ph.HE]} \BibitemShut {NoStop}%
\bibitem [{\citenamefont {{Silverman}}\ \emph {et~al.}(2011)\citenamefont
  {{Silverman}}, \citenamefont {{Ganeshalingam}}, \citenamefont {{Li}},
  \citenamefont {{Filippenko}}, \citenamefont {{Miller}},\ and\ \citenamefont
  {{Poznanski}}}]{2011MNRAS.410..585S}%
  \BibitemOpen
  \bibfield  {author} {\bibinfo {author} {\bibfnamefont {J.~M.}\ \bibnamefont
  {{Silverman}}}, \bibinfo {author} {\bibfnamefont {M.}~\bibnamefont
  {{Ganeshalingam}}}, \bibinfo {author} {\bibfnamefont {W.}~\bibnamefont
  {{Li}}}, \bibinfo {author} {\bibfnamefont {A.~V.}\ \bibnamefont
  {{Filippenko}}}, \bibinfo {author} {\bibfnamefont {A.~A.}\ \bibnamefont
  {{Miller}}}, \ and\ \bibinfo {author} {\bibfnamefont {D.}~\bibnamefont
  {{Poznanski}}},\ }\href {\doibase 10.1111/j.1365-2966.2010.17474.x}
  {\bibfield  {journal} {\bibinfo  {journal} {\mnras}\ }\textbf {\bibinfo
  {volume} {410}},\ \bibinfo {pages} {585} (\bibinfo {year} {2011})},\ \Eprint
  {http://arxiv.org/abs/1003.2417} {arXiv:1003.2417 [astro-ph.HE]} \BibitemShut
  {NoStop}%
\bibitem [{\citenamefont {{Filippenko}}\ \emph {et~al.}(1992)\citenamefont
  {{Filippenko}}, \citenamefont {{Richmond}}, \citenamefont {{Branch}},
  \citenamefont {{Gaskell}}, \citenamefont {{Herbst}}, \citenamefont {{Ford}},
  \citenamefont {{Treffers}}, \citenamefont {{Matheson}}, \citenamefont {{Ho}},
  \citenamefont {{Dey}}, \citenamefont {{Sargent}}, \citenamefont {{Small}},\
  and\ \citenamefont {{van Breugel}}}]{1992AJ....104.1543F}%
  \BibitemOpen
  \bibfield  {author} {\bibinfo {author} {\bibfnamefont {A.~V.}\ \bibnamefont
  {{Filippenko}}}, \bibinfo {author} {\bibfnamefont {M.~W.}\ \bibnamefont
  {{Richmond}}}, \bibinfo {author} {\bibfnamefont {D.}~\bibnamefont
  {{Branch}}}, \bibinfo {author} {\bibfnamefont {M.}~\bibnamefont {{Gaskell}}},
  \bibinfo {author} {\bibfnamefont {W.}~\bibnamefont {{Herbst}}}, \bibinfo
  {author} {\bibfnamefont {C.~H.}\ \bibnamefont {{Ford}}}, \bibinfo {author}
  {\bibfnamefont {R.~R.}\ \bibnamefont {{Treffers}}}, \bibinfo {author}
  {\bibfnamefont {T.}~\bibnamefont {{Matheson}}}, \bibinfo {author}
  {\bibfnamefont {L.~C.}\ \bibnamefont {{Ho}}}, \bibinfo {author}
  {\bibfnamefont {A.}~\bibnamefont {{Dey}}}, \bibinfo {author} {\bibfnamefont
  {W.~L.~W.}\ \bibnamefont {{Sargent}}}, \bibinfo {author} {\bibfnamefont
  {T.~A.}\ \bibnamefont {{Small}}}, \ and\ \bibinfo {author} {\bibfnamefont
  {W.~J.~M.}\ \bibnamefont {{van Breugel}}},\ }\href {\doibase 10.1086/116339}
  {\bibfield  {journal} {\bibinfo  {journal} {\aj}\ }\textbf {\bibinfo {volume}
  {104}},\ \bibinfo {pages} {1543} (\bibinfo {year} {1992})}\BibitemShut
  {NoStop}%
\bibitem [{\citenamefont {{Turatto}}\ \emph {et~al.}(1998)\citenamefont
  {{Turatto}}, \citenamefont {{Piemonte}}, \citenamefont {{Benetti}},
  \citenamefont {{Cappellaro}}, \citenamefont {{Mazzali}}, \citenamefont
  {{Danziger}},\ and\ \citenamefont {{Patat}}}]{1998AJ....116.2431T}%
  \BibitemOpen
  \bibfield  {author} {\bibinfo {author} {\bibfnamefont {M.}~\bibnamefont
  {{Turatto}}}, \bibinfo {author} {\bibfnamefont {A.}~\bibnamefont
  {{Piemonte}}}, \bibinfo {author} {\bibfnamefont {S.}~\bibnamefont
  {{Benetti}}}, \bibinfo {author} {\bibfnamefont {E.}~\bibnamefont
  {{Cappellaro}}}, \bibinfo {author} {\bibfnamefont {P.~A.}\ \bibnamefont
  {{Mazzali}}}, \bibinfo {author} {\bibfnamefont {I.~J.}\ \bibnamefont
  {{Danziger}}}, \ and\ \bibinfo {author} {\bibfnamefont {F.}~\bibnamefont
  {{Patat}}},\ }\href {\doibase 10.1086/300622} {\bibfield  {journal} {\bibinfo
   {journal} {\aj}\ }\textbf {\bibinfo {volume} {116}},\ \bibinfo {pages}
  {2431} (\bibinfo {year} {1998})},\ \Eprint
  {http://arxiv.org/abs/astro-ph/9808013} {arXiv:astro-ph/9808013 [astro-ph]}
  \BibitemShut {NoStop}%
\bibitem [{\citenamefont {{Modjaz}}\ \emph {et~al.}(2001)\citenamefont
  {{Modjaz}}, \citenamefont {{Li}}, \citenamefont {{Filippenko}}, \citenamefont
  {{King}}, \citenamefont {{Leonard}}, \citenamefont {{Matheson}},
  \citenamefont {{Treffers}},\ and\ \citenamefont
  {{Riess}}}]{2001PASP..113..308M}%
  \BibitemOpen
  \bibfield  {author} {\bibinfo {author} {\bibfnamefont {M.}~\bibnamefont
  {{Modjaz}}}, \bibinfo {author} {\bibfnamefont {W.}~\bibnamefont {{Li}}},
  \bibinfo {author} {\bibfnamefont {A.~V.}\ \bibnamefont {{Filippenko}}},
  \bibinfo {author} {\bibfnamefont {J.~Y.}\ \bibnamefont {{King}}}, \bibinfo
  {author} {\bibfnamefont {D.~C.}\ \bibnamefont {{Leonard}}}, \bibinfo {author}
  {\bibfnamefont {T.}~\bibnamefont {{Matheson}}}, \bibinfo {author}
  {\bibfnamefont {R.~R.}\ \bibnamefont {{Treffers}}}, \ and\ \bibinfo {author}
  {\bibfnamefont {A.~G.}\ \bibnamefont {{Riess}}},\ }\href {\doibase
  10.1086/319338} {\bibfield  {journal} {\bibinfo  {journal} {\pasp}\ }\textbf
  {\bibinfo {volume} {113}},\ \bibinfo {pages} {308} (\bibinfo {year}
  {2001})},\ \Eprint {http://arxiv.org/abs/astro-ph/0008012}
  {arXiv:astro-ph/0008012 [astro-ph]} \BibitemShut {NoStop}%
\bibitem [{\citenamefont {{Taubenberger}}\ \emph {et~al.}(2008)\citenamefont
  {{Taubenberger}}, \citenamefont {{Hachinger}}, \citenamefont {{Pignata}},
  \citenamefont {{Mazzali}}, \citenamefont {{Contreras}}, \citenamefont
  {{Valenti}}, \citenamefont {{Pastorello}}, \citenamefont {{Elias-Rosa}},
  \citenamefont {{B{\"a}rnbantner}}, \citenamefont {{Barwig}}, \citenamefont
  {{Benetti}}, \citenamefont {{Dolci}}, \citenamefont {{Fliri}}, \citenamefont
  {{Folatelli}}, \citenamefont {{Freedman}}, \citenamefont {{Gonzalez}},
  \citenamefont {{Hamuy}}, \citenamefont {{Krzeminski}}, \citenamefont
  {{Morrell}}, \citenamefont {{Navasardyan}}, \citenamefont {{Persson}},
  \citenamefont {{Phillips}}, \citenamefont {{Ries}}, \citenamefont {{Roth}},
  \citenamefont {{Suntzeff}}, \citenamefont {{Turatto}},\ and\ \citenamefont
  {{Hillebrandt}}}]{2008MNRAS.385...75T}%
  \BibitemOpen
  \bibfield  {author} {\bibinfo {author} {\bibfnamefont {S.}~\bibnamefont
  {{Taubenberger}}}, \bibinfo {author} {\bibfnamefont {S.}~\bibnamefont
  {{Hachinger}}}, \bibinfo {author} {\bibfnamefont {G.}~\bibnamefont
  {{Pignata}}}, \bibinfo {author} {\bibfnamefont {P.~A.}\ \bibnamefont
  {{Mazzali}}}, \bibinfo {author} {\bibfnamefont {C.}~\bibnamefont
  {{Contreras}}}, \bibinfo {author} {\bibfnamefont {S.}~\bibnamefont
  {{Valenti}}}, \bibinfo {author} {\bibfnamefont {A.}~\bibnamefont
  {{Pastorello}}}, \bibinfo {author} {\bibfnamefont {N.}~\bibnamefont
  {{Elias-Rosa}}}, \bibinfo {author} {\bibfnamefont {O.}~\bibnamefont
  {{B{\"a}rnbantner}}}, \bibinfo {author} {\bibfnamefont {H.}~\bibnamefont
  {{Barwig}}}, \bibinfo {author} {\bibfnamefont {S.}~\bibnamefont {{Benetti}}},
  \bibinfo {author} {\bibfnamefont {M.}~\bibnamefont {{Dolci}}}, \bibinfo
  {author} {\bibfnamefont {J.}~\bibnamefont {{Fliri}}}, \bibinfo {author}
  {\bibfnamefont {G.}~\bibnamefont {{Folatelli}}}, \bibinfo {author}
  {\bibfnamefont {W.~L.}\ \bibnamefont {{Freedman}}}, \bibinfo {author}
  {\bibfnamefont {S.}~\bibnamefont {{Gonzalez}}}, \bibinfo {author}
  {\bibfnamefont {M.}~\bibnamefont {{Hamuy}}}, \bibinfo {author} {\bibfnamefont
  {W.}~\bibnamefont {{Krzeminski}}}, \bibinfo {author} {\bibfnamefont
  {N.}~\bibnamefont {{Morrell}}}, \bibinfo {author} {\bibfnamefont
  {H.}~\bibnamefont {{Navasardyan}}}, \bibinfo {author} {\bibfnamefont {S.~E.}\
  \bibnamefont {{Persson}}}, \bibinfo {author} {\bibfnamefont {M.~M.}\
  \bibnamefont {{Phillips}}}, \bibinfo {author} {\bibfnamefont
  {C.}~\bibnamefont {{Ries}}}, \bibinfo {author} {\bibfnamefont
  {M.}~\bibnamefont {{Roth}}}, \bibinfo {author} {\bibfnamefont {N.~B.}\
  \bibnamefont {{Suntzeff}}}, \bibinfo {author} {\bibfnamefont
  {M.}~\bibnamefont {{Turatto}}}, \ and\ \bibinfo {author} {\bibfnamefont
  {W.}~\bibnamefont {{Hillebrandt}}},\ }\href {\doibase
  10.1111/j.1365-2966.2008.12843.x} {\bibfield  {journal} {\bibinfo  {journal}
  {\mnras}\ }\textbf {\bibinfo {volume} {385}},\ \bibinfo {pages} {75}
  (\bibinfo {year} {2008})},\ \Eprint {http://arxiv.org/abs/0711.4548}
  {arXiv:0711.4548 [astro-ph]} \BibitemShut {NoStop}%
\bibitem [{\citenamefont {{Das}}\ and\ \citenamefont
  {{Mukhopadhyay}}(2015{\natexlab{b}})}]{2015IJMPD..2444026D}%
  \BibitemOpen
  \bibfield  {author} {\bibinfo {author} {\bibfnamefont {U.}~\bibnamefont
  {{Das}}}\ and\ \bibinfo {author} {\bibfnamefont {B.}~\bibnamefont
  {{Mukhopadhyay}}},\ }\href {\doibase 10.1142/S0218271815440265} {\bibfield
  {journal} {\bibinfo  {journal} {International Journal of Modern Physics D}\
  }\textbf {\bibinfo {volume} {24}},\ \bibinfo {eid} {1544026} (\bibinfo {year}
  {2015}{\natexlab{b}})},\ \Eprint {http://arxiv.org/abs/1506.02779}
  {arXiv:1506.02779 [astro-ph.HE]} \BibitemShut {NoStop}%
\bibitem [{\citenamefont {{Carvalho}}\ \emph {et~al.}(2017)\citenamefont
  {{Carvalho}}, \citenamefont {{Lobato}}, \citenamefont {{Moraes}},
  \citenamefont {{Arba{\~n}il}}, \citenamefont {{Otoniel}}, \citenamefont
  {{Marinho}},\ and\ \citenamefont {{Malheiro}}}]{2017EPJC...77..871C}%
  \BibitemOpen
  \bibfield  {author} {\bibinfo {author} {\bibfnamefont {G.~A.}\ \bibnamefont
  {{Carvalho}}}, \bibinfo {author} {\bibfnamefont {R.~V.}\ \bibnamefont
  {{Lobato}}}, \bibinfo {author} {\bibfnamefont {P.~H.~R.~S.}\ \bibnamefont
  {{Moraes}}}, \bibinfo {author} {\bibfnamefont {J.~D.~V.}\ \bibnamefont
  {{Arba{\~n}il}}}, \bibinfo {author} {\bibfnamefont {E.}~\bibnamefont
  {{Otoniel}}}, \bibinfo {author} {\bibfnamefont {R.~M.}\ \bibnamefont
  {{Marinho}}}, \ and\ \bibinfo {author} {\bibfnamefont {M.}~\bibnamefont
  {{Malheiro}}},\ }\href {\doibase 10.1140/epjc/s10052-017-5413-5} {\bibfield
  {journal} {\bibinfo  {journal} {European Physical Journal C}\ }\textbf
  {\bibinfo {volume} {77}},\ \bibinfo {eid} {871} (\bibinfo {year} {2017})},\
  \Eprint {http://arxiv.org/abs/1706.03596} {arXiv:1706.03596 [gr-qc]}
  \BibitemShut {NoStop}%
\bibitem [{\citenamefont {{Utami}}\ and\ \citenamefont
  {{Sulaksono}}(2021)}]{2021AIPC.2320e0029U}%
  \BibitemOpen
  \bibfield  {author} {\bibinfo {author} {\bibfnamefont {K.~M.}\ \bibnamefont
  {{Utami}}}\ and\ \bibinfo {author} {\bibfnamefont {A.}~\bibnamefont
  {{Sulaksono}}},\ }in\ \href {\doibase 10.1063/5.0038106} {\emph {\bibinfo
  {booktitle} {American Institute of Physics Conference Series}}},\ \bibinfo
  {series} {American Institute of Physics Conference Series}, Vol.\ \bibinfo
  {volume} {2320}\ (\bibinfo {year} {2021})\ p.\ \bibinfo {pages}
  {050029}\BibitemShut {NoStop}%
\bibitem [{\citenamefont {{S.~J.}}(2021)}]{2021PhRvD.103b4022S}%
  \BibitemOpen
  \bibfield  {author} {\bibinfo {author} {\bibfnamefont {G.~G.}\ \bibnamefont
  {{S.~J.}}},\ }\href {\doibase 10.1103/PhysRevD.103.024022} {\bibfield
  {journal} {\bibinfo  {journal} {\prd}\ }\textbf {\bibinfo {volume} {103}},\
  \bibinfo {eid} {024022} (\bibinfo {year} {2021})},\ \Eprint
  {http://arxiv.org/abs/2003.04304} {arXiv:2003.04304 [gr-qc]} \BibitemShut
  {NoStop}%
\bibitem [{\citenamefont {{De Felice}}\ and\ \citenamefont
  {{Tsujikawa}}(2010)}]{2010LRR....13....3D}%
  \BibitemOpen
  \bibfield  {author} {\bibinfo {author} {\bibfnamefont {A.}~\bibnamefont {{De
  Felice}}}\ and\ \bibinfo {author} {\bibfnamefont {S.}~\bibnamefont
  {{Tsujikawa}}},\ }\href {\doibase 10.12942/lrr-2010-3} {\bibfield  {journal}
  {\bibinfo  {journal} {Living Reviews in Relativity}\ }\textbf {\bibinfo
  {volume} {13}},\ \bibinfo {eid} {3} (\bibinfo {year} {2010})},\ \Eprint
  {http://arxiv.org/abs/1002.4928} {arXiv:1002.4928 [gr-qc]} \BibitemShut
  {NoStop}%
\bibitem [{\citenamefont {{Starobinsky}}(1980)}]{1980PhLB...91...99S}%
  \BibitemOpen
  \bibfield  {author} {\bibinfo {author} {\bibfnamefont {A.~A.}\ \bibnamefont
  {{Starobinsky}}},\ }\href {\doibase 10.1016/0370-2693(80)90670-X} {\bibfield
  {journal} {\bibinfo  {journal} {Physics Letters B}\ }\textbf {\bibinfo
  {volume} {91}},\ \bibinfo {pages} {99} (\bibinfo {year} {1980})}\BibitemShut
  {NoStop}%
\bibitem [{\citenamefont {{Kozak}}\ and\ \citenamefont
  {{Borowiec}}(2019)}]{2019EPJC...79..335K}%
  \BibitemOpen
  \bibfield  {author} {\bibinfo {author} {\bibfnamefont {A.}~\bibnamefont
  {{Kozak}}}\ and\ \bibinfo {author} {\bibfnamefont {A.}~\bibnamefont
  {{Borowiec}}},\ }\href {\doibase 10.1140/epjc/s10052-019-6836-y} {\bibfield
  {journal} {\bibinfo  {journal} {European Physical Journal C}\ }\textbf
  {\bibinfo {volume} {79}},\ \bibinfo {eid} {335} (\bibinfo {year} {2019})},\
  \Eprint {http://arxiv.org/abs/1808.05598} {arXiv:1808.05598 [hep-th]}
  \BibitemShut {NoStop}%
\bibitem [{\citenamefont {{Afonso}}\ \emph {et~al.}(2018)\citenamefont
  {{Afonso}}, \citenamefont {{Olmo}},\ and\ \citenamefont
  {{Rubiera-Garcia}}}]{2018PhRvD..97b1503A}%
  \BibitemOpen
  \bibfield  {author} {\bibinfo {author} {\bibfnamefont {V.~I.}\ \bibnamefont
  {{Afonso}}}, \bibinfo {author} {\bibfnamefont {G.~J.}\ \bibnamefont
  {{Olmo}}}, \ and\ \bibinfo {author} {\bibfnamefont {D.}~\bibnamefont
  {{Rubiera-Garcia}}},\ }\href {\doibase 10.1103/PhysRevD.97.021503} {\bibfield
   {journal} {\bibinfo  {journal} {\prd}\ }\textbf {\bibinfo {volume} {97}},\
  \bibinfo {eid} {021503} (\bibinfo {year} {2018})},\ \Eprint
  {http://arxiv.org/abs/1801.10406} {arXiv:1801.10406 [gr-qc]} \BibitemShut
  {NoStop}%
\bibitem [{\citenamefont {{Sotiriou}}\ and\ \citenamefont
  {{Faraoni}}(2010)}]{2010RvMP...82..451S}%
  \BibitemOpen
  \bibfield  {author} {\bibinfo {author} {\bibfnamefont {T.~P.}\ \bibnamefont
  {{Sotiriou}}}\ and\ \bibinfo {author} {\bibfnamefont {V.}~\bibnamefont
  {{Faraoni}}},\ }\href {\doibase 10.1103/RevModPhys.82.451} {\bibfield
  {journal} {\bibinfo  {journal} {Reviews of Modern Physics}\ }\textbf
  {\bibinfo {volume} {82}},\ \bibinfo {pages} {451} (\bibinfo {year} {2010})},\
  \Eprint {http://arxiv.org/abs/0805.1726} {arXiv:0805.1726 [gr-qc]}
  \BibitemShut {NoStop}%
\bibitem [{\citenamefont {{Szyd{\l}owski}}\ \emph {et~al.}(2017)\citenamefont
  {{Szyd{\l}owski}}, \citenamefont {{Stachowski}},\ and\ \citenamefont
  {{Borowiec}}}]{2017EPJC...77..603S}%
  \BibitemOpen
  \bibfield  {author} {\bibinfo {author} {\bibfnamefont {M.}~\bibnamefont
  {{Szyd{\l}owski}}}, \bibinfo {author} {\bibfnamefont {A.}~\bibnamefont
  {{Stachowski}}}, \ and\ \bibinfo {author} {\bibfnamefont {A.}~\bibnamefont
  {{Borowiec}}},\ }\href {\doibase 10.1140/epjc/s10052-017-5181-2} {\bibfield
  {journal} {\bibinfo  {journal} {European Physical Journal C}\ }\textbf
  {\bibinfo {volume} {77}},\ \bibinfo {eid} {603} (\bibinfo {year} {2017})},\
  \Eprint {http://arxiv.org/abs/1707.01948} {arXiv:1707.01948 [gr-qc]}
  \BibitemShut {NoStop}%
\bibitem [{\citenamefont {{Stachowski}}\ \emph {et~al.}(2017)\citenamefont
  {{Stachowski}}, \citenamefont {{Szyd{\l}owski}},\ and\ \citenamefont
  {{Borowiec}}}]{2017EPJC...77..406S}%
  \BibitemOpen
  \bibfield  {author} {\bibinfo {author} {\bibfnamefont {A.}~\bibnamefont
  {{Stachowski}}}, \bibinfo {author} {\bibfnamefont {M.}~\bibnamefont
  {{Szyd{\l}owski}}}, \ and\ \bibinfo {author} {\bibfnamefont {A.}~\bibnamefont
  {{Borowiec}}},\ }\href {\doibase 10.1140/epjc/s10052-017-4981-8} {\bibfield
  {journal} {\bibinfo  {journal} {European Physical Journal C}\ }\textbf
  {\bibinfo {volume} {77}},\ \bibinfo {eid} {406} (\bibinfo {year} {2017})},\
  \Eprint {http://arxiv.org/abs/1608.03196} {arXiv:1608.03196 [gr-qc]}
  \BibitemShut {NoStop}%
\bibitem [{\citenamefont {{Afonso}}\ \emph {et~al.}(2019)\citenamefont
  {{Afonso}}, \citenamefont {{Olmo}}, \citenamefont {{Orazi}},\ and\
  \citenamefont {{Rubiera-Garcia}}}]{2019PhRvD..99d4040A}%
  \BibitemOpen
  \bibfield  {author} {\bibinfo {author} {\bibfnamefont {V.~I.}\ \bibnamefont
  {{Afonso}}}, \bibinfo {author} {\bibfnamefont {G.~J.}\ \bibnamefont
  {{Olmo}}}, \bibinfo {author} {\bibfnamefont {E.}~\bibnamefont {{Orazi}}}, \
  and\ \bibinfo {author} {\bibfnamefont {D.}~\bibnamefont {{Rubiera-Garcia}}},\
  }\href {\doibase 10.1103/PhysRevD.99.044040} {\bibfield  {journal} {\bibinfo
  {journal} {\prd}\ }\textbf {\bibinfo {volume} {99}},\ \bibinfo {eid} {044040}
  (\bibinfo {year} {2019})},\ \Eprint {http://arxiv.org/abs/1810.04239}
  {arXiv:1810.04239 [gr-qc]} \BibitemShut {NoStop}%
\bibitem [{\citenamefont {{Wojnar}}(2018)}]{2018EPJC...78..421W}%
  \BibitemOpen
  \bibfield  {author} {\bibinfo {author} {\bibfnamefont {A.}~\bibnamefont
  {{Wojnar}}},\ }\href {\doibase 10.1140/epjc/s10052-018-5900-3} {\bibfield
  {journal} {\bibinfo  {journal} {European Physical Journal C}\ }\textbf
  {\bibinfo {volume} {78}},\ \bibinfo {eid} {421} (\bibinfo {year} {2018})},\
  \Eprint {http://arxiv.org/abs/1712.01943} {arXiv:1712.01943 [gr-qc]}
  \BibitemShut {NoStop}%
\bibitem [{\citenamefont {{Wojnar}}(2019)}]{2019EPJC...79...51W}%
  \BibitemOpen
  \bibfield  {author} {\bibinfo {author} {\bibfnamefont {A.}~\bibnamefont
  {{Wojnar}}},\ }\href {\doibase 10.1140/epjc/s10052-019-6555-4} {\bibfield
  {journal} {\bibinfo  {journal} {European Physical Journal C}\ }\textbf
  {\bibinfo {volume} {79}},\ \bibinfo {eid} {51} (\bibinfo {year} {2019})},\
  \Eprint {http://arxiv.org/abs/1808.04188} {arXiv:1808.04188 [gr-qc]}
  \BibitemShut {NoStop}%
\bibitem [{\citenamefont {{Mana}}\ \emph {et~al.}(2015)\citenamefont {{Mana}},
  \citenamefont {{Fatibene}},\ and\ \citenamefont
  {{Ferraris}}}]{2015JCAP...10..040M}%
  \BibitemOpen
  \bibfield  {author} {\bibinfo {author} {\bibfnamefont {A.}~\bibnamefont
  {{Mana}}}, \bibinfo {author} {\bibfnamefont {L.}~\bibnamefont {{Fatibene}}},
  \ and\ \bibinfo {author} {\bibfnamefont {M.}~\bibnamefont {{Ferraris}}},\
  }\href {\doibase 10.1088/1475-7516/2015/10/040} {\bibfield  {journal}
  {\bibinfo  {journal} {\jcap}\ }\textbf {\bibinfo {volume} {2015}},\ \bibinfo
  {eid} {040} (\bibinfo {year} {2015})},\ \Eprint
  {http://arxiv.org/abs/1505.06575} {arXiv:1505.06575 [gr-qc]} \BibitemShut
  {NoStop}%
\bibitem [{\citenamefont {{Sergyeyev}}\ and\ \citenamefont
  {{Wojnar}}(2020)}]{2020EPJC...80..313S}%
  \BibitemOpen
  \bibfield  {author} {\bibinfo {author} {\bibfnamefont {A.}~\bibnamefont
  {{Sergyeyev}}}\ and\ \bibinfo {author} {\bibfnamefont {A.}~\bibnamefont
  {{Wojnar}}},\ }\href {\doibase 10.1140/epjc/s10052-020-7876-z} {\bibfield
  {journal} {\bibinfo  {journal} {European Physical Journal C}\ }\textbf
  {\bibinfo {volume} {80}},\ \bibinfo {eid} {313} (\bibinfo {year} {2020})},\
  \Eprint {http://arxiv.org/abs/1901.10448} {arXiv:1901.10448 [gr-qc]}
  \BibitemShut {NoStop}%
\bibitem [{\citenamefont {{Chandrasekhar}}(1964)}]{1964ApJ...140..417C}%
  \BibitemOpen
  \bibfield  {author} {\bibinfo {author} {\bibfnamefont {S.}~\bibnamefont
  {{Chandrasekhar}}},\ }\href {\doibase 10.1086/147938} {\bibfield  {journal}
  {\bibinfo  {journal} {\apj}\ }\textbf {\bibinfo {volume} {140}},\ \bibinfo
  {pages} {417} (\bibinfo {year} {1964})}\BibitemShut {NoStop}%
\bibitem [{\citenamefont {{Pretel}}\ \emph {et~al.}(2021)\citenamefont
  {{Pretel}}, \citenamefont {{Jor{\'a}s}}, \citenamefont {{Reis}},\ and\
  \citenamefont {{Arba{\~n}il}}}]{2021JCAP...04..064M}%
  \BibitemOpen
  \bibfield  {author} {\bibinfo {author} {\bibfnamefont {J.~M.~Z.}\
  \bibnamefont {{Pretel}}}, \bibinfo {author} {\bibfnamefont {S.~E.}\
  \bibnamefont {{Jor{\'a}s}}}, \bibinfo {author} {\bibfnamefont {R.~R.~R.}\
  \bibnamefont {{Reis}}}, \ and\ \bibinfo {author} {\bibfnamefont {J.~D.~V.}\
  \bibnamefont {{Arba{\~n}il}}},\ }\href {\doibase
  10.1088/1475-7516/2021/04/064} {\bibfield  {journal} {\bibinfo  {journal}
  {\jcap}\ }\textbf {\bibinfo {volume} {2021}},\ \bibinfo {eid} {064} (\bibinfo
  {year} {2021})},\ \Eprint {http://arxiv.org/abs/2012.03342} {arXiv:2012.03342
  [gr-qc]} \BibitemShut {NoStop}%
\bibitem [{\citenamefont {{Pretel}}\ \emph {et~al.}(2020)\citenamefont
  {{Pretel}}, \citenamefont {{Jor{\'a}s}},\ and\ \citenamefont
  {{Reis}}}]{2020JCAP...11..048P}%
  \BibitemOpen
  \bibfield  {author} {\bibinfo {author} {\bibfnamefont {J.~M.~Z.}\
  \bibnamefont {{Pretel}}}, \bibinfo {author} {\bibfnamefont {S.~E.}\
  \bibnamefont {{Jor{\'a}s}}}, \ and\ \bibinfo {author} {\bibfnamefont
  {R.~R.~R.}\ \bibnamefont {{Reis}}},\ }\href {\doibase
  10.1088/1475-7516/2020/11/048} {\bibfield  {journal} {\bibinfo  {journal}
  {\jcap}\ }\textbf {\bibinfo {volume} {2020}},\ \bibinfo {eid} {048} (\bibinfo
  {year} {2020})},\ \Eprint {http://arxiv.org/abs/2008.00536} {arXiv:2008.00536
  [gr-qc]} \BibitemShut {NoStop}%
\bibitem [{\citenamefont {{Wojnar}}(2020{\natexlab{a}})}]{2020arXiv200100388W}%
  \BibitemOpen
  \bibfield  {author} {\bibinfo {author} {\bibfnamefont {A.}~\bibnamefont
  {{Wojnar}}},\ }\href@noop {} {\bibfield  {journal} {\bibinfo  {journal} {Acta
  Phys.Polon.Supp.}\ }\textbf {\bibinfo {volume} {13}},\ \bibinfo {eid} {249}
  (\bibinfo {year} {2020}{\natexlab{a}})},\ \Eprint
  {http://arxiv.org/abs/2001.00388} {arXiv:2001.00388 [gr-qc]} \BibitemShut
  {NoStop}%
\bibitem [{\citenamefont {{Hansen}}\ \emph {et~al.}(2004)\citenamefont
  {{Hansen}}, \citenamefont {{Kawaler}},\ and\ \citenamefont
  {{Trimble}}}]{2004sipp.book.....H}%
  \BibitemOpen
  \bibfield  {author} {\bibinfo {author} {\bibfnamefont {C.~J.}\ \bibnamefont
  {{Hansen}}}, \bibinfo {author} {\bibfnamefont {S.~D.}\ \bibnamefont
  {{Kawaler}}}, \ and\ \bibinfo {author} {\bibfnamefont {V.}~\bibnamefont
  {{Trimble}}},\ }\href@noop {} {\emph {\bibinfo {title} {{Stellar interiors :
  physical principles, structure, and evolution}}}}\ (\bibinfo {year}
  {2004})\BibitemShut {NoStop}%
\bibitem [{\citenamefont {{N{\"a}f}}\ and\ \citenamefont
  {{Jetzer}}(2010)}]{2010PhRvD..81j4003N}%
  \BibitemOpen
  \bibfield  {author} {\bibinfo {author} {\bibfnamefont {J.}~\bibnamefont
  {{N{\"a}f}}}\ and\ \bibinfo {author} {\bibfnamefont {P.}~\bibnamefont
  {{Jetzer}}},\ }\href {\doibase 10.1103/PhysRevD.81.104003} {\bibfield
  {journal} {\bibinfo  {journal} {\prd}\ }\textbf {\bibinfo {volume} {81}},\
  \bibinfo {eid} {104003} (\bibinfo {year} {2010})},\ \Eprint
  {http://arxiv.org/abs/1004.2014} {arXiv:1004.2014 [gr-qc]} \BibitemShut
  {NoStop}%
\bibitem [{\citenamefont {{Mas{\'o}-Ferrando}}\ \emph
  {et~al.}(2021)\citenamefont {{Mas{\'o}-Ferrando}}, \citenamefont
  {{Sanchis-Gual}}, \citenamefont {{Font}},\ and\ \citenamefont
  {{Olmo}}}]{2021CQGra..38s4003M}%
  \BibitemOpen
  \bibfield  {author} {\bibinfo {author} {\bibfnamefont {A.}~\bibnamefont
  {{Mas{\'o}-Ferrando}}}, \bibinfo {author} {\bibfnamefont {N.}~\bibnamefont
  {{Sanchis-Gual}}}, \bibinfo {author} {\bibfnamefont {J.~A.}\ \bibnamefont
  {{Font}}}, \ and\ \bibinfo {author} {\bibfnamefont {G.~J.}\ \bibnamefont
  {{Olmo}}},\ }\href {\doibase 10.1088/1361-6382/ac1fd0} {\bibfield  {journal}
  {\bibinfo  {journal} {Classical and Quantum Gravity}\ }\textbf {\bibinfo
  {volume} {38}},\ \bibinfo {eid} {194003} (\bibinfo {year} {2021})},\ \Eprint
  {http://arxiv.org/abs/2103.15705} {arXiv:2103.15705 [gr-qc]} \BibitemShut
  {NoStop}%
\bibitem [{\citenamefont {{Olmo}}(2005)}]{2005PhRvL..95z1102O}%
  \BibitemOpen
  \bibfield  {author} {\bibinfo {author} {\bibfnamefont {G.~J.}\ \bibnamefont
  {{Olmo}}},\ }\href {\doibase 10.1103/PhysRevLett.95.261102} {\bibfield
  {journal} {\bibinfo  {journal} {\prl}\ }\textbf {\bibinfo {volume} {95}},\
  \bibinfo {eid} {261102} (\bibinfo {year} {2005})},\ \Eprint
  {http://arxiv.org/abs/gr-qc/0505101} {arXiv:gr-qc/0505101 [astro-ph]}
  \BibitemShut {NoStop}%
\bibitem [{\citenamefont {{Avelino}}(2012)}]{2012JCAP...11..022A}%
  \BibitemOpen
  \bibfield  {author} {\bibinfo {author} {\bibfnamefont {P.~P.}\ \bibnamefont
  {{Avelino}}},\ }\href {\doibase 10.1088/1475-7516/2012/11/022} {\bibfield
  {journal} {\bibinfo  {journal} {\jcap}\ }\textbf {\bibinfo {volume} {2012}},\
  \bibinfo {eid} {022} (\bibinfo {year} {2012})},\ \Eprint
  {http://arxiv.org/abs/1207.4730} {arXiv:1207.4730 [astro-ph.CO]} \BibitemShut
  {NoStop}%
\bibitem [{\citenamefont {{Beltr{\'a}n Jim{\'e}nez}}\ \emph
  {et~al.}(2018)\citenamefont {{Beltr{\'a}n Jim{\'e}nez}}, \citenamefont
  {{Heisenberg}}, \citenamefont {{Olmo}},\ and\ \citenamefont
  {{Rubiera-Garcia}}}]{2018PhR...727....1B}%
  \BibitemOpen
  \bibfield  {author} {\bibinfo {author} {\bibfnamefont {J.}~\bibnamefont
  {{Beltr{\'a}n Jim{\'e}nez}}}, \bibinfo {author} {\bibfnamefont
  {L.}~\bibnamefont {{Heisenberg}}}, \bibinfo {author} {\bibfnamefont {G.~J.}\
  \bibnamefont {{Olmo}}}, \ and\ \bibinfo {author} {\bibfnamefont
  {D.}~\bibnamefont {{Rubiera-Garcia}}},\ }\href {\doibase
  10.1016/j.physrep.2017.11.001} {\bibfield  {journal} {\bibinfo  {journal}
  {\physrep}\ }\textbf {\bibinfo {volume} {727}},\ \bibinfo {pages} {1}
  (\bibinfo {year} {2018})},\ \Eprint {http://arxiv.org/abs/1704.03351}
  {arXiv:1704.03351 [gr-qc]} \BibitemShut {NoStop}%
\bibitem [{\citenamefont {Glendenning}(2010)}]{glendenning2010}%
  \BibitemOpen
  \bibfield  {author} {\bibinfo {author} {\bibfnamefont {N.}~\bibnamefont
  {Glendenning}},\ }\href@noop {} {\emph {\bibinfo {title} {Special and General
  Relativity: With Applications to White Dwarfs, Neutron Stars and Black
  Holes}}},\ Astronomy and Astrophysics Library\ (\bibinfo  {publisher}
  {Springer New York},\ \bibinfo {year} {2010})\BibitemShut {NoStop}%
\bibitem [{200(2000)}]{2000csnp.conf.....G}%
  \BibitemOpen
  \href@noop {} {\emph {\bibinfo {title} {Compact stars : nuclear physics}}}\
  (\bibinfo {year} {2000})\BibitemShut {NoStop}%
\bibitem [{\citenamefont {{Sharma}}\ and\ \citenamefont
  {{Mukhopadhyay}}(2021)}]{2021arXiv210501702S}%
  \BibitemOpen
  \bibfield  {author} {\bibinfo {author} {\bibfnamefont {A.}~\bibnamefont
  {{Sharma}}}\ and\ \bibinfo {author} {\bibfnamefont {B.}~\bibnamefont
  {{Mukhopadhyay}}},\ }\href@noop {} {\bibfield  {journal} {\bibinfo  {journal}
  {Scientific Voyage}\ }\textbf {\bibinfo {volume} {2}},\ \bibinfo {eid} {20}
  (\bibinfo {year} {2021})},\ \Eprint {http://arxiv.org/abs/2105.01702}
  {arXiv:2105.01702 [gr-qc]} \BibitemShut {NoStop}%
\bibitem [{\citenamefont {{Otoniel}}\ \emph {et~al.}(2019)\citenamefont
  {{Otoniel}}, \citenamefont {{Franzon}}, \citenamefont {{Carvalho}},
  \citenamefont {{Malheiro}}, \citenamefont {{Schramm}},\ and\ \citenamefont
  {{Weber}}}]{2019ApJ...879...46O}%
  \BibitemOpen
  \bibfield  {author} {\bibinfo {author} {\bibfnamefont {E.}~\bibnamefont
  {{Otoniel}}}, \bibinfo {author} {\bibfnamefont {B.}~\bibnamefont
  {{Franzon}}}, \bibinfo {author} {\bibfnamefont {G.~A.}\ \bibnamefont
  {{Carvalho}}}, \bibinfo {author} {\bibfnamefont {M.}~\bibnamefont
  {{Malheiro}}}, \bibinfo {author} {\bibfnamefont {S.}~\bibnamefont
  {{Schramm}}}, \ and\ \bibinfo {author} {\bibfnamefont {F.}~\bibnamefont
  {{Weber}}},\ }\href {\doibase 10.3847/1538-4357/ab24d1} {\bibfield  {journal}
  {\bibinfo  {journal} {\apj}\ }\textbf {\bibinfo {volume} {879}},\ \bibinfo
  {eid} {46} (\bibinfo {year} {2019})}\BibitemShut {NoStop}%
\bibitem [{\citenamefont {{Olmo}}\ \emph {et~al.}(2020)\citenamefont {{Olmo}},
  \citenamefont {{Rubiera-Garcia}},\ and\ \citenamefont
  {{Wojnar}}}]{2020PhR...876....1O}%
  \BibitemOpen
  \bibfield  {author} {\bibinfo {author} {\bibfnamefont {G.~J.}\ \bibnamefont
  {{Olmo}}}, \bibinfo {author} {\bibfnamefont {D.}~\bibnamefont
  {{Rubiera-Garcia}}}, \ and\ \bibinfo {author} {\bibfnamefont
  {A.}~\bibnamefont {{Wojnar}}},\ }\href {\doibase
  10.1016/j.physrep.2020.07.001} {\bibfield  {journal} {\bibinfo  {journal}
  {\physrep}\ }\textbf {\bibinfo {volume} {876}},\ \bibinfo {pages} {1}
  (\bibinfo {year} {2020})},\ \Eprint {http://arxiv.org/abs/1912.05202}
  {arXiv:1912.05202 [gr-qc]} \BibitemShut {NoStop}%
\bibitem [{\citenamefont {{Wojnar}}(2020{\natexlab{b}})}]{2020PhRvD.102l4045W}%
  \BibitemOpen
  \bibfield  {author} {\bibinfo {author} {\bibfnamefont {A.}~\bibnamefont
  {{Wojnar}}},\ }\href {\doibase 10.1103/PhysRevD.102.124045} {\bibfield
  {journal} {\bibinfo  {journal} {\prd}\ }\textbf {\bibinfo {volume} {102}},\
  \bibinfo {eid} {124045} (\bibinfo {year} {2020}{\natexlab{b}})},\ \Eprint
  {http://arxiv.org/abs/2007.13451} {arXiv:2007.13451 [gr-qc]} \BibitemShut
  {NoStop}%
\bibitem [{\citenamefont {{Sakstein}}(2015)}]{2015PhRvL.115t1101S}%
  \BibitemOpen
  \bibfield  {author} {\bibinfo {author} {\bibfnamefont {J.}~\bibnamefont
  {{Sakstein}}},\ }\href {\doibase 10.1103/PhysRevLett.115.201101} {\bibfield
  {journal} {\bibinfo  {journal} {\prl}\ }\textbf {\bibinfo {volume} {115}},\
  \bibinfo {eid} {201101} (\bibinfo {year} {2015})},\ \Eprint
  {http://arxiv.org/abs/1510.05964} {arXiv:1510.05964 [astro-ph.CO]}
  \BibitemShut {NoStop}%
\bibitem [{\citenamefont {{Olmo}}\ \emph {et~al.}(2019)\citenamefont {{Olmo}},
  \citenamefont {{Rubiera-Garcia}},\ and\ \citenamefont
  {{Wojnar}}}]{2019PhRvD.100d4020O}%
  \BibitemOpen
  \bibfield  {author} {\bibinfo {author} {\bibfnamefont {G.~J.}\ \bibnamefont
  {{Olmo}}}, \bibinfo {author} {\bibfnamefont {D.}~\bibnamefont
  {{Rubiera-Garcia}}}, \ and\ \bibinfo {author} {\bibfnamefont
  {A.}~\bibnamefont {{Wojnar}}},\ }\href {\doibase 10.1103/PhysRevD.100.044020}
  {\bibfield  {journal} {\bibinfo  {journal} {\prd}\ }\textbf {\bibinfo
  {volume} {100}},\ \bibinfo {eid} {044020} (\bibinfo {year} {2019})},\ \Eprint
  {http://arxiv.org/abs/1906.04629} {arXiv:1906.04629 [gr-qc]} \BibitemShut
  {NoStop}%
\bibitem [{\citenamefont {{Wojnar}}(2021{\natexlab{b}})}]{2021PhRvD.103d4037W}%
  \BibitemOpen
  \bibfield  {author} {\bibinfo {author} {\bibfnamefont {A.}~\bibnamefont
  {{Wojnar}}},\ }\href {\doibase 10.1103/PhysRevD.103.044037} {\bibfield
  {journal} {\bibinfo  {journal} {\prd}\ }\textbf {\bibinfo {volume} {103}},\
  \bibinfo {eid} {044037} (\bibinfo {year} {2021}{\natexlab{b}})},\ \Eprint
  {http://arxiv.org/abs/2009.10983} {arXiv:2009.10983 [gr-qc]} \BibitemShut
  {NoStop}%
\bibitem [{\citenamefont {{Rosyadi}}\ \emph {et~al.}(2019)\citenamefont
  {{Rosyadi}}, \citenamefont {{Sulaksono}}, \citenamefont {{Kassim}},\ and\
  \citenamefont {{Yusof}}}]{2019EPJC...79.1030R}%
  \BibitemOpen
  \bibfield  {author} {\bibinfo {author} {\bibfnamefont {A.~S.}\ \bibnamefont
  {{Rosyadi}}}, \bibinfo {author} {\bibfnamefont {A.}~\bibnamefont
  {{Sulaksono}}}, \bibinfo {author} {\bibfnamefont {H.~A.}\ \bibnamefont
  {{Kassim}}}, \ and\ \bibinfo {author} {\bibfnamefont {N.}~\bibnamefont
  {{Yusof}}},\ }\href {\doibase 10.1140/epjc/s10052-019-7560-3} {\bibfield
  {journal} {\bibinfo  {journal} {European Physical Journal C}\ }\textbf
  {\bibinfo {volume} {79}},\ \bibinfo {eid} {1030} (\bibinfo {year}
  {2019})}\BibitemShut {NoStop}%
\bibitem [{\citenamefont {{Chowdhury}}\ and\ \citenamefont
  {{Sarkar}}(2021)}]{2021JCAP...05..040C}%
  \BibitemOpen
  \bibfield  {author} {\bibinfo {author} {\bibfnamefont {S.}~\bibnamefont
  {{Chowdhury}}}\ and\ \bibinfo {author} {\bibfnamefont {T.}~\bibnamefont
  {{Sarkar}}},\ }\href {\doibase 10.1088/1475-7516/2021/05/040} {\bibfield
  {journal} {\bibinfo  {journal} {\jcap}\ }\textbf {\bibinfo {volume} {2021}},\
  \bibinfo {eid} {040} (\bibinfo {year} {2021})},\ \Eprint
  {http://arxiv.org/abs/2008.12264} {arXiv:2008.12264 [gr-qc]} \BibitemShut
  {NoStop}%
\bibitem [{\citenamefont {{Benito}}\ and\ \citenamefont
  {{Wojnar}}(2021)}]{2021PhRvD.103f4032B}%
  \BibitemOpen
  \bibfield  {author} {\bibinfo {author} {\bibfnamefont {M.}~\bibnamefont
  {{Benito}}}\ and\ \bibinfo {author} {\bibfnamefont {A.}~\bibnamefont
  {{Wojnar}}},\ }\href {\doibase 10.1103/PhysRevD.103.064032} {\bibfield
  {journal} {\bibinfo  {journal} {\prd}\ }\textbf {\bibinfo {volume} {103}},\
  \bibinfo {eid} {064032} (\bibinfo {year} {2021})},\ \Eprint
  {http://arxiv.org/abs/2101.02146} {arXiv:2101.02146 [gr-qc]} \BibitemShut
  {NoStop}%
\bibitem [{\citenamefont {{Moore}}\ \emph {et~al.}(2015)\citenamefont
  {{Moore}}, \citenamefont {{Cole}},\ and\ \citenamefont
  {{Berry}}}]{2015CQGra..32a5014M}%
  \BibitemOpen
  \bibfield  {author} {\bibinfo {author} {\bibfnamefont {C.~J.}\ \bibnamefont
  {{Moore}}}, \bibinfo {author} {\bibfnamefont {R.~H.}\ \bibnamefont {{Cole}}},
  \ and\ \bibinfo {author} {\bibfnamefont {C.~P.~L.}\ \bibnamefont {{Berry}}},\
  }\href {\doibase 10.1088/0264-9381/32/1/015014} {\bibfield  {journal}
  {\bibinfo  {journal} {Classical and Quantum Gravity}\ }\textbf {\bibinfo
  {volume} {32}},\ \bibinfo {eid} {015014} (\bibinfo {year} {2015})},\ \Eprint
  {http://arxiv.org/abs/1408.0740} {arXiv:1408.0740 [gr-qc]} \BibitemShut
  {NoStop}%
\bibitem [{\citenamefont {{Huang}}\ \emph {et~al.}(2020)\citenamefont
  {{Huang}}, \citenamefont {{Hu}}, \citenamefont {{Korol}}, \citenamefont
  {{Li}}, \citenamefont {{Liang}}, \citenamefont {{Lu}}, \citenamefont
  {{Wang}}, \citenamefont {{Yu}},\ and\ \citenamefont
  {{Mei}}}]{2020PhRvD.102f3021H}%
  \BibitemOpen
  \bibfield  {author} {\bibinfo {author} {\bibfnamefont {S.-J.}\ \bibnamefont
  {{Huang}}}, \bibinfo {author} {\bibfnamefont {Y.-M.}\ \bibnamefont {{Hu}}},
  \bibinfo {author} {\bibfnamefont {V.}~\bibnamefont {{Korol}}}, \bibinfo
  {author} {\bibfnamefont {P.-C.}\ \bibnamefont {{Li}}}, \bibinfo {author}
  {\bibfnamefont {Z.-C.}\ \bibnamefont {{Liang}}}, \bibinfo {author}
  {\bibfnamefont {Y.}~\bibnamefont {{Lu}}}, \bibinfo {author} {\bibfnamefont
  {H.-T.}\ \bibnamefont {{Wang}}}, \bibinfo {author} {\bibfnamefont
  {S.}~\bibnamefont {{Yu}}}, \ and\ \bibinfo {author} {\bibfnamefont
  {J.}~\bibnamefont {{Mei}}},\ }\href {\doibase 10.1103/PhysRevD.102.063021}
  {\bibfield  {journal} {\bibinfo  {journal} {\prd}\ }\textbf {\bibinfo
  {volume} {102}},\ \bibinfo {eid} {063021} (\bibinfo {year} {2020})},\ \Eprint
  {http://arxiv.org/abs/2005.07889} {arXiv:2005.07889 [astro-ph.HE]}
  \BibitemShut {NoStop}%
\bibitem [{\citenamefont {{Capozziello}}\ \emph {et~al.}(2014)\citenamefont
  {{Capozziello}}, \citenamefont {{Lobo}},\ and\ \citenamefont
  {{Mimoso}}}]{2014PhLB..730..280C}%
  \BibitemOpen
  \bibfield  {author} {\bibinfo {author} {\bibfnamefont {S.}~\bibnamefont
  {{Capozziello}}}, \bibinfo {author} {\bibfnamefont {F.~S.~N.}\ \bibnamefont
  {{Lobo}}}, \ and\ \bibinfo {author} {\bibfnamefont {J.~P.}\ \bibnamefont
  {{Mimoso}}},\ }\href {\doibase 10.1016/j.physletb.2014.01.066} {\bibfield
  {journal} {\bibinfo  {journal} {Physics Letters B}\ }\textbf {\bibinfo
  {volume} {730}},\ \bibinfo {pages} {280} (\bibinfo {year} {2014})},\ \Eprint
  {http://arxiv.org/abs/1312.0784} {arXiv:1312.0784 [gr-qc]} \BibitemShut
  {NoStop}%
\bibitem [{\citenamefont {{Mimoso}}\ \emph {et~al.}(2015)\citenamefont
  {{Mimoso}}, \citenamefont {{Lobo}},\ and\ \citenamefont
  {{Capozziello}}}]{2015JPhCS.600a2047M}%
  \BibitemOpen
  \bibfield  {author} {\bibinfo {author} {\bibfnamefont {J.~P.}\ \bibnamefont
  {{Mimoso}}}, \bibinfo {author} {\bibfnamefont {F.~S.~N.}\ \bibnamefont
  {{Lobo}}}, \ and\ \bibinfo {author} {\bibfnamefont {S.}~\bibnamefont
  {{Capozziello}}},\ }in\ \href {\doibase 10.1088/1742-6596/600/1/012047}
  {\emph {\bibinfo {booktitle} {Journal of Physics Conference Series}}},\
  \bibinfo {series} {Journal of Physics Conference Series}, Vol.\ \bibinfo
  {volume} {600}\ (\bibinfo {year} {2015})\ p.\ \bibinfo {pages} {012047},\
  \Eprint {http://arxiv.org/abs/1412.6670} {arXiv:1412.6670 [gr-qc]}
  \BibitemShut {NoStop}%
\bibitem [{\citenamefont {{Wojnar}}\ and\ \citenamefont
  {{Velten}}(2016)}]{2016EPJC...76..697W}%
  \BibitemOpen
  \bibfield  {author} {\bibinfo {author} {\bibfnamefont {A.}~\bibnamefont
  {{Wojnar}}}\ and\ \bibinfo {author} {\bibfnamefont {H.}~\bibnamefont
  {{Velten}}},\ }\href {\doibase 10.1140/epjc/s10052-016-4549-z} {\bibfield
  {journal} {\bibinfo  {journal} {European Physical Journal C}\ }\textbf
  {\bibinfo {volume} {76}},\ \bibinfo {eid} {697} (\bibinfo {year} {2016})},\
  \Eprint {http://arxiv.org/abs/1604.04257} {arXiv:1604.04257 [gr-qc]}
  \BibitemShut {NoStop}%
\bibitem [{\citenamefont {{Velten}}\ \emph {et~al.}(2016)\citenamefont
  {{Velten}}, \citenamefont {{Oliveira}},\ and\ \citenamefont
  {{Wojnar}}}]{2016arXiv160103000V}%
  \BibitemOpen
  \bibfield  {author} {\bibinfo {author} {\bibfnamefont {H.}~\bibnamefont
  {{Velten}}}, \bibinfo {author} {\bibfnamefont {A.~M.}\ \bibnamefont
  {{Oliveira}}}, \ and\ \bibinfo {author} {\bibfnamefont {A.}~\bibnamefont
  {{Wojnar}}},\ }\href@noop {} {\bibfield  {journal} {\bibinfo  {journal}
  {arXiv e-prints}\ ,\ \bibinfo {eid} {arXiv:1601.03000}} (\bibinfo {year}
  {2016})},\ \Eprint {http://arxiv.org/abs/1601.03000} {arXiv:1601.03000
  [astro-ph.CO]} \BibitemShut {NoStop}%
\end{thebibliography}%

\end{document}